\documentclass[preprint,12pt,authoryear]{elsarticle}

\usepackage{amssymb}

\usepackage{amsmath}
\usepackage{mathrsfs}
\DeclareMathOperator*{\argmin}{arg\,min}

\usepackage{makecell}
\usepackage{multirow}
\usepackage{subcaption}

\usepackage{array}

\usepackage{float}

\usepackage{lineno}

\begin{document}

\begin{frontmatter}

\title{Reduced-Order Surrogates for Forced Flexible Mesh Coastal-Ocean Models} 

\author[DHI,DTU]{Freja Høgholm Petersen} 
\author[DHI]{Jesper Sandvig Mariegaard}
\author[DHI]{Rocco Palmitessa}
\author[DTU]{Allan P. Engsig-Karup}

\affiliation[DHI]{organization={DHI},
            addressline={Agern Allé 5}, 
            postcode={2970}, 
            city={Hørsholm},
            country={Denmark}}
\affiliation[DTU]{organization={DTU},
            deparment={Department of Applied Mathematics and Computer Science,},
            addressline={Bygning 324}, 
            postcode={2800}, 
            city={Kgs. Lyngby},
            country={Denmark}}

\begin{abstract}
While proper orthogonal decomposition (POD)-based surrogates are widely explored for hydrodynamic applications, the use of Koopman autoencoders for real-world coastal-ocean modelling remains relatively limited. This paper introduces a flexible Koopman autoencoder formulation that incorporates meteorological forcings and boundary conditions, and systematically compares its performance against POD-based surrogates. The Koopman autoencoder employs a learned linear temporal operator in latent space, enabling eigenvalue regularization to promote temporal stability. This strategy is evaluated alongside temporal unrolling techniques for achieving stable and accurate long-term predictions.

The models are assessed on three test cases spanning distinct dynamical regimes, with prediction horizons up to one year at 30-minute temporal resolution. Across all cases, the reduced order surrogates with temporal unrolling achieve high accuracy with relative root-mean-squared-errors of 0.0068--0.14 and $R^2$-values of 0.61--0.995, where prediction errors are largest for current velocities, and smallest for water surface elevations. In two of the three cases, the Koopman Autoencoder have higher accuracy than the POD-based surrogates. Comparing to in-situ observations, the surrogate yields -0.64\% to 12\% increase in water surface elevation prediction error when compared to prediction errors of the physics-based model. These error levels, corresponding to a few centimeters, are acceptable for many practical applications, while inference speed-ups of 300--1400x enables workflows such as ensemble forecasting and long climate simulations for coastal-ocean modelling.
\end{abstract}

\begin{highlights}
\item Koopman autoencoder extended to forced coastal-ocean systems and compared to POD
\item Surrogates evaluated on three real-world hydrodynamic coastal-ocean domains
\item Temporal unrolling improves multi-step prediction accuracy
\item Koopman surrogates yield -0.64--12\% error change vs MIKE 21 compared to observations
\item Surrogates are 300--1400x faster than physics-based model
\end{highlights}


\end{frontmatter}

\section{Introduction}
Physics-based ocean modelling is a crucial tool in protecting coastal regions from increasing threats from climate change, including sea-level rise and extreme weather events. Accurate prediction of water levels and currents is critical for short-term emergency response planning and long-term coastal infrastructure design \citep{VIEIRA1993301,DeltaresEarlyWarning}. Operational forecasting systems typically rely on physics-based hydrodynamic models such as ADCIRC \citep{Luettich1992ADCIRC}, Delft3D \citep{Lesser2004Delft3D}, or MIKE 21 \citep{MIKE21documentation}. Although these models provide high-fidelity simulations, they remain computationally expensive because of the fine spatial resolution and small time steps required to capture hydrodynamic processes. Despite advances in hardware and high-performance computing, this limits the applicability to long climate scenarios and large ensemble simulations, motivating the development of equation-free machine learning based surrogate models that are capable of reproducing the dynamics at a fraction of the computational expense.

While recent advances in deep learning, including neural operators and end-to-end emulation frameworks, have demonstrated impressive performance on idealized or moderately sized domains, their application to large-scale, operational coastal-ocean models remains challenging. In particular, long-horizon stability, robustness under realistic atmospheric and boundary forcing, and interpretability remain key concerns for practical deployment. As a result, equation-free reduced-order surrogate models that explicitly leverage the structure of the underlying dynamics and allow for controlled stability properties continue to be of high relevance in operational ocean modelling.

In this study, we develop and evaluate reduced-order surrogate models for long-term prediction of two-dimensional sea-surface elevation and currents in forced coastal-ocean systems, using three realistic hydrodynamic test cases. Our aim is to achieve substantial speed-up while retaining the physics-based model's skill, when validated against observations. We systematically compare Koopman autoencoder formulations with proper orthogonal decomposition (POD) based surrogates, and we investigate measures to promote temporal stability in the latent-space dynamics.

\subsection{Related work}
Machine learning has been applied extensively to ocean modelling, ranging from early wave-height forecasting using neural networks \citep{deo_real_1998} to recent neural-operator-based digital twins for coastal modelling \citep{jiang2021digitaltwinearth}. While neural-operator-based approaches demonstrate strong potential for end-to-end emulation, reduced-order surrogates remain attractive for large-scale modelling due to their lower training cost, explicit separation of spatial and temporal components, and fast inference. 

Reduced-order models have demonstrated significant potential for accelerating coastal and estuarine simulations. \citet{nogueira_reduced_2021} employed a truncated POD-basis to compress over 40,000 grid cells to a latent space of dimension five and then used neural networks to propagate the reduced state. Their surrogate achieved forecasts in a matter of seconds compared with 1.5 hours for the original model. While highly effective, the authors point to limitations of using POD-based surrogates for convection-dominant problems, and they suggest nonlinear autoencoders for future work. More recently, \citet{rivera_casillas_neural_2025} developed a neural-operator-based surrogate (MITONet) for shallow-water dynamics, reporting 300-600x speed-ups. While MITONet resembles the auto-encoder architectures in this study, the use of neural operators in the latent space greatly increases the number of hyper-parameters to tune, which affects the predictive skills and applicability of the surrogate. However, their study of nonlinear neural autoencoders and temporal operator learning indicates potential of end-to-end training of reduced-order surrogates as opposed to the POD-based surrogates. 

Koopman operator theory provides a complementary framework in which nonlinear dynamics are represented by a linear operator acting on an infinite-dimensional observable space. Koopman-inspired approaches aim to learn the observable space in which nonlinear dynamics can be approximated by a linear operator. Convolutional Koopman autoencoders (CKAEs) have been explored for sea-surface temperature and height forecasting \citep{rice2021analyzingkoopmanapproachesphysicsinformed, brettin_learning_2025}. The latter study found that while POD generally provides stronger reconstruction accuracy, CKAEs can yield superior predictive performance and more stable latent-space propagators on a daily time-scale. Such KAE techniques remain under-explored in the context of hydrodynamic applications and hence is investigated in this study.

A critical distinction separates prior Koopman surrogate work from hydrodynamic applications: most existing formulations assume \textit{internal} dynamics, where the system evolves without external inputs. Coastal hydrodynamics, by contrast, are \textit{forced} systems driven by time-varying wind fields, atmospheric pressure, and boundary conditions e.g. from coarser global models. \citet{proctor_generalizing_2018} extended Koopman operator theory to systems with control inputs, and they establish a formulation for inputs that have their own dynamics. \citet{shi_deep_2022} introduced a deep-learning architecture that uses a separate neural network to handle control inputs in their Koopman operator. To our knowledge, forced system formulations have not yet been applied to coastal-ocean surrogate models operating under realistic meteorological forcing.

Beyond architecture, recent work has explored loss formulations that promote desirable operator properties, such as forward–backward consistency \citep{azencot_forecasting_2020}, temporal consistency \citep{Nayak2025tcKAE}, asymptotic stability constraints \citep{lortie2024forwardbackwardextendeddmdasymptotic}, or trajectory-unrolled training \citep{morton2019deepvariationalkoopmanmodels,otto2019linearly,list2024differentiabilityunrolledtrainingneural}. In this work we examine two such strategies, stability regularization and temporal unrolling, and evaluate their impact on both POD-based and Koopman-based surrogates.

\subsection{Contributions}
This study contributes to advancing the state of reduced-order surrogate modelling for coastal hydrodynamics in four ways: 1) We develop and evaluate a Koopman autoencoder surrogate formulation for forced coastal hydrodynamics, explicitly incorporating atmospheric and boundary forcing relevant to operational ocean models. 2) We provide a systematic comparison of Koopman-based and POD-based reduced-order surrogates for sea-surface elevation and depth-averaged currents to illuminate tradeoffs between performance and skill. 3) We assess the impact of stability-promoting training strategies, including eigenvalue regularization and temporal unrolling, on long-term predictive skill and robustness. 4) We demonstrate the generalizability and limitations of these approaches through application to three real-world hydrodynamic cases, highlighting practical trade-offs between model complexity, accuracy, and stability, as well as cost of implementation.

The remainder of the paper presents the surrogate formulations, and training strategies in Section \ref{sec:Methodology}, the three hydrodynamic test cases in Section \ref{sec:Data}, followed by a comparative evaluation of accuracy, stability, and speed against both physics-based simulations used for training and observational data in Section \ref{sec:results}. Section \ref{sec:discussion} discusses the results and proposes future directions followed by a conclusion in Section \ref{sec:conclusion}.

\section{Methodology}\label{sec:Methodology}
\subsection{Flowmap formulation}\label{sec:Flowmap}
The physics-based models considered in this study are numerical hydrodynamic models that solve the shallow-water equations (SWE), i.e. a set of equations derived by depth-averaging the Navier–Stokes equations, on a spatial domain $\Omega$. With a spatial discretization on an unstructured mesh $\Omega_h \approx \Omega$, the mesh element values of the model state variables (e.g. surface elevations and horizontal velocities) are collected in $\boldsymbol{x}\in \mathbb{R}^{N_x}$. Under forcings $\boldsymbol{u}\in \mathbb{R}^{N_u}$, the state evolves in discrete time according to a nonlinear temporal mapping $\mathcal{F}_{\boldsymbol{\beta}}: \mathbb{R}^{N_x}\times \mathbb{R}^{N_u} \rightarrow \mathbb{R}^{N_x}$,
\begin{align}
    \boldsymbol{x}_{k+1} = \mathcal{F}_{\boldsymbol{\beta}}(\boldsymbol{x}_k,\boldsymbol{u}_{k}),\quad k=1,...,N_T,
    \label{eq:physbased_flowmap}
\end{align}
where $k$ denotes the discrete time index of time-increments $\Delta t$, assumed constant, $N_T$ is the total number of time steps, and $\boldsymbol{\beta}$ represents the model parameters, such as bottom-friction coefficients in coastal-ocean applications. Throughout this study, the external forcing $\boldsymbol{u}$ (e.g. wind, pressure and boundary conditions) is assumed known over the prediction horizon and prescribed from external data sources. While the underlying physics-based model may use dynamic time-stepping, the formulation \eqref{eq:physbased_flowmap} corresponds to the form of the output data, where the time-step is assumed constant.

The surrogate model in this study can be described as an approximation to the flowmap, denoted $\tilde{\mathcal{F}}_{\boldsymbol{\beta}}\approx \mathcal{F}_{\boldsymbol{\beta}}$. The dimensionality of both the states and forcings is reduced through an autoencoder mapping, and the latent variables are propagated temporally, 
\begin{align}\label{eq:general_surrogate}
    \boldsymbol{x}_{k+1} \approx \Tilde{\mathcal{F}}_{\boldsymbol{\beta}}(\boldsymbol{x}_k,\boldsymbol{u}_{k},\boldsymbol{u}_{k+1}) = \boldsymbol{\Phi}_D\left( \mathcal{M}\left( \boldsymbol{\Phi}_E(\boldsymbol{x}_k), \boldsymbol{\Psi}_E(\boldsymbol{u}_{k}),\boldsymbol{\Psi}_E(\boldsymbol{u}_{k+1})\right) \right),\\\nonumber
    \quad k=1,...,N_T,
\end{align}
where $\boldsymbol{\Phi}_E: \mathbb{R}^{N_x} \rightarrow \mathbb{R}^{\tilde{N}_x}$ and $\boldsymbol{\Phi}_D: \mathbb{R}^{\tilde{N}_x} \rightarrow \mathbb{R}^{N_x}$ are the encoder and decoder for the states, with $\tilde{N}_x\ll N_x$, $\boldsymbol{\Psi}_E: \mathbb{R}^{N_u} \rightarrow \mathbb{R}^{\tilde{N}_u}, \tilde{N} \ll N$, is an encoder for the forcings, and $\mathcal{M}: \mathbb{R}^{\tilde{N}_x}\times \mathbb{R}^{\tilde{N}_u}\times \mathbb{R}^{\tilde{N}_u} \rightarrow \mathbb{R}^{\tilde{N}_x}$ is the temporal propagator in the latent space of reduced dimension. The forcings both at time $k$ and $k+1$ are used in the temporal propagator for improved accuracy, leading to the indicated size of $\mathcal{M}$. The state variables and the forcings are encoded and decoded separately, because the properties of the states and the forcings are different, posing different requirements for the encoders and decoder.

Because the surrogate uses identical inputs and outputs as the physics-based model, it can work as a direct replacement in ensemble forecasting, probabilistic studies, and long climate simulations. In the general formulation \eqref{eq:general_surrogate}, the autoencoders and the propagator can be linear or nonlinear. In the following sections, we will look into two special cases of this surrogate, namely a Koopman autoencoder and a POD-based surrogate.

\subsection{Koopman autoencoder surrogate}
Koopman operator theory is concerned with establishing a Koopman invariant subspace of infinite dimensions, where nonlinear dynamics are represented by a linear infinite-dimensional operator. This idea has inspired the use of neural networks for discovering finite-dimensional Koopman-invariant subspaces, where the dynamic is linear \citep{takeishi_learning_2017,lusch_deep_2018,azencot_forecasting_2020}. Having a globally linear latent operator as opposed to locally linearizing nonlinear dynamics allows for exact analyses in applications such as control problems \citep{proctor_generalizing_2018,kaiser_data_driven_2021}. In the current application, the advantage of a linear operator is the straight-forward analysis of asymptotic stability through computation of eigenvalues. 
Section \ref{sec:Koopman_unforced} introduces the finite-dimensional approximation of the Koopman operator, Section \ref{sec:Koopman_forced} introduces the approximation for forced systems, and Section \ref{sec:Koopman_practice} links this to the general surrogate formulation \eqref{eq:general_surrogate}.

\subsubsection{Koopman operators for autonomous systems}\label{sec:Koopman_unforced}
For a nonlinear, discrete-time dynamical system, $\boldsymbol{x}_{k+1} = \mathbf{F}(\boldsymbol{x}_{k})$, where $\boldsymbol{x}$ exists on a manifold $M\subset \mathbb{R}^{m}$, Koopman operator theory \citep{Koopman1931,mezic_spectral_2005,brunton_koopman_2016} describes an infinite-dimensional linear operator, $\mathcal{K}: \mathscr{F}\rightarrow \mathscr{F}$, that acts on a set $\mathscr{F}$ of observable function $g: M\rightarrow\mathbb{C}$,
\begin{align}\label{eq:KoopmanUpdate}
    \mathcal{K}g(\boldsymbol{x}_k) = g(\mathbf{F}(\boldsymbol{x}_k)) = g(\boldsymbol{x}_{k+1}).
\end{align}
For some dynamical systems, a finite-dimensional Koopman invariant subspace exists on which the dynamics are exactly linear and representable by the Koopman matrix, $\boldsymbol{K}$. In the general case, we can approximate a transformation from state space to the Koopman observable space, where dynamics are linear. One approach is the Dynamic Mode Decomposition (DMD) \citep{ROWLEY_2009,SCHMID_2010}, which assumes a linear approximation to $g$. Related methods, such as extended DMD (eDMD), approximate nonlinear observables \citep{WilliamsEDMD}. A growing area of research uses neural networks to discover the transformation. 

\subsubsection{Koopman operators for forced systems}\label{sec:Koopman_forced}
Koopman theory can be extended for dynamical systems with control inputs \citep{proctor_generalizing_2018}. For open-loop control, where the input is a time-varying exogenous forcing term with internal dynamic, $\boldsymbol{u}_{k+1} = F_u(\boldsymbol{u}_k)$, the observable function $g: M \times M_u \rightarrow \mathbb{C}$ acts on the state and the input. For the flowmap \eqref{eq:physbased_flowmap}, the Koopman step is
\begin{align}\label{eq:Koopmanwcontrol}
    \mathcal{K}_fg(\boldsymbol{x}_k,\boldsymbol{u}_k) = g(F(\boldsymbol{x}_k,\boldsymbol{u}_k),F_u(\boldsymbol{u}_k)) = g(\boldsymbol{x}_{k+1},\boldsymbol{u}_{k+1}).
\end{align}
Since the forcing has its own dynamic, which is assumed known, the Koopman operator for the forced system, $\mathcal{K}_f$, only propagates observables of the state. Similar to DMD, Dynamic mode decomposition with control (DMDc) approximates the Koopman operator through assuming a linear $g$ \citep{ProctorDMDc}. Another approach is to approximate the observable functions with neural networks. 

\subsubsection{Neural Koopman autoencoder}\label{sec:Koopman_practice}
In physical systems, data for the states and forcings is typically available in the form of measurements or discretized on a mesh, such that $\boldsymbol{x}\in \mathbb{R}^{N_x}$ and $\boldsymbol{u}\in \mathbb{R}^{N_u}$. Consider a vector-valued observable $\boldsymbol{g}:\mathbb{R}^{N_x}\times \mathbb{R}^{N_u} \rightarrow \mathbb{R}^{\tilde{N}}$
\begin{align*}
    \boldsymbol{g}(\boldsymbol{x},\boldsymbol{u}) = \begin{bmatrix}
        g_1(\boldsymbol{x},\boldsymbol{u})\\
        g_2(\boldsymbol{x},\boldsymbol{u})\\
        \vdots\\
        g_{\tilde{N}}(\boldsymbol{x},\boldsymbol{u})
    \end{bmatrix}.
\end{align*}
In the case where $\tilde{N}$ is small, this corresponds to the autoencoder surrogate \eqref{eq:general_surrogate}, where $(\Phi_E,\Psi_E) \equiv \boldsymbol{g}$, $\Phi_D \equiv \boldsymbol{g}^{-1}$, and $\mathcal{M}(\cdot)\equiv \boldsymbol{K}_f(\cdot)$,  where $\boldsymbol{g}^{-1}$ approximately recovers $\boldsymbol{x}$ from the latent representation. 
In \eqref{eq:general_surrogate}, the forcing encoder, $\Psi_E$, encodes two time steps of the forcings, instead of just one. Hence, the Koopman autoencoder surrogate learns the observable functions through the autoencoder structure, $(\Phi_E,\Psi_E,\Phi_D)$, which map to and from a reduced dimensional space, where the dynamics are linear described by $\boldsymbol{K}_f$. The key assumption behind autoencoders is that the data lies near a low-dimensional manifold so that it can be well represented by a few dominant structures. 

\begin{figure}[H]
    \centering
    \begin{subfigure}[t]{0.46\textwidth}
        \centering
        \includegraphics[width=1.0\textwidth]{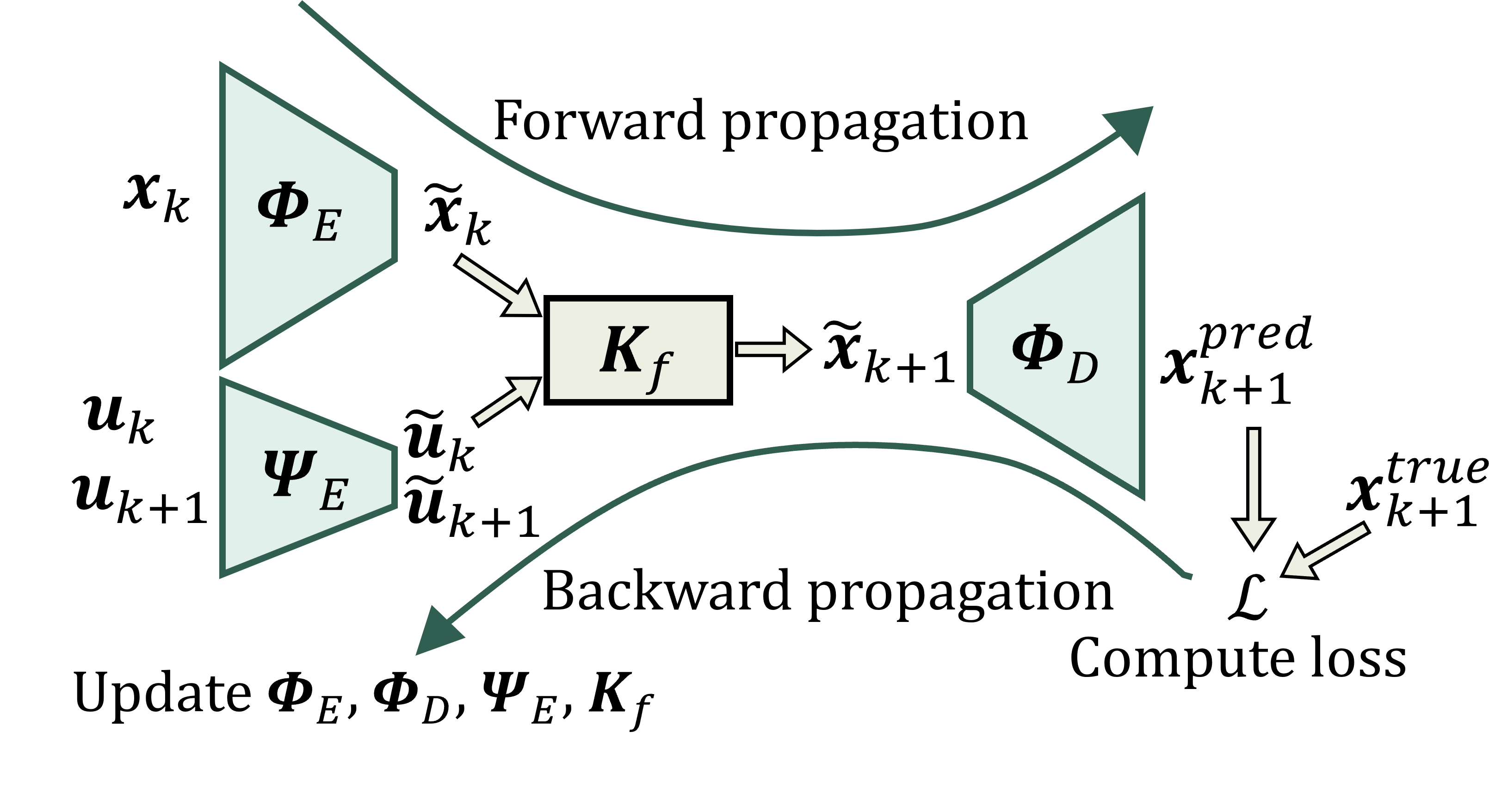}
        \caption{KAE surrogate training}\label{fig:KAEsurTrain}
    \end{subfigure}
    \begin{subfigure}[t]{0.48\textwidth}
        \centering
        \includegraphics[width=1.0\textwidth]{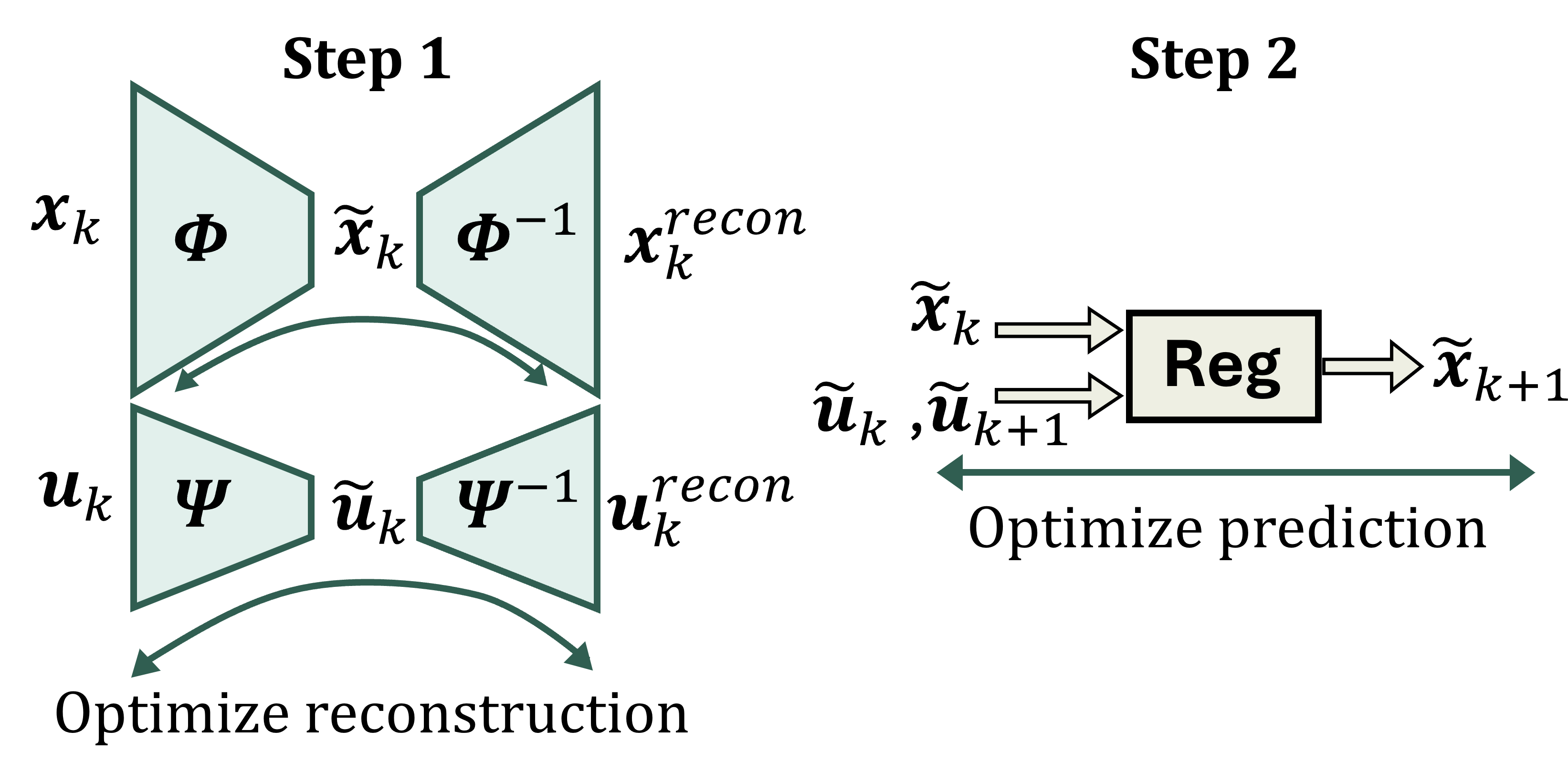}
        \caption{POD surrogate training}\label{fig:podSurTrain}
    \end{subfigure}
    \caption[Surrogate training]{(a) Training of the Koopman-based Autoencoder in an end to end setting, and (b) training of POD surrogate in two steps: 1) Optimize the reconstruction loss for encoding and decoding, 2) optimize prediction loss.}\label{fig:surrogatetraining}
\end{figure}

The encoder/decoder structure and $\boldsymbol{K}_f$ are parameterized using neural networks, and determined by solving
\begin{align}\label{eq:min_koopman}\argmin_{\boldsymbol{\Phi}_E,\boldsymbol{\Phi}_D,\boldsymbol{\Psi}_E,\boldsymbol{K}_f} \mathcal{L} &= \mathcal{L}_{pred} + \mathcal{L}_{recon} \\\nonumber &= \text{MSE}(\boldsymbol{x}^{pred}_{k+1},\boldsymbol{x}^{true}_{k+1})+\text{MSE}(\boldsymbol{x}^{recon}_{k},\boldsymbol{x}^{true}_{k}),
\end{align}
where MSE is the mean squared error defined as $\text{MSE}(A,B) = 1/(N T) \sum_{i,j}(A-B)^2_{i,j}, A,B\in\mathbb{R}^{N\times T}$, $N$ is the number of spatial elements and $T$ is the number of time steps, and 
\begin{align*}
    \boldsymbol{x}^{pred}_{k+1} &= \boldsymbol{\Phi}_D \left[ \boldsymbol{K}_f \left( \boldsymbol{\Phi}_E(\boldsymbol{x}_k), \boldsymbol{\Psi}_E(\boldsymbol{u}_{k}),\boldsymbol{\Psi}_E(\boldsymbol{u}_{k+1})\right) \right],\\
    \boldsymbol{x}^{recon}_{k} &= \boldsymbol{\Phi}_D\left( \boldsymbol{\Phi}_E(\boldsymbol{x}_k)\right)
\end{align*} 
denote the propagation through the KAE and the state reconstruction with the autoencoder, respectively. The loss-term separates the reconstruction and the prediction loss, such that the the autoencoders only reconstruct the data and avoid unintentionally encoding dynamics of the variables. The training of the encoder, decoder and Koopman matrix is done simultaneously through end-to-end training as visualized in Figure \ref{fig:KAEsurTrain}.

\subsection{POD-based surrogate model}

Proper Orthogonal Decomposition (POD), also commonly referred to as Empirical Orthogonal Functions (EOFs) in oceanography and as Principal Component Analysis (PCA) in statistics and machine learning, is a well-established technique for reduced-order modelling of geophysical flows. POD determines an orthogonal basis whose directions maximize the variance of the data, representing a time-series of snapshots of state information, upon projection onto the basis. The spatial POD modes correspond to the left singular directions of the covariance matrix, $\boldsymbol{X} \boldsymbol{X}^T$, when $\boldsymbol{X}$ contains discrete data points,
\begin{align*}
    \boldsymbol{X} = \begin{bmatrix}
        x_{1,1} & x_{1,2} & \cdots & x_{1,N_T}\\
        x_{2,1} & x_{2,2} & \cdots & x_{2,N_T}\\
        \vdots & \vdots & \ddots & \vdots \\ 
        x_{N_x,1} & x_{N_x,2} & \cdots & x_{N_x,N_T}\\
    \end{bmatrix}.
\end{align*}
A latent space can be obtained by truncating the resulting spatial POD basis, such that only the first $r$ singular vectors are used to span the latent space. Generally, the singular value decomposition (SVD) used for obtaining the matrix decomposition of the snapshot matrix $\boldsymbol{X}$ is computationally expensive, so the randomized SVD-algorithm \cite{halko2010findingstructurerandomnessprobabilistic} is used in this study for computational efficiency.

A reduced-order surrogate model for hydrodynamic simulations can be constructed using POD for compression and a regression model for temporal propagation. Figure  \ref{fig:podSurTrain} illustrates the two components of the POD-based surrogate being trained separately. Compared to the KAE, this means that the POD-model makes no assumptions about the dynamic in the latent space. 

While the POD is identical to the DMD basis, this study investigates two different regression models in the latent space, besides a linear regression (LR): a multi-layer perceptron (MLP) and a gated recurrent unit network (GRU), which is a type of RNN. The POD+MLP and POD+GRU models, allow for nonlinear temporal propagation, since the POD does not assume a latent space with linear dynamic.

\subsubsection{Temporal propagators}
The linear regression has the form
\begin{align}\label{eq:linreg}
    \Tilde{\boldsymbol{x}}_{k+1} = \boldsymbol{A} \begin{bmatrix}
       \Tilde{\boldsymbol{x}}_{k}\\
       \Tilde{\boldsymbol{u}}_{k}\\
       \Tilde{\boldsymbol{u}}_{k+1}
    \end{bmatrix},
\end{align}
where $\boldsymbol{A} \in \mathbb{R}^{\Tilde{N}_x \times (\Tilde{N}_x +2\Tilde{N}_u )}$ is the coefficient matrix of a linear flowmap formulation, and the $\Tilde{}$ denotes latent variables after POD projection. $\boldsymbol{A}$ can be determined using an ordinary least squares (OLS) fit that solves the optimization problem with the closed-form solution
\begin{align}\label{eq:OLS_fit}
    &\argmin_{\boldsymbol{A}} ||\Tilde{\boldsymbol{X}}'-\boldsymbol{AB}||^2,\nonumber\\
    \boldsymbol{A} &= (\boldsymbol{B}^T\boldsymbol{B})^{-1}\boldsymbol{B}^T\Tilde{\boldsymbol{X}}', \quad \boldsymbol{B}=\begin{bmatrix}
        \Tilde{\boldsymbol{X}}\\
        \Tilde{\boldsymbol{U}}\\
        \Tilde{\boldsymbol{U}}'
    \end{bmatrix},
\end{align}
where $\Tilde{\boldsymbol{U}} = [\Tilde{\boldsymbol{u}}_1,\Tilde{\boldsymbol{u}}_3,...,\Tilde{\boldsymbol{u}}_{N_T-1}]$, $\Tilde{\boldsymbol{X}} = [\Tilde{\boldsymbol{x}}_1,\Tilde{\boldsymbol{x}}_2,...,\Tilde{\boldsymbol{x}}_{N_T-1}]$ and $\Tilde{\boldsymbol{X}}' = [\Tilde{\boldsymbol{x}}_2,\Tilde{\boldsymbol{x}}_3,...,\Tilde{\boldsymbol{x}}_{N_T}]$. The prime denotes a time-shifted matrix, and hence, the linear regression optimizes the one-step prediction error. 

Apart from the OLS, another option is to use a simple neural network without hidden layers or activation functions. It can be fitted using backpropagation and an optimization algorithm, which aims at minimizing the MSE loss:
\begin{align}\label{eq:min_LR}
    \argmin_{\boldsymbol{A}} \mathcal{L}_{pred} = \text{MSE}(\Tilde{\boldsymbol{X}}',\boldsymbol{A} \boldsymbol{B}).
\end{align}
The purpose of this form of the linear regression is that it is possible to add different regularization terms to the loss as described in Section \ref{sec:loss_reg}.

An MLP is a feed-forward neural network with a number of hidden layers and hidden units and activation functions. A popular choice of activation is rectified linear unit (ReLU), which is used in this study. Smooth activation functions (e.g. SiLU) were tested with similar performance, but they may be preferable for physics-informed extensions. Similar to the linear regression, the MLP is trained on one-step predictions given the previous state and the next forcing values. 

The GRU \citep{cho2014propertiesneuralmachinetranslation} is a type of recurrent neural network with update and reset gates for learning long-term dependencies. Compared to the vanilla recurrent neural network, it tackles the challenge of vanishing gradients, while being a more light-weight version of the otherwise popular long-short-term-memory (LSTM) network \citep{hochreiter_long_1997}. The GRU is trained on one-step predictions, but it learns dependencies across a longer sequence rather than relying only on the immediately preceding time step.

\subsection{Loss regularization and temporal unrolling}\label{sec:loss_reg}
The LR and MLP are usually trained based on one-step predictions. However, the surrogates are supposed to be applied for long-term predictions of lengths from a few days to several years. One-step training can yield propagators whose errors compound rapidly, leading to numerical instability or drift. We investigate two complementary remedies: eigenvalue regularization to enforce stability constraints on the linear propagators, and temporal unrolling to expose training to multi-step error accumulation. This is only applied for the Koopman autoencoders, POD+LR and POD+MLP, but not the GRU, since the GRU is already using longer sequences of past timeseries.

For eigenvalue regularization, a loss-term is added to the objective functions \eqref{eq:min_koopman} and \eqref{eq:min_LR}, such that the loss is penalized if  any eigenvalue of the system matrix is larger than 1. Specifically, the linear propagator in \eqref{eq:linreg} has the form $\boldsymbol{A} = [\boldsymbol{A}_{X,X'},\boldsymbol{A}_{U,X'},\boldsymbol{A}_{U',X'}]$, where the subscript $(X,X')$ indicates that $X'$ is explained by input $X$ and similarly for $U$ on $X'$ and $U'$ on $X'$. Similarly for the Koopman approximation $K_f$. The eigenvalue regularization is then based on the eigenvalues of the square matrix $\boldsymbol{A}_{X,X'}$, describing the \textit{inner dynamics}, such that the optimization problem for the KAE and POD-based surrogates is 
\begin{align*}
\argmin_{\boldsymbol{\theta}} \mathcal{L} &= \alpha_{pred}\mathcal{L}_{pred} + \alpha_{recon}\mathcal{L}_{recon} + \alpha_{eig}\sum_{i}^{\Tilde{N_x}}ReLU(|\lambda(\boldsymbol{A}_{X,X'})_i|-1),
\end{align*}
where $\alpha_i\in[0,1]$, $i=[pred,recon,eig]$, are regularization weights, and $\theta$ are the surrogate parameters. For the POD-based surrogates, $\alpha_{recon} = 0$. For the current study, the regularization weights are set to either 0 (inactive) or 1 (active) to promote equal importance between the objectives. The eigenvalue regularization is zero when all eigenvalues within the unit circle, and it is equal to the eigenvalue magnitude when outside or on the unit circle. Eigenvalues within unit indicates asymptotic stability in the temporal propagator, and as long as the external forcing input is bounded, the dynamic of the temporal propagator will be bounded. We note that for systems with sustained oscillatory modes, such as tidal dynamics, eigenvalues near the unit circle $(|\lambda|\approx 1)$ are physically meaningful. The regularization above does not penalize eigenvalues below unity, preserving the model's ability to represent neutrally stable oscillations.

The eigenvalue regularization promotes temporal stability, but it does not guarantee accuracy, and it only works in linear temporal models. Temporal unrolling is implemented for both linear propagators and the MLP. During training, the loss is computed based on several \textit{unrolled} time-steps. That is, the temporal propagator is applied auto-regressively for a number of time-steps, and the model parameters are updated based on gradients backpropagated through all of the unrolled time-steps. For the temporal mapping $\mathcal{M}$ in \eqref{eq:general_surrogate}, this means that the prediction loss is computed as a sum of errors over the unrolled trajectory in latent space,
\begin{align*}
    \mathcal{L}_{pred} = \sum_{s=1}^{N_{TU}} \text{MSE}(\mathcal{M}^s(\tilde{\boldsymbol{x}}_{k},\tilde{\boldsymbol{u}}_{k},\tilde{\boldsymbol{u}}_{k+1}),\tilde{\boldsymbol{x}}_{k+s}),
\end{align*}
where $N_{TU}$ is the number of unrolled time steps, and $\mathcal{M}^s$ corresponds to $s$ consecutive evaluations of the temporal propagator given an initial value on the trajectory.

The assumption is that exposing the parameter optimization to longer sequences will allow it to explore more of the transient dynamics, which will reduce the data-distribution shift between training and test data.
The challenge is to balance the number of unrolled time-steps, since the backpropagated gradients can diverge with too long unrolled sequences \citep{list2024differentiabilityunrolledtrainingneural}.

\subsection{Computational complexity}\label{sec:complexity}
The POD-based surrogates and the KAE-surrogates differ largely in terms of computational complexity. The POD computation with the randomized SVD has computational complexity $\mathcal{O}(N_xN_T\log(k))$ for finding the $k$ dominant components in a matrix of size $N_x\times N_T$ \citep{halko2010findingstructurerandomnessprobabilistic}. The OLS fit \eqref{eq:OLS_fit} has complexity $\mathcal{O}(N_T \tilde{N}^2_{x,u})$ (for least squares fit with QR-factorization) where $\tilde{N}_{x,u} = \tilde{N}_x+2\tilde{N}_u$ is the latent space dimension for the forcings and states together. This work is carried out in the offline phase, and in the online phase, the compression and decompression of the state and forcings requires $\mathcal{O}(N_x\tilde{N}_x + N_u\tilde{N}_u)$ floating point operations (FLOP), and a linear operator requires $\mathcal{O}(\tilde{N}_x\tilde{N}_{x,u})$ FLOP for a one step prediction. Since $\tilde{N}_x \ll N_x$ and $\tilde{N}_u \ll N_u$, the overall complexity of a one step prediction with POD-based surrogate simplifies to $\mathcal{O}(N_x + N_u)$. The complexity is the same for a MLP-based temporal propagator, except for the offline phase being more computationally expensive due to iterative optimization of parameters. Similarly for the GRU, whose complexity depends on the number of layers and hidden dimension of the units, which are negligible in the reduced space.

For the KAE, the offline phase is an iterative process, where the optimization of parameters in the autoencoder is based on updates using loss gradients. In each iteration (epoch), the the optimization requires forward passes through the network for computation of losses and backward passes for computation of gradients. The forward pass of one training sample in a KAE with purely linear autoencoder has complexity $\mathcal{O}(N_x+N_u)$, and the backpropagation is $\mathcal{O}(N_x+N_u)$. The overall time complexity then depends on the number of training samples $N_{T}$ and the number of epochs needed for the training $N_{epoch}$, such that the overall complexity of the offline phase is $\mathcal{O}(N_{epoch}N_{T}(N_x+N_u))$, ignoring the operations in the latent space since the latent space is assumed very small relative to $N_x$ and $N_u$. For nonlinear layers, e.g. of size $(N_{xl1},N_{xl2},...,)$ in the state autoencoder, the complexity for a forward and backward pass is $\mathcal{O}(N_{epoch}N_{T}(N_xN_{xl1} + N_{xl1}N_{xl2}+...+N_u))$. Increasing layers simply lead to increased complexity. Since many epochs are often required, the computational complexity of the offline phase of the KAE is much more expensive than that of the POD. The use of batch-parallelizations can improve the wall-clock time. The online phase of the KAE for arbitrary layers in the autoencoders, corresponds to a forward pass, namely $\mathcal{O}(N_xN_{xl1} + N_{xl1}N_{xl2}+...+N_uN_{ul1}+N_{ul1}N_{ul2}+...)$ for one time-step. 

Temporal unrolling and eigenvalue regularization adds to the time complexity. For POD-based models, these are negligible compared to the compression and decompression parts, since they are performed in the latent space. The eigenvalue decomposition is $\mathcal{O}(\tilde{N}_x^3)$. For KAE, the eigenvalue decomposition is also only computed in the latent space and therefore negligible, however, the temporal unrolling has to be back-propagated through the encoders. The temporal unrolling itself is $\mathcal{O}(\tilde{N}_x\tilde{N}_{x,u})$, but the backpropagation through time treats each time-step as a layer in the neural network, meaning that the backpropagation has complexity $\mathcal{O}(N_{TU}(N_x+N_u))$ for a linear autoencoder, where $N_{TU}$ is the number of unrolled time-steps. Increasing the number of layers increases the time-complexity similar to above. Space complexity also scales with $N_{TU}$.

\subsection{Variable-separated autoencoders}
For the present application, the states and forcings consist of several variables, which have different dynamics. Specifically, the state vector $\boldsymbol{x}_k$ consists of three state variables: surface elevations ($S$), and the depth-averaged current velocity consisting of its eastward component $U$, and northward component $V$. The forcing vector $\boldsymbol{u}_k$ consists of boundary conditions ($S_{BC}$, $U_{BC}$, $V_{BC}$), that is, $S$, $U$ and $V$ imposed on element faces on the open boundary, and 10-metre wind velocity components ($W_U$, $W_V$) and mean sea level pressure ($P$). 

\begin{figure}[h]
\centering
\includegraphics[width=0.6\textwidth]{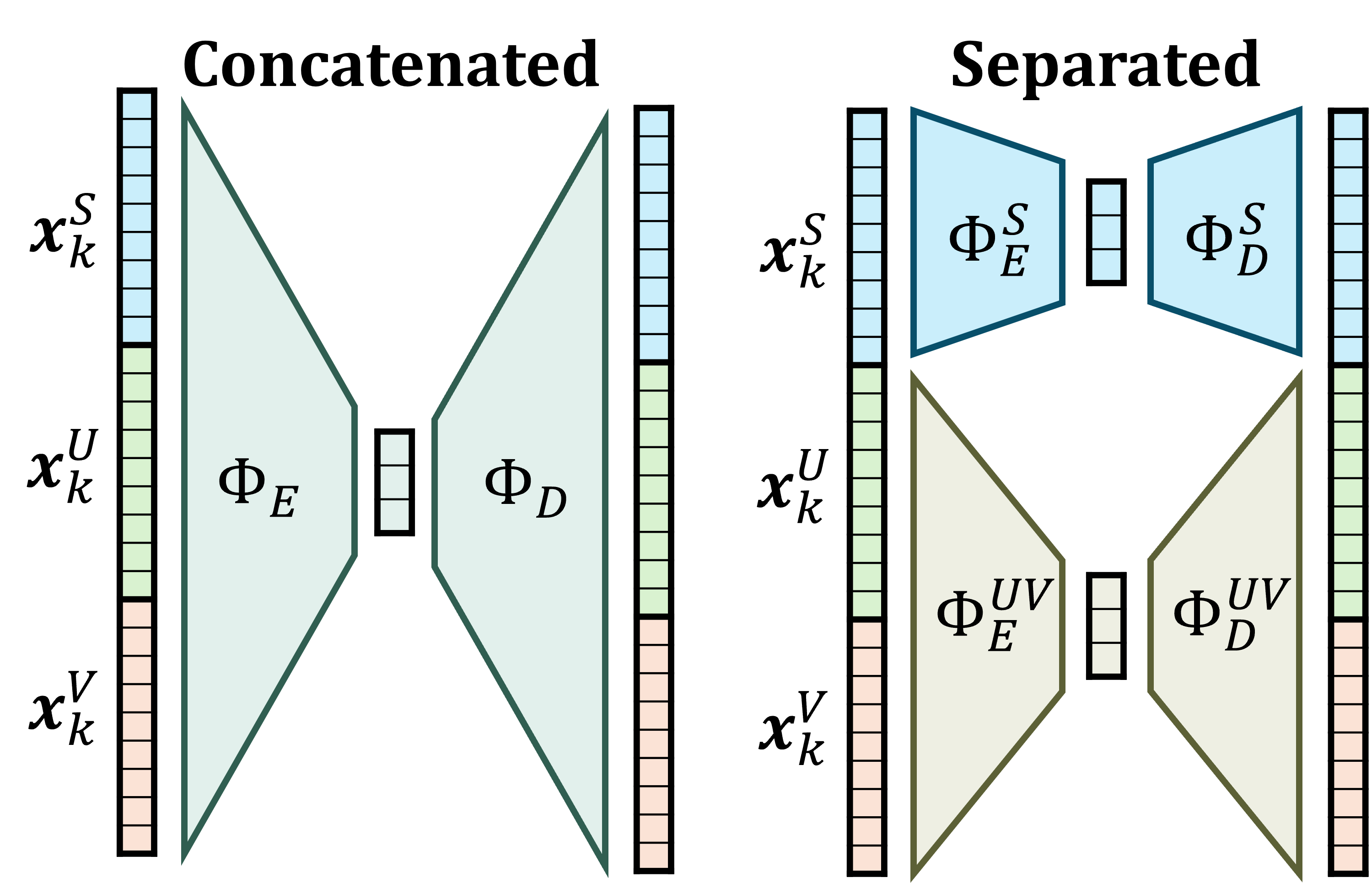}
\caption[Autoencoders]{Graphical representation of autoencoders.}\label{fig:fig1_Autoencoders}
\end{figure}

Since the dynamics of the different variables are remarkably different, e.g. $S$ is very smooth, while $U$ and $V$ are relatively non-smooth, it can make sense to introduce separate autoencoders. Figure \ref{fig:fig1_Autoencoders} shows an example of a concatenated autoencoder structure where all state variables are encoded together, and a separated architecture where $S$ is encoded separately from $U$ and $V$. Similar adjustments of the autoencoders can be made for the forcing variables. Differentiating between dynamics, allows for different choices of architectures e.g. mixing POD with neural networks, or using convolutional neural networks for gridded data. Further, the separated architecture uses fewer parameters for the same latent space dimension. Section \ref{sec:results}, will present the chosen autoencoder configurations.

\subsection{Summary and implementation of methods}
Surrogates are abbreviated in the following way: PODLR (POD with OLS linear regression), PODLRt (POD with backpropagation based training of linear regression), PODMLP (POD with multi-layer-perceptron), PODGRU (POD with gated recurrent unit) and KAE (Koopman autoencoder). Further, the suffix 'eig' and 'TU' can be added, to denote eigenvalue regularization and temporal unrolling, respectively. 

All experiments are performed in Python 3.12. The POD is implemented using the Scikit-learn \citep{scikit_learn} (version 1.6) implementation of principal component analysis. The temporal propagators for the PODLR and PODGRU are implemented using Darts \citep{darts}, which is a useful library for timeseries prediction.
The remaining propagators and neural network based autoencoders are implemented using Pytorch \citep{paszke2019pytorchimperativestylehighperformance}. All models are trained on an NVIDIA L4 GPU with 24 GB memory and 8 CPUs through the Lightning AI platform \citep{LightningAI}. The neural networks are trained using the Adam or AdamW optimizer \citep{kingma2017adam,AdamW,AdamWpytorch}. 

For each surrogate architecture, a hyper-parameter optimization is performed using Optuna \citep{optuna_2019}, which uses a TPE (Tree-structured Parzen Estimator) algorithm for automatically sampling hyper-parameters during the optimization. The optimization aims at selecting the hyper-parameters that minimizes the validation error in an autoregressive forecast. The resulting parameters are shown in \ref{app:hyperparams}. Only the best-performing models per class are reported. 

\section{Data and experimental setup}\label{sec:Data}
This section describes the simulation data used to train and evaluate surrogate models. We employ three coastal-ocean domains spanning distinct dynamical regimes, allowing assessment of surrogate generalizability across conditions encountered in operational coastal modelling.

All surrogates are trained using physics-based simulation data generated numerically from the MIKE 21 Flow Model FM \citep{MIKE21documentation}. MIKE 21 is a widely used commercial hydrodynamic model that solves the two-dimensional shallow water equations (SWE) using a cell-centered finite volume discretization on a flexible mesh. The model outputs surface elevation ($S$) and depth-averaged velocity components ($U, V$) at 30-minute intervals on all mesh elements. Forcing inputs include 10-metre wind velocity components and mean sea level pressure as well as $S$, $U$ and $V$ on lateral open boundaries.

\subsection{Test cases}\label{sec:testCases}
Table \ref{tab:case_overview} summarizes the three cases representative for coastal-ocean applications of realistic complexity. The number of mesh elements indicates the spatial dimension of the data, meaning that the surrogate will have spatial dimension $N_x = 3\cdot N_{elements}$, since there are three state variables represented on the mesh. The forcing data is summarized in Table \ref{tab:forcings_overview} in \ref{app:obs_overview}. 

\begin{figure*}
    \centering
    \begin{subfigure}[t]{0.22\textwidth}
        \centering
        \includegraphics[height=4.6cm]{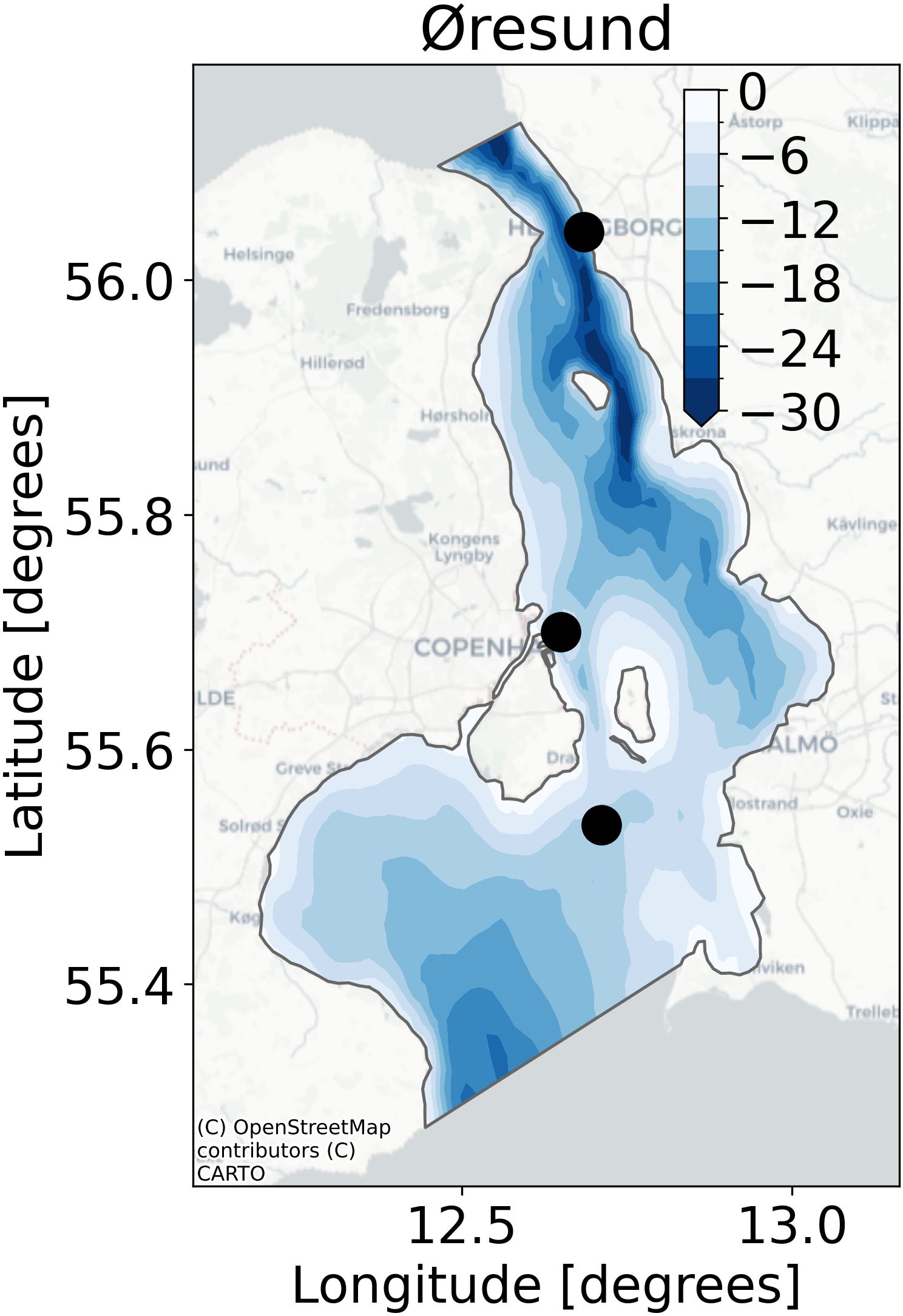}
        \caption{}\label{fig:geometry_on_map_Oresund}
    \end{subfigure}
    \begin{subfigure}[t]{0.36\textwidth}
        \centering
        \includegraphics[height=4.6cm]{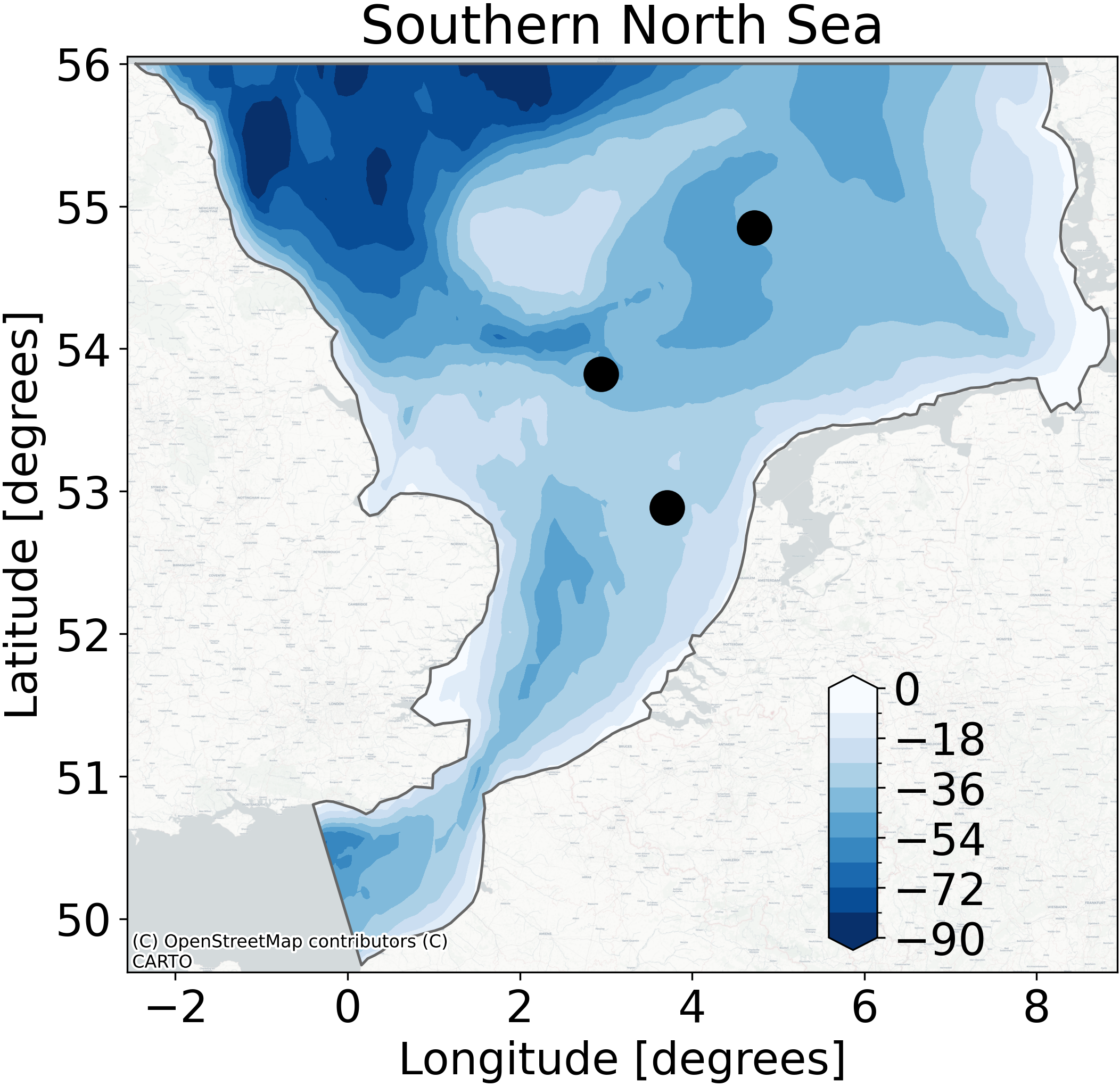}
        \caption{}
    \end{subfigure}
    \begin{subfigure}[t]{0.28\textwidth}
        \centering
        \includegraphics[height=4.6cm]{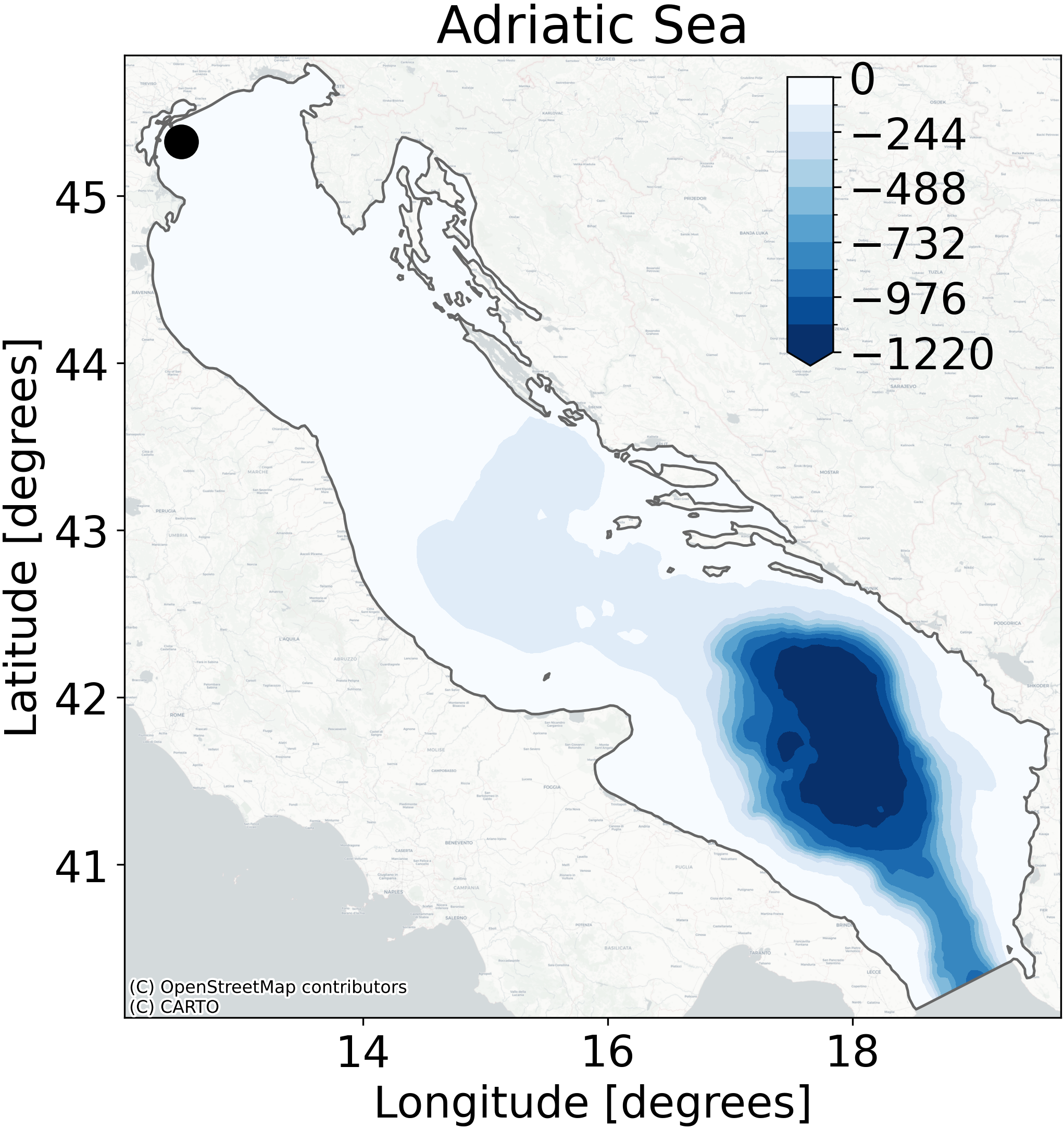}
        \caption{}
    \end{subfigure}
    \caption[Test case domain maps]{The domains and the bathymetries of the three cases, Øresund (a), Southern North Sea (b), and Adriatic Sea (c). The dots indicate the location of measuring stations.}\label{fig:geometry_on_map_all}
\end{figure*}

The bathymetries of the cases are visualized in Figure \ref{fig:geometry_on_map_all}, where the black dots mark measuring stations, which will be used for comparing the surrogate model results to observation data. An overview of the measurement stations is found in Table \ref{tab:observations_sources} in \ref{app:obs_overview}. The model setups and output data are publicly available for two of the cases, Øresund and Southern North Sea \citep{dhi_2024_14160710,dhi_2025_14929387}. 

\begin{figure}[H]
    \centering
    \begin{subfigure}[t]{1\textwidth}
        \centering
        \includegraphics[width=0.7\textwidth]{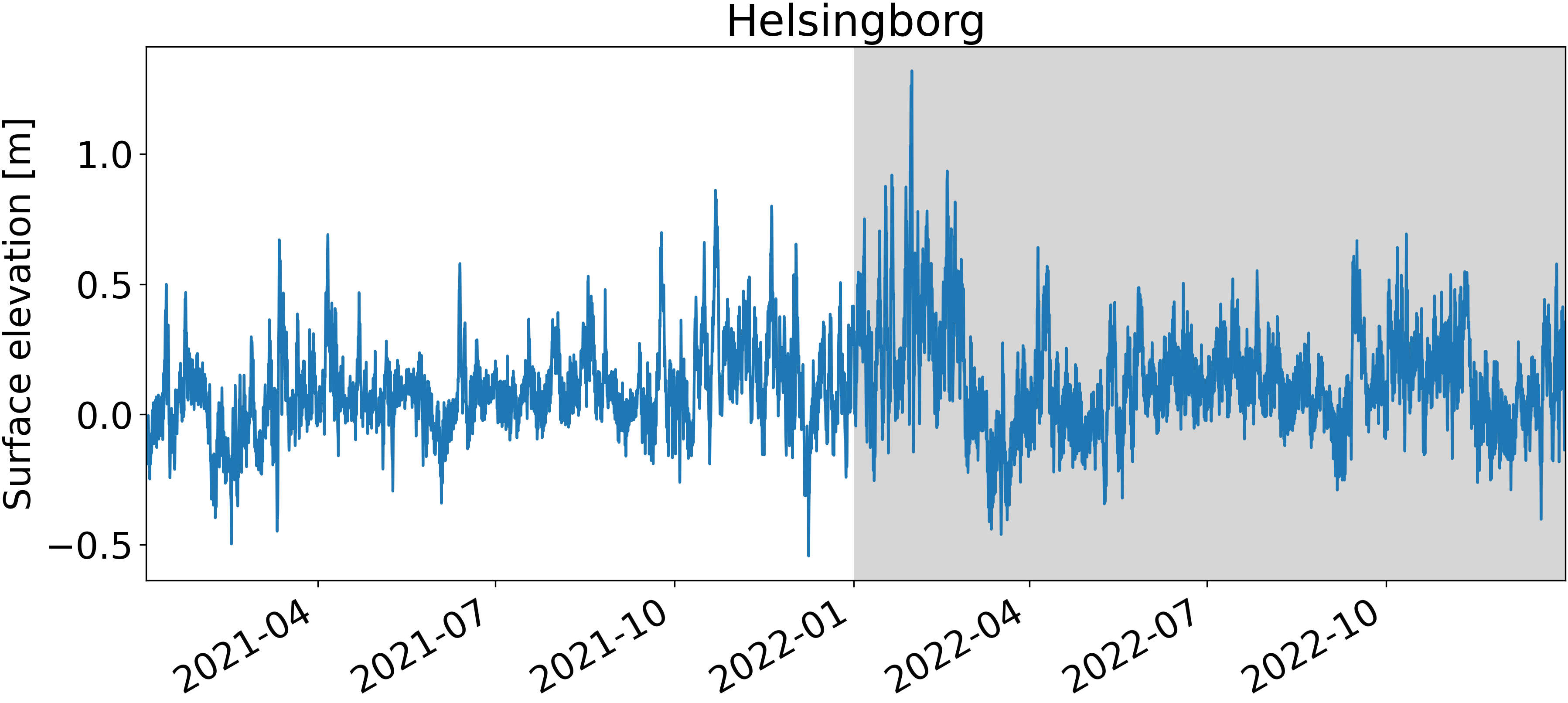}
        \caption{The Øresund data in the measuring station Helsingborg.}
    \end{subfigure}
    \begin{subfigure}[t]{1\textwidth}
        \centering
        \includegraphics[width=0.7\textwidth]{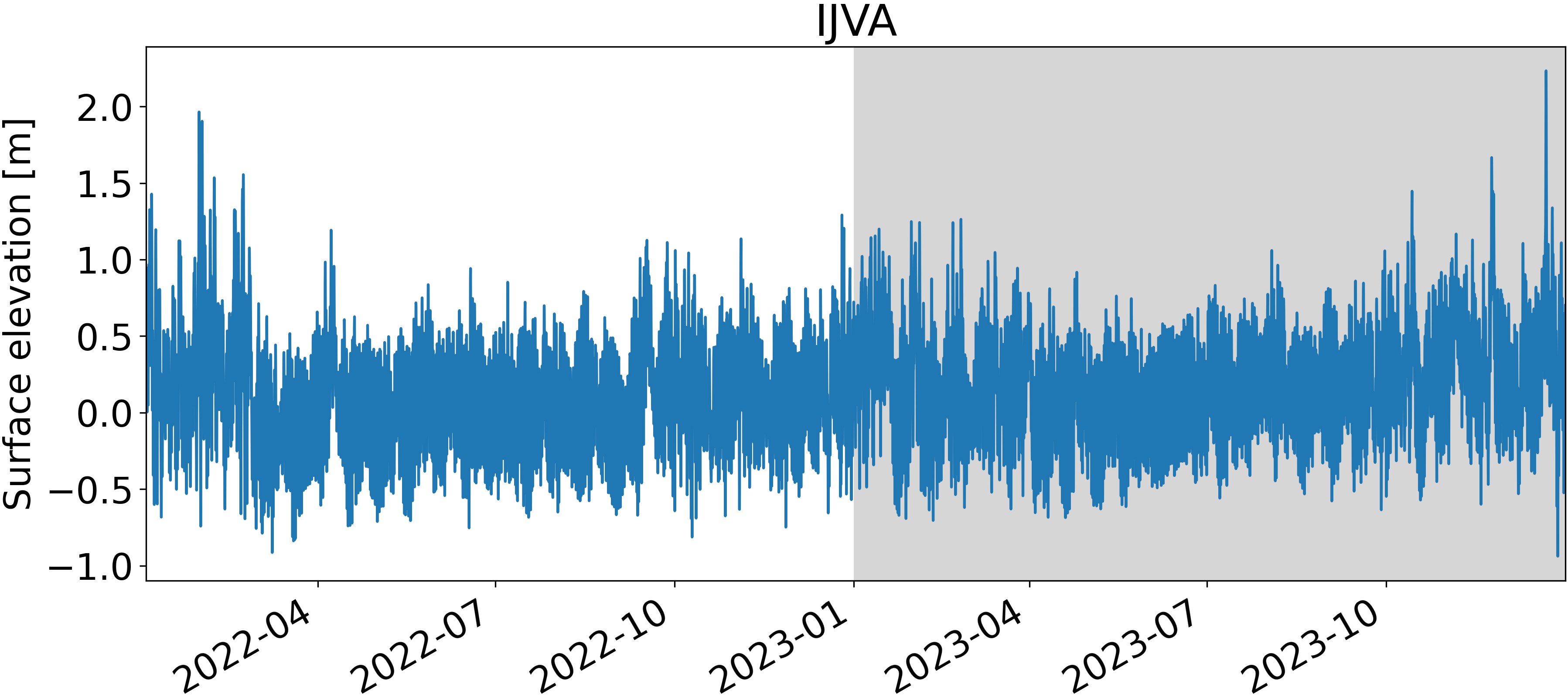}
        \caption{The Southern North Sea data in the IJVA measuring station.}
    \end{subfigure}
    \begin{subfigure}[t]{1\textwidth}
        \centering
        \includegraphics[width=0.7\textwidth]{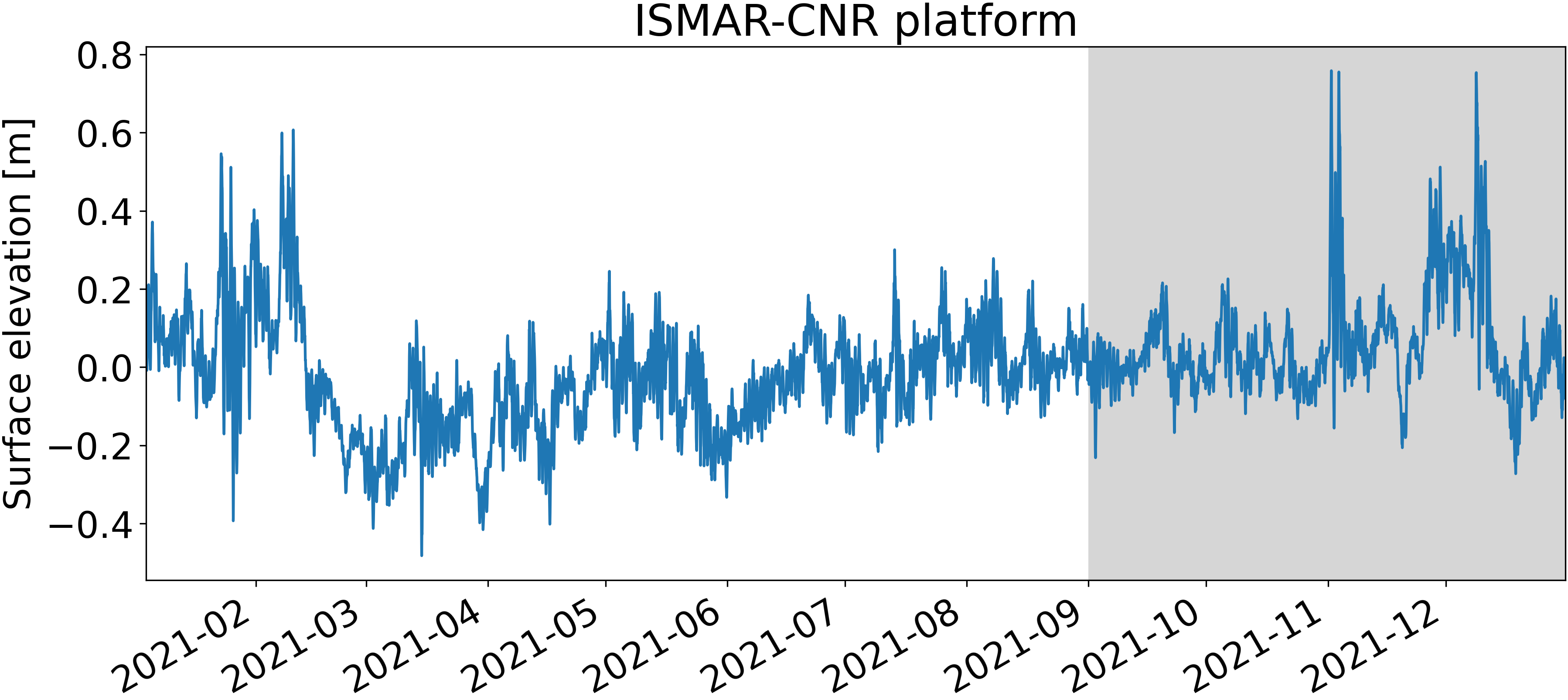}
        \caption{The Adriatic Sea data in the measuring station ISMAR-CNR platform.}
    \end{subfigure}
    \caption[Surface elevation over time]{Example timeseries of the surface elevation data used for training the surrogates.}\label{fig:EDA_single_point}
\end{figure}

The temporal train and test split is visualized in Figure \ref{fig:EDA_single_point}, where the time series of the surface elevation is shown for a single spatial point and the testing period is marked by a gray background. During training, the last 10\% of the time-steps in the training data is used as validation data for early stopping. With this choice of training and test split, the models are exposed to periods with a large variation and to calmer periods. 

\begin{table}[h]
\centering
\caption{Overview of simulation domains and time periods available for the three cases.}\label{tab:case_overview}
\begin{tabular}{c|c|c|c}
\textbf{} & \textbf{Øresund} & \textbf{Southern North Sea} & \textbf{Adriatic Sea} \\ \hline
Train period & Jan-Dec 2021 & Jan-Dec 2022 & Jan-Aug 2021\\
Test period & Jan-Dec 2022 & Jan-Dec 2023 & Sep-Dec 2021\\
Mesh elements & 3320 & 8533 & 28158 \\
Area $[km^2]$ &  2000 & 240 000 &  130 000
\end{tabular}
\end{table}

\subsubsection{Case 1: Øresund}
The Øresund strait between Denmark and Sweden constitutes the first case study \cite{dhi_2024_14160710}. The area has a minor tidal range (10-20 cm) and the dynamics are dominated by the surge introduced through the boundaries from wind- and pressure-driven flow, as well as density driven circulation and seiching in the Baltic Sea. 

\subsubsection{Case 2: Southern North Sea}
The second test case is a model of the Southern North Sea \citep{dhi_2025_14929387} spanning from the English Channel along the Southern and Eastern English coast, part of the Northern French coast, and the coasts of Belgium, the Netherlands, Northwest Germany, and Southwest Denmark. This large shelf sea exhibits strong semi-diurnal tides, with spring ranges exceeding 5 metres in the eastern English Channel. The model represents a \textit{simplified} version of the work performed by DHI for the Rijksdienst voor Ondernemend Nederland (RVO) \citep{SouthNorthSeaReport}. The domain has an open boundary to the North and an open boundary to the South in the English channel.

\subsubsection{Case 3: Adriatic Sea}
The third test case is the Adriatic Sea located between the Italian Peninsula and the Balkan Peninsula. The test case is based on the operational model used for a Decision Support System that protects Venice against flooding \citep{DHIvenice}. The semi-enclosed basin dynamics combine moderate tidal forcing, strong wind-driven circulation from the Bora and Sirocco winds, and resonant basin oscillations (seiching) that can amplify water levels at the northern end. This phenomenon causes Venice's acqua alta flooding events \citep{DHIvenice}. 

In the present simulation setup, tidal forcing has been removed to isolate the meteorologically-driven dynamics. This detided configuration focuses the surrogate learning task on the more challenging aperiodic wind and pressure response, without the regular tidal signal that aids prediction in the North Sea case. The Adriatic case thus provides the most demanding test of surrogate capability, combining complex resonant dynamics, the largest mesh, and a shorter training period.

\subsection{Evaluation metrics}
The surrogate aims at emulating the physics-based model and to evaluate its performance on the test period, the correlation coefficient, $R^2$, the root-mean-squared-error (RMSE) as well as the RMSE relative to the value range are used. Hence, the MIKE 21 simulation data is assumed the ground truth when evaluating the $R^2$ by
\begin{equation*}
    R^2 = 1-\frac{SS_{res}}{SS_{tot}},\label{eq:R2}
\end{equation*}
where the sum of squared residuals $SS_{res} = \sum_{i=1}^{N_x} { \sum_{k=1}^{N_T}({x_{k,i}-\hat{x}_{k,i})^2}}$ and the total sum of squares $SS_{tot} = \sum_{i=1}^{N_x} { \sum_{k=1}^{N_T}({x_{k,i}-\bar{x})^2}}$, with $\bar{x}$ being the mean value. Similarly for the spatio-temporal RMSE,
\begin{equation*}
    RMSE(\boldsymbol{\hat{x}},\boldsymbol{x}) = \sqrt{ \frac{1}{N_T N_x} \sum_{k=1}^{N_T} { \sum_{i=1}^{N_x}{w_i(x_{k,i}-\hat{x}_{k,i}})^2 }},\label{eq:RMSE}
\end{equation*}
where $w_i$ consists of the spatial mesh element areas.

The relative RMSE is normalized by the spatial range of values of the physics-based simulation,
\begin{align}\label{eq:relRMSE}
    Rel. \text{ }RMSE(\boldsymbol{\hat{x}},\boldsymbol{x}) &= \frac{1}{N_x}\sum_{i=1}^{N_x}\frac{\sqrt{ \frac{1}{N_T} \sum_{k=1}^{N_T} { w_i(x_{k,i}-\hat{x}_{k,i}})^2 }}{\max_{k\in{1,...,N_T}}x_{k,i}-\min_{k\in{1,...,N_T}}x_{k,i}}.
\end{align}
That is, the relative RMSE is the mean of the RMSE across time in each mesh element divided by the range of values in that specific element, in order to take local variability into account. Since the relative RMSE is an aggregated metric, the spread of errors is also evaluated by computing quantiles, median and 1 and 99th percentiles of the relative absolute errors across space and time. For computational efficiency, histograms with one million bins are used to approximate the percentiles.

At observation stations, we compare both MIKE 21 and surrogate predictions against measurements and report the percentage increase in RMSE when using the surrogate. A surrogate meeting an objective of 90\% skill retention would have RMSE increase < 10\% relative to the physics-based model. This framing recognizes that surrogate accuracy is bounded by the physics model's own structural errors; the goal is to preserve predictive skill rather than match simulation outputs exactly.

\subsection{Fairness and practical interpretation of model comparisons}
All surrogate model architectures are compared for the same test case using a fixed latent space dimensionality and are trained and evaluated on identical computational hardware. This ensures fair comparisons of accuracy, training time, and inference time across architectures. The physics-based ocean model used for data generation is run on a different machine and employs a different time-stepping scheme. Consequently, timing comparisons between the physics-based model and the surrogate models are not intended as controlled hardware benchmarks, but rather as a practical comparison reflecting typical execution environments: high-fidelity ocean simulations on specialized computing resources versus surrogate models trained and deployed on readily available machine-learning hardware.

\section{Results}\label{sec:results}
\subsection{Latent space dimension}\label{sec:latentdim}
The latent space dimension and the autoencoder composition cf. Figure \ref{fig:fig1_Autoencoders} is determined by analysis of reconstruction-prediction error trade-offs observed. Increasing dimension improves reconstruction, but it does not necessarily improve prediction. This trade-off is assessed qualitatively using the PODLR surrogate for deciding the latent space dimensions. An overview of the chosen dimensions is found in \ref{app:reconstruction_errs}.

\begin{figure*}[h]
    \centering
    \begin{subfigure}[t]{0.4\textwidth}
        \centering
        \includegraphics[height=4.5cm]{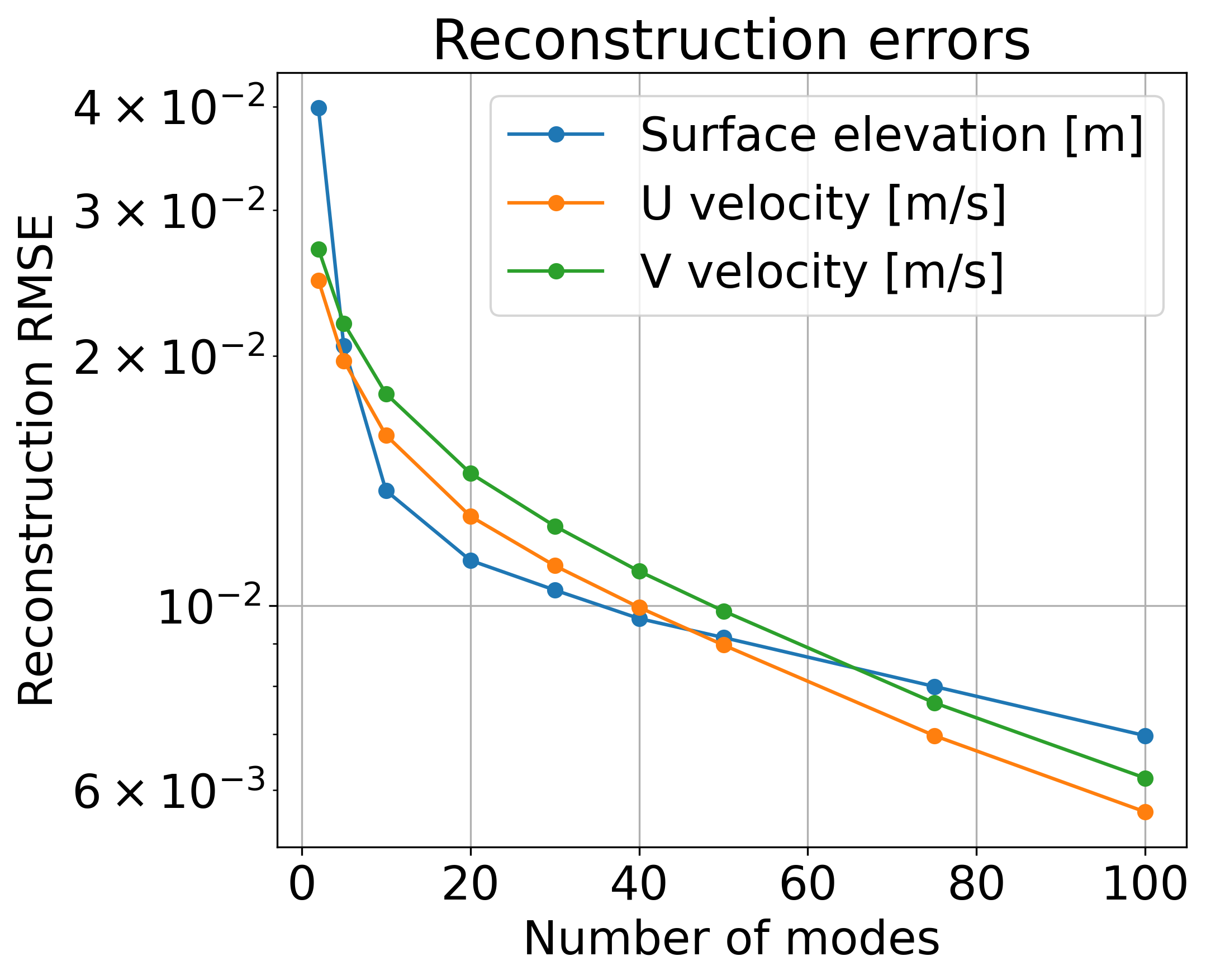}
        \caption{Concatenated: (SUV)}
    \end{subfigure}
    \begin{subfigure}[t]{0.4\textwidth}
        \centering
        \includegraphics[height=4.5cm]{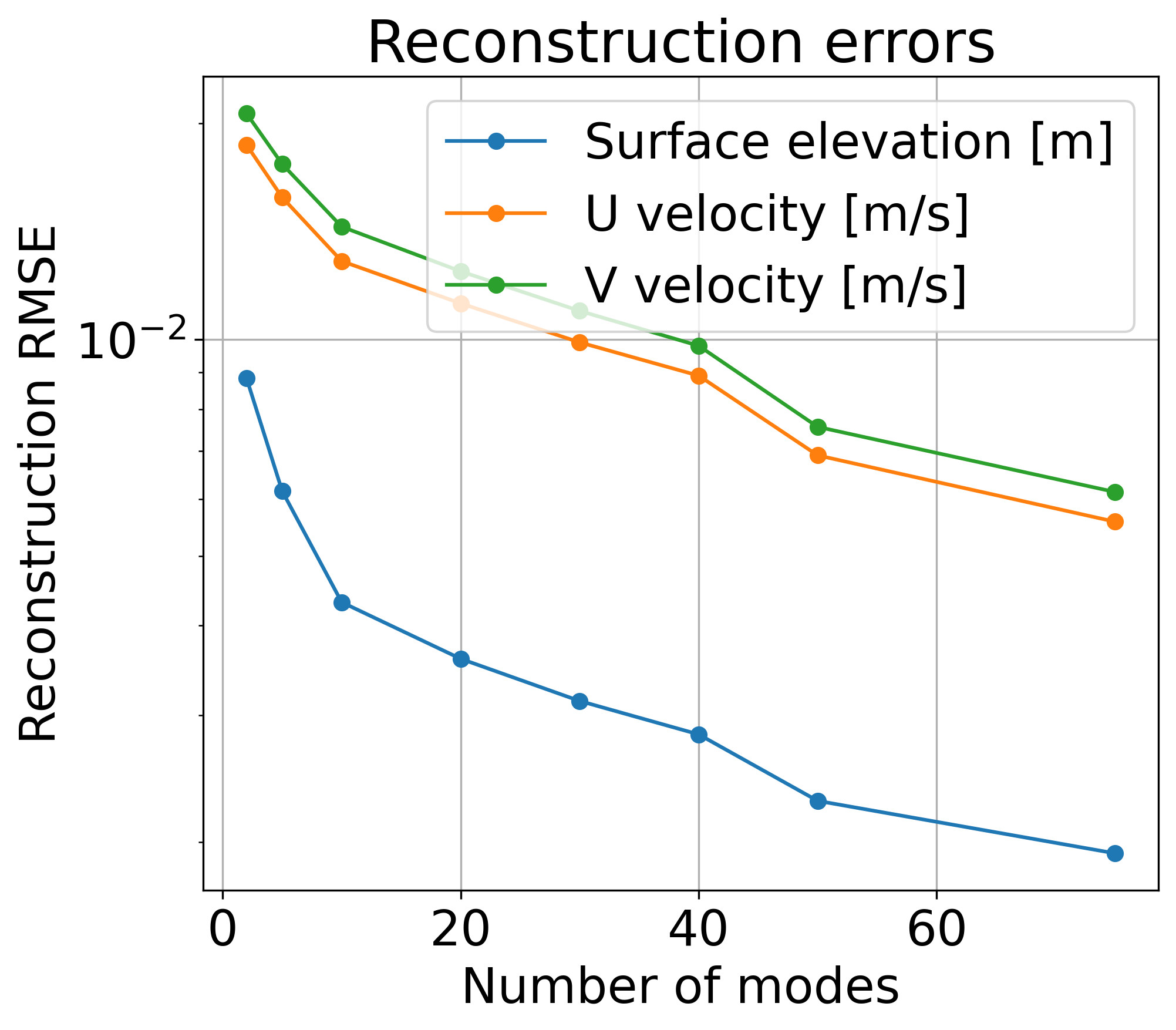}
        \caption{Separated: $S$, ($UV$)}
    \end{subfigure}
    \caption[Prediction RMSE vs. latent dimension]{Reconstruction RMSEs for POD in the Adriatic test case.}\label{fig:Adr_recon_plots}
\end{figure*}

The best reconstruction is achieved with a concatenated state autoencoder structure for Øresund, and a separated autoencoder for Adriatic Sea and Southern North Sea. The reason is evident in Figure \ref{fig:Adr_recon_plots}, which shows the reconstruction errors of the state variables in the Adriatic Sea case with a concatenated POD on the left and a separated POD autoencoder on the right, where $U$ and $V$ are separated from $S$. In the concatenated POD, the reconstruction error of $S$ follows that of $U$ and $V$, while in the separated POD, the reconstruction of $S$ is very accurate even for few modes. Based on this specific figure, it is decided to use 10 modes for $S$ and 50 modes for ($U,V$), yielding reconstruction errors of around $4\cdot 10^{-3}$ and $7\cdot 10^{-3}$ for the two groups, respectively. In comparison, a concatenated POD with latent dimension of 60 would yield errors around $7\cdot 10^{-3}$ for all variables. Similar conclusions are drawn for the reconstruction errors of the forcings, where the autoencoder is concatenated in the Øresund case, and the BCs and wind-related forcings are separated in Adriatic and Southern North Sea. 

\begin{figure*}[h]
    \centering
    \begin{subfigure}[t]{0.4\textwidth}
        \centering
        \includegraphics[height=4cm]{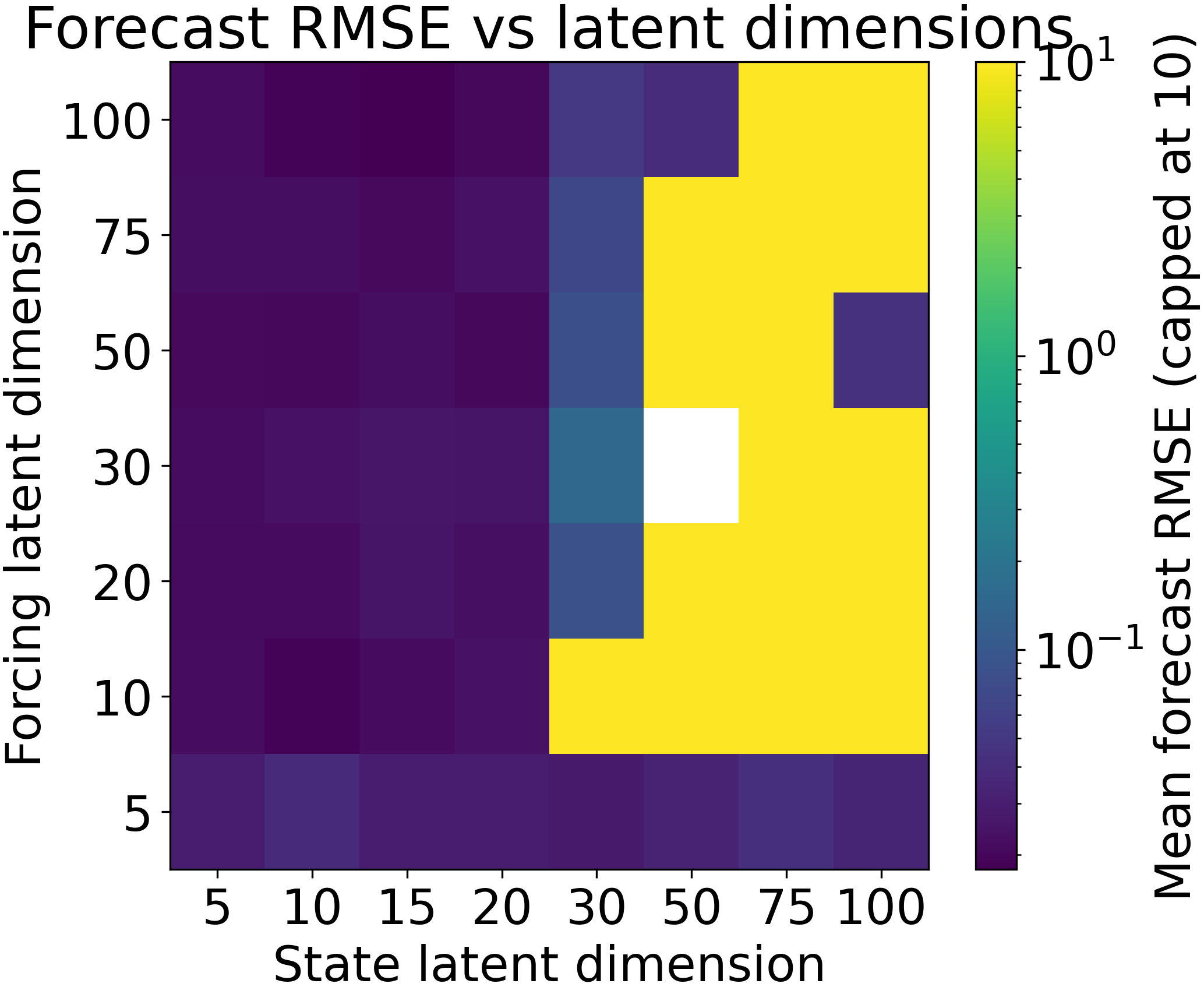}
        \caption{PODLR}
    \end{subfigure}
    \begin{subfigure}[t]{0.4\textwidth}
        \centering
        \includegraphics[height=4cm]{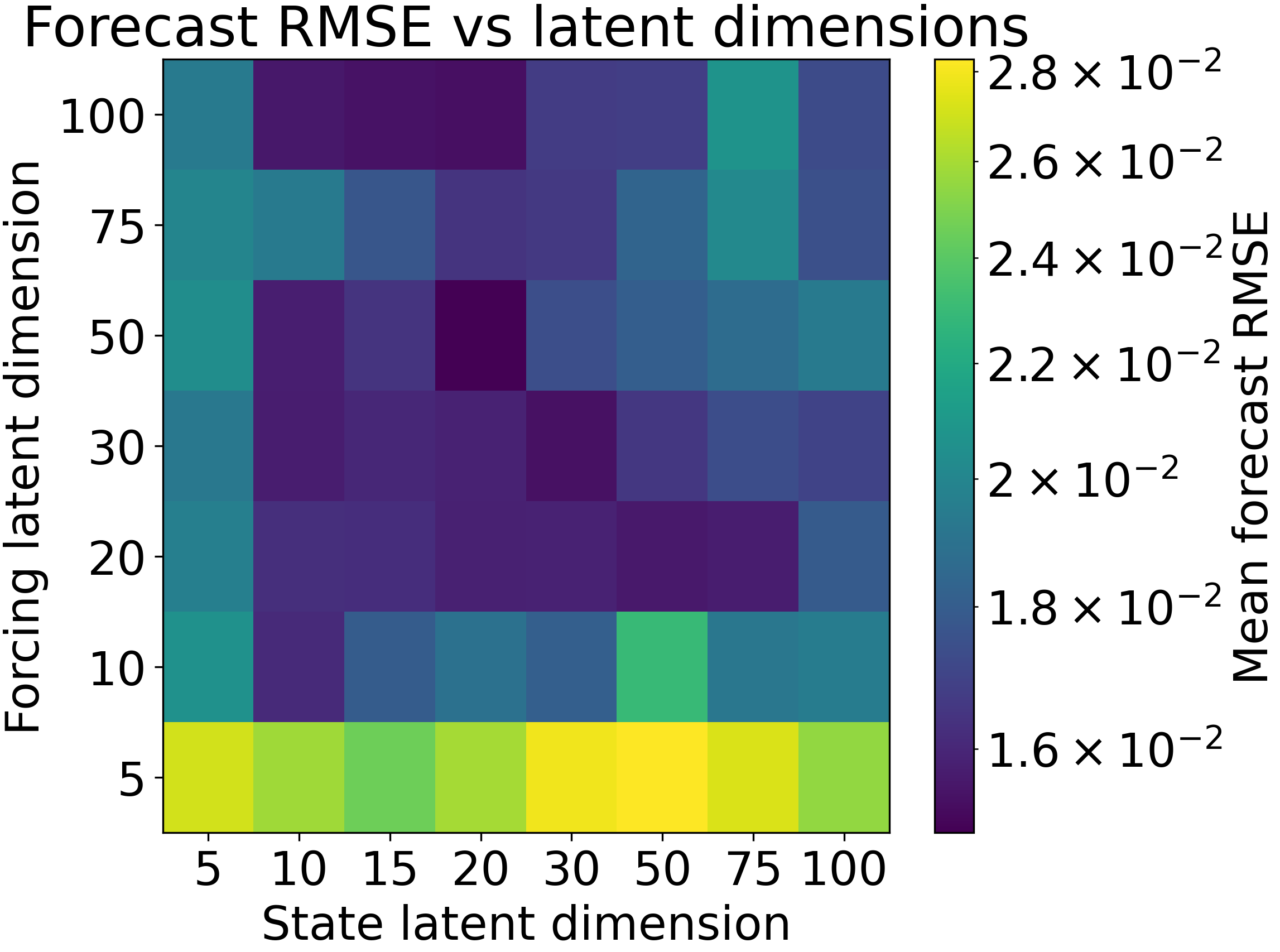}
        \caption{PODLRt (TU)}
    \end{subfigure}
    \caption[Prediction RMSE vs. latent dimension]{Prediction RMSE in a validation period for the Øresund case. }\label{fig:Øresund_latent_dim_rmse_heat}
\end{figure*}

For analyzing the effect on latent space dimension on the prediction error, Figure \ref{fig:Øresund_latent_dim_rmse_heat} shows the prediction RMSE in a validation period for the Øresund test case for different dimensions of the state and forcing latent spaces. The leftmost figure shows the PODLR model, and the rightmost is the PODLRt (TU) model. While the reconstruction errors monotonically decreases with increasing dimension, the prediction errors actually explode in the PODLR. The reason is that the linear regression is unstable for the increasing number of input variables. Applying temporal unrolling solves the instability issues, but the same pattern is evident here: there is an optimal number of latent dimensions, which is not simply the largest possible, but somewhere around 20 for the state and 50 for the forcings. 

\subsection{Surrogate prediction performance}

\begin{figure*}
    \centering
    \begin{subfigure}[t]{0.23\textwidth}
        \centering
        \includegraphics[width=1.0\textwidth]{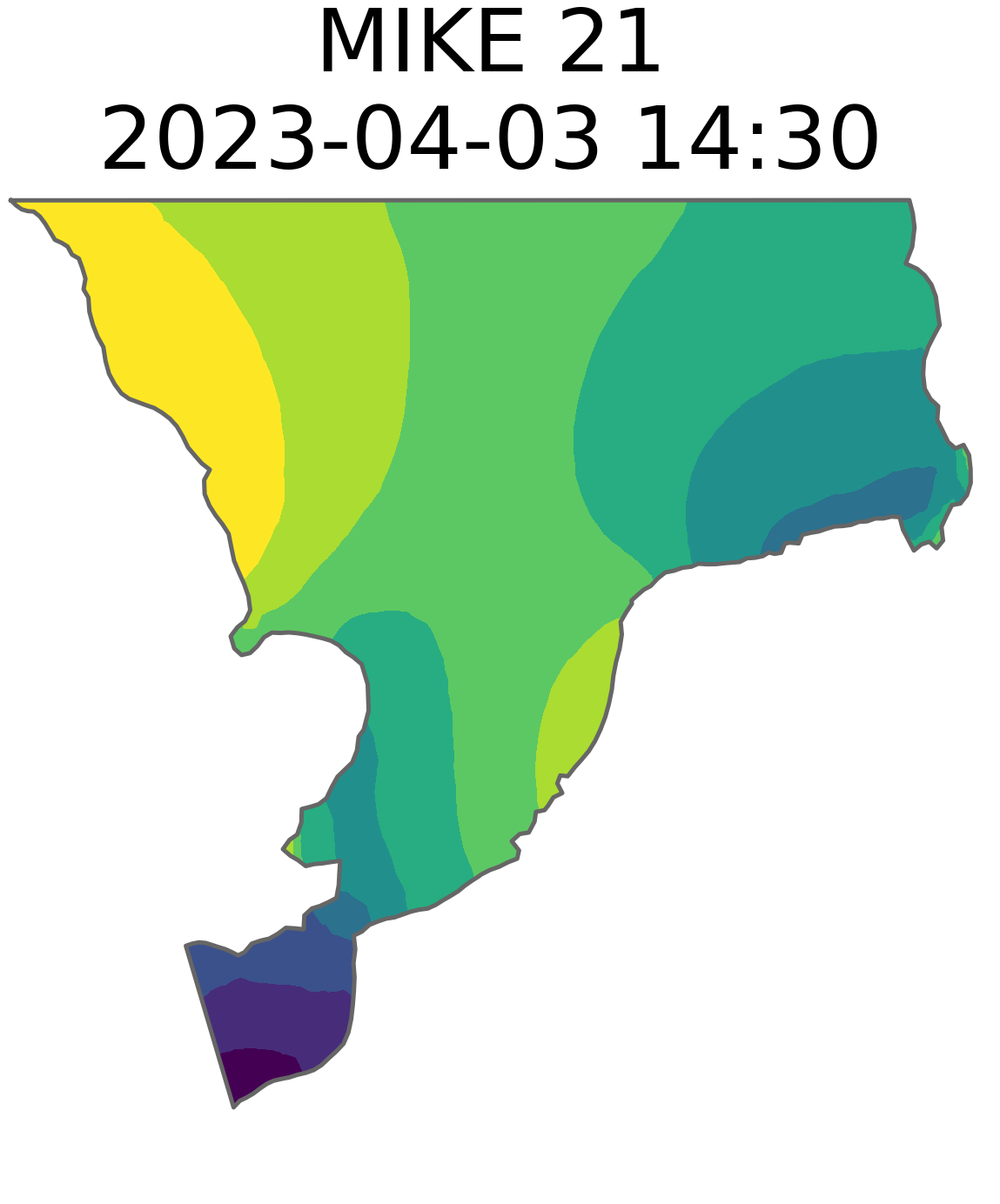}
    \end{subfigure}
    \begin{subfigure}[t]{0.32\textwidth}
        \centering
        \includegraphics[width=1.0\textwidth]{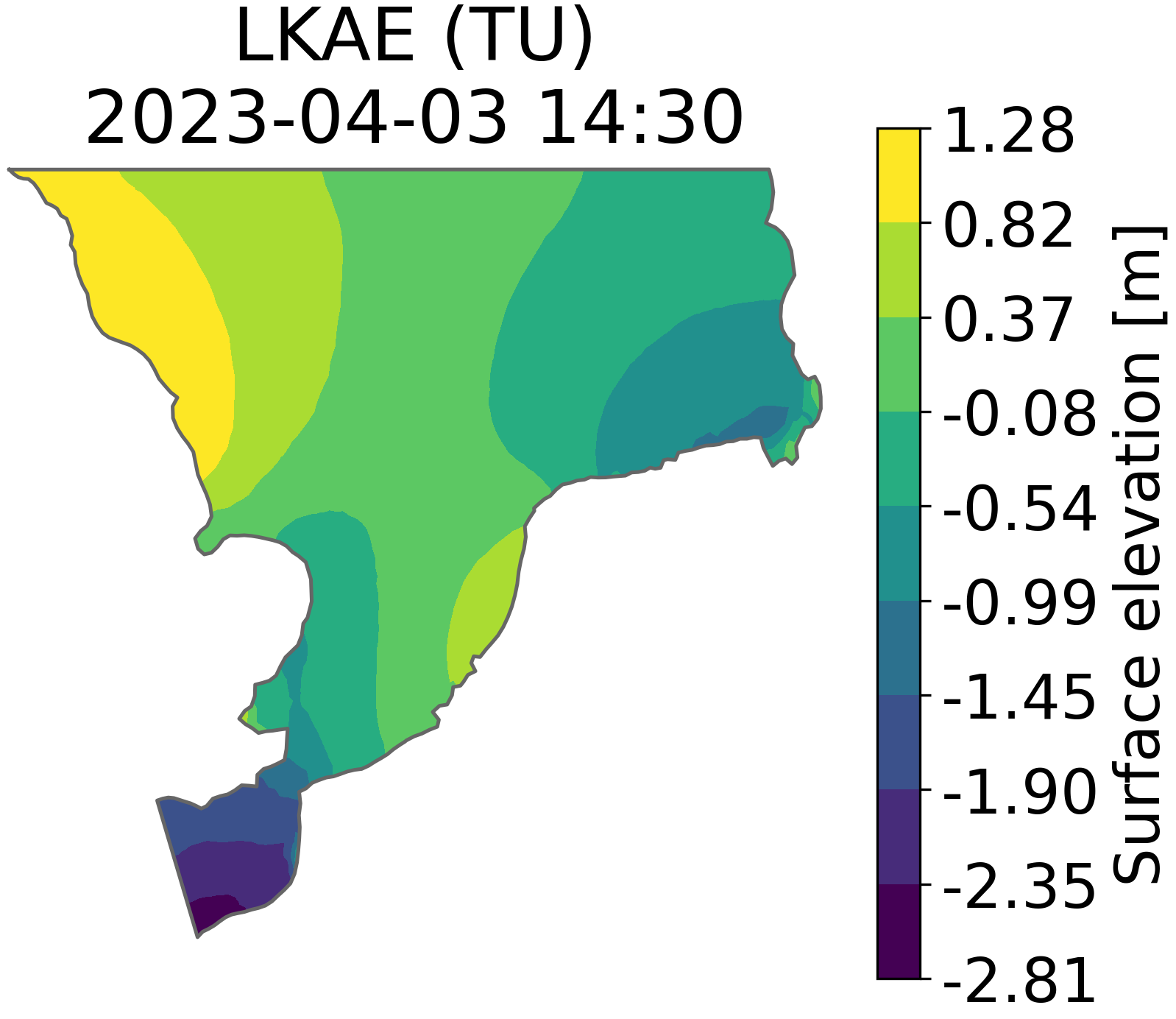}
    \end{subfigure}
    \begin{subfigure}[t]{0.32\textwidth}
        \centering \includegraphics[width=1.0\textwidth]{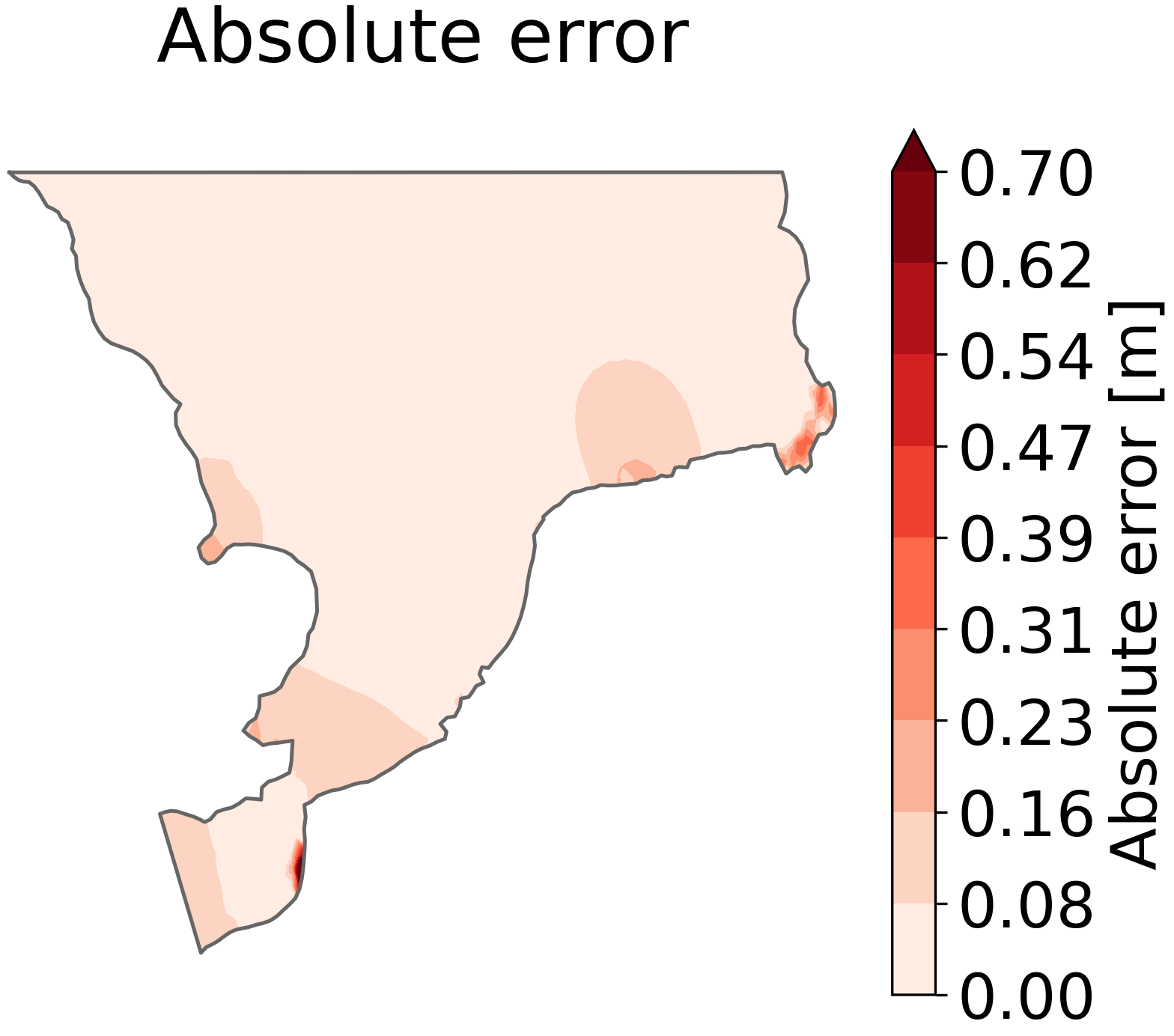}
    \end{subfigure}
    \caption[Prediction SNS]{Prediction of surface elevations with the MIKE 21 and the linear KAE (TU) in the time-step corresponding to median RMSE.}\label{fig:median_err_tstep}
\end{figure*}

Figure \ref{fig:median_err_tstep} shows an example of a surface elevation prediction made for the Southern North Sea with the KAE (TU) model. The prediction is based on auto-regressive temporal steps and the shown time-step corresponds to the median prediction RMSE, hence, the figures reflect the median performance. Generally, the KAE estimates the MIKE 21 simulation output very well, with the largest errors located by the coast in the South and in the East. 

\begin{figure}[h]
\centering
\includegraphics[width=1\textwidth]{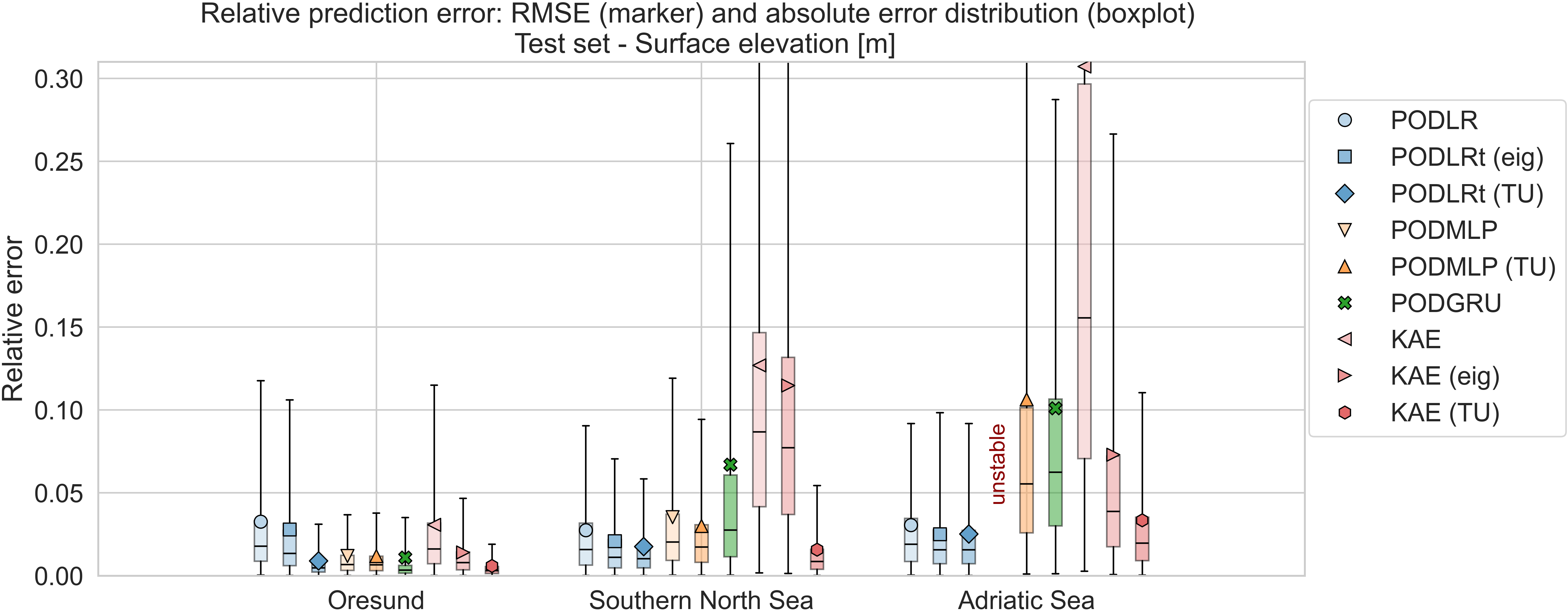}
\caption[Surrogate RMSEs]{Test errors of the surface elevation relative to the range of values. The boxplots show the quantiles and median of absolute errors with 1st and 99th percentile whiskers. Markers are RMSE.}\label{fig:com_test_SE}
\end{figure}

\begin{figure}[h]
\centering
\includegraphics[width=0.65\textwidth]{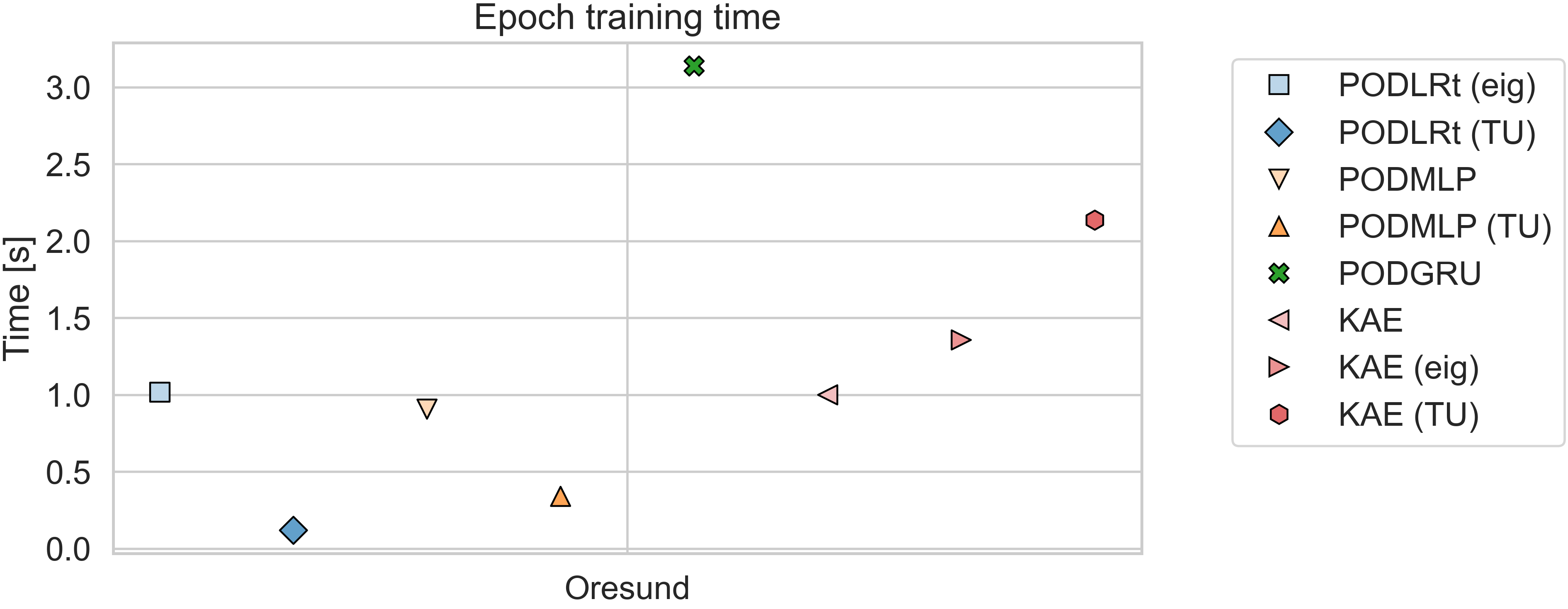}
\caption[Train times]{Time spent per epoch when training models. Total time for each model, in the order of the plot: 124, 21.5, 369, 63.7, 104, 454, 500, 1550 seconds. PODLR takes 10 seconds (not shown in plot).}\label{fig:com_time}
\end{figure}

Figure \ref{fig:com_test_SE} shows the relative prediction RMSE together with the distribution of relative absolute errors of the different surrogate models based on an autoregressive prediction in the test period. Results for the surface elevations are shown here, and for the current velocity in \ref{app:surr_comparisons}. Conclusions drawn for Figure \ref{fig:com_test_SE} are in line with the current velocity results. Figure \ref{fig:com_time} shows the wall-clock time spent per epoch when training the surrogate model for the Øresund case. The patterns are similar in the other two cases. 

Eigenvalue regularization improves the predictive performance to some or to a large extent for all models and test cases. For the KAE for Adriatic Sea, the Surface elevation RMSE decreases from 0.307 to 0.07 with eigenvalue regularization, and the distribution of errors change similarly. In other cases, e.g. PODLR for Southern North Sea, the effect of eigenvalue regularization is less distinct. 

Across all models, temporal unrolling has a positive effect on the accuracy and/or stability of the surrogate. For instance, the PODMLP is unstable in the Adriatic Sea case, but temporal unrolling dramatically improves the model to have similar performance to the PODGRU. The effect on the KAE for Southern North Sea is likewise remarkable, where the relative RMSE reduces from 0.13 to 0.016 for the Surface elevation and from 0.16 to 0.029 on average for the velocities, respectively. Hence, temporal unrolling improves the models by an order of magnitude in some cases. Across test cases and state variables, the best performing surrogates with temporal unrolling have relative RMSEs from 0.0068 to 0.14.

Generally, it is observed that the PODGRU has good performance due to its recurrent nature. However, the other architectures with temporal unrolling are able to compete with the GRU, and in most cases they perform slightly (or much) better, despite only being dependent on the immediately preceding timestep. It is noticed, however, that all of the model architectures (with temporal unrolling) perform quite similarly, and the results cannot be used to draw conclusions about significant differences between models. When it comes to the upper percentiles, the KAE perform slightly better than the POD-based surrogates for Øresund and Southern North Sea, while the PODLR have lower percentile values for the Adriatic Sea.

The computational complexity analysis in Section \ref{sec:complexity} explains the training-time differences observed in Figure \ref{fig:com_time} and reveals a clear offline–online trade-off. POD-based surrogates, except PODGRU, offer very low offline cost, while Koopman autoencoders require higher training effort but deliver improved stability and long-horizon accuracy. The relatively high epoch training time for PODGRU is due to the implemented in-batch chunking in Darts, which results more expensive epochs relative to other POD-models and the KAEs \citep{DartsTFM}. The figure also shows that the temporal unrolling implementation for the POD-based models is more effective than the regular implementation, which is explained by a more efficient batch implementation. As expected from the computational complexity, the effect of applying eigenvalue regularization is minor. Since online inference time-complexities are similar across surrogate classes, model choice is primarily governed by available training resources and required robustness in deployment.

\subsection{Analysis of selected surrogates}
The best-performing model in terms of relative RMSE for surface elevation is the KAE (TU) for Øresund and Southern North Sea, and PODLRt (TU) for Adriatic Sea, and their $R^2$-performance in the test-period is shown in Table \ref{tab:best_R2}. For Øresund, the 99th percentile of the relative errors is below 0.05 across state variables, and the $R^2$ are very close to 1. For Southern North Sea, the 99th percentile is below 0.1 across states, and the $R^2$ are also very high. For the Adriatic Sea, however, the 99th percentile relative errors range from 0.027 for the surface elevation up to 0.34 for the U-component of the currents, and the $R^2$ is further from 1, suggesting that some dynamics in the Adriatic Sea are more difficult to represent with the proposed architectures.

\begin{table}[h]
\centering
\caption{$R^2$ values in the test-period. The closer $R^2$ is to 1, the better.}\label{tab:best_R2}
\begin{tabular}{cc|c|c|c}
& & \makecell{\textbf{Øresund}\\KAE (TU)}&\makecell{\textbf{S. North Sea}\\KAE (TU)} & \makecell{\textbf{Adr. Sea}\\PODLRt (TU)}\\\hline
\multirow{3}{*}{$R^2$}  & $S$  & 0.996 & 0.993 & 0.955\\
                        & $U$   & 0.993 & 0.986 & 0.701\\
                        & $V$  & 0.995 & 0.986 & 0.614
\end{tabular}
\end{table}

\begin{figure*}[h]
    \centering
    \begin{subfigure}[t]{0.25\textwidth}
        \centering
        \includegraphics[height=4cm]{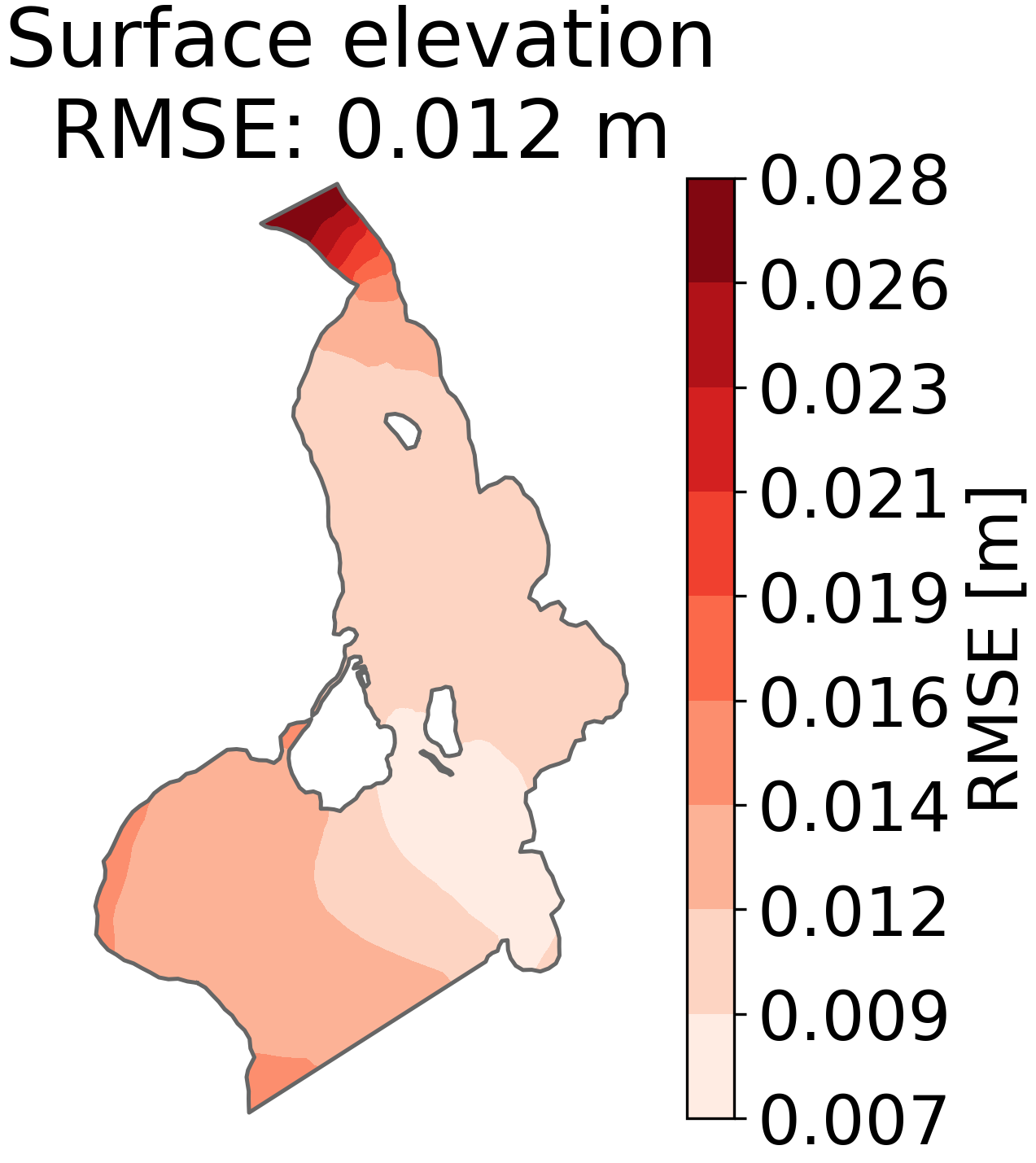}
        \caption{Øresund}
    \end{subfigure}
    \begin{subfigure}[t]{0.35\textwidth}
        \centering
        \includegraphics[height=4cm]{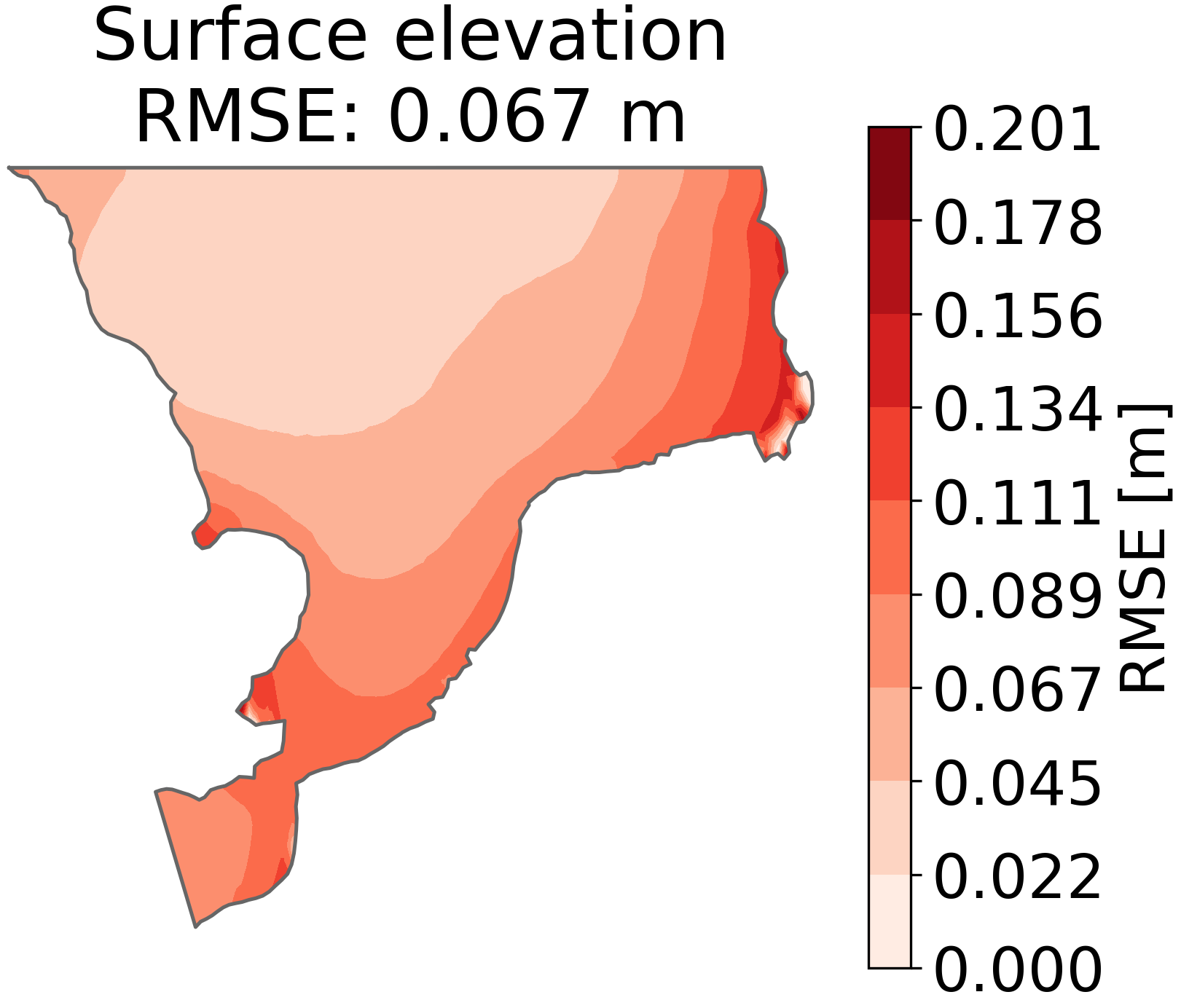}
        \caption{Southern North Sea}
    \end{subfigure}
    \begin{subfigure}[t]{0.32\textwidth}
        \centering \includegraphics[height=4cm]{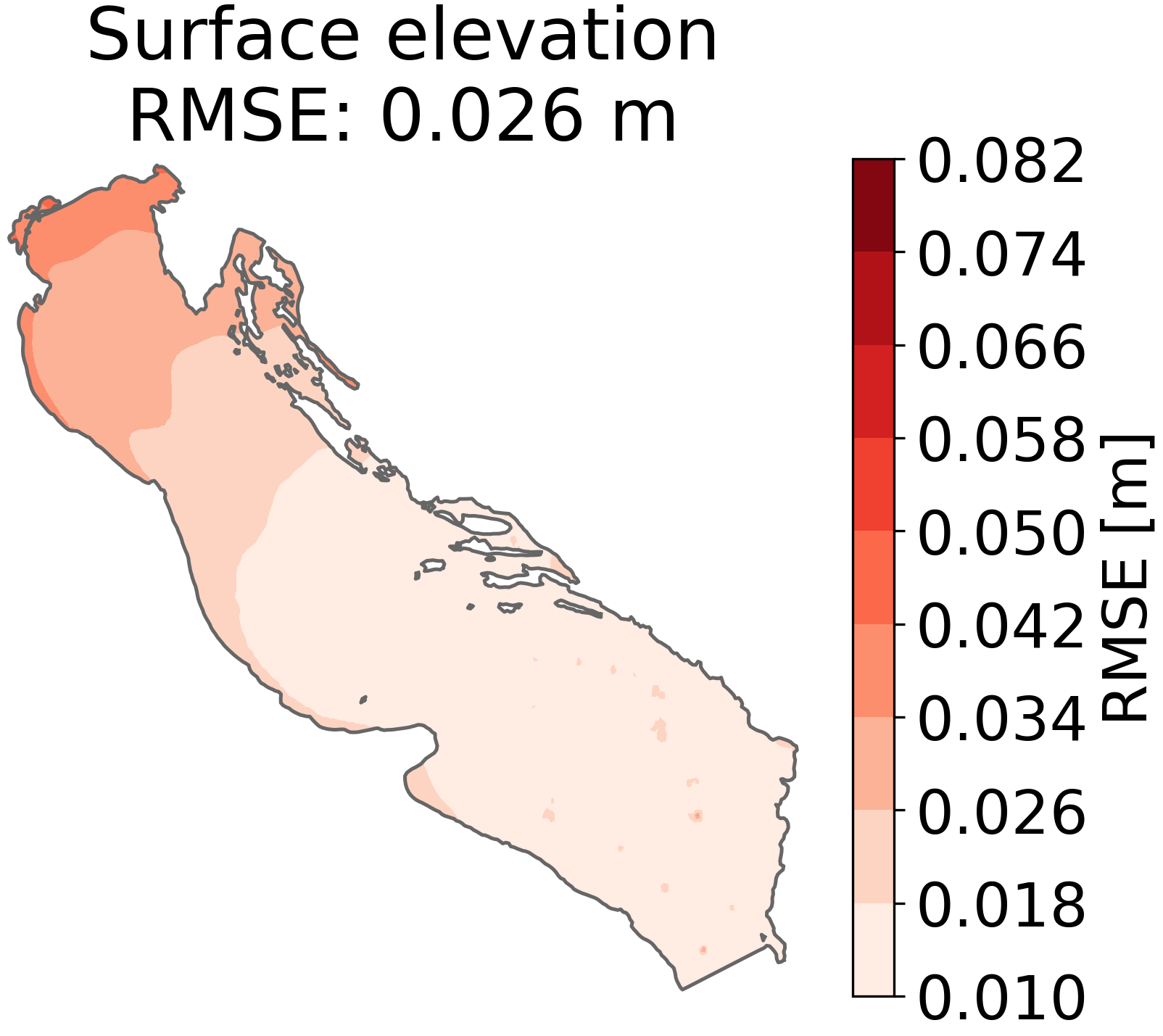}
        \caption{Adriatic Sea}
    \end{subfigure}
    \caption[RMSE maps]{RMSE across the test period for each test case.}\label{fig:RMSE_maps}
\end{figure*}

Complementary to the aggregated metrics, the RMSE is computed for each spatial element and visualized in Figure \ref{fig:RMSE_maps} and in \ref{app:rmsemaps}. In Øresund, the largest errors are in the North, yet they are only on the scale of a few centimeters. In the Southern North Sea, the largest errors range up to around 20 cm, and they are located close to the Eastern part of the domain, which is the Wadden Sea off the West-coast of Denmark and Northern Germany. This area is known for its very shallow waters, and the especially the low water levels are not as well captured by the surrogate as the deeper waters. In the Adriatic Sea, the largest errors are located in the Northwestern end of the basin, as expected due to the seiching occurring in this area. Yet, the RMSEs are relatively small, ranging up to 8.5 cm, indicating that on average the surface elevation predictions are very accurate.

\begin{figure*}[h]
    \centering
    \begin{subfigure}[t]{0.32\textwidth}
        \centering
        \includegraphics[width=1.0\textwidth]{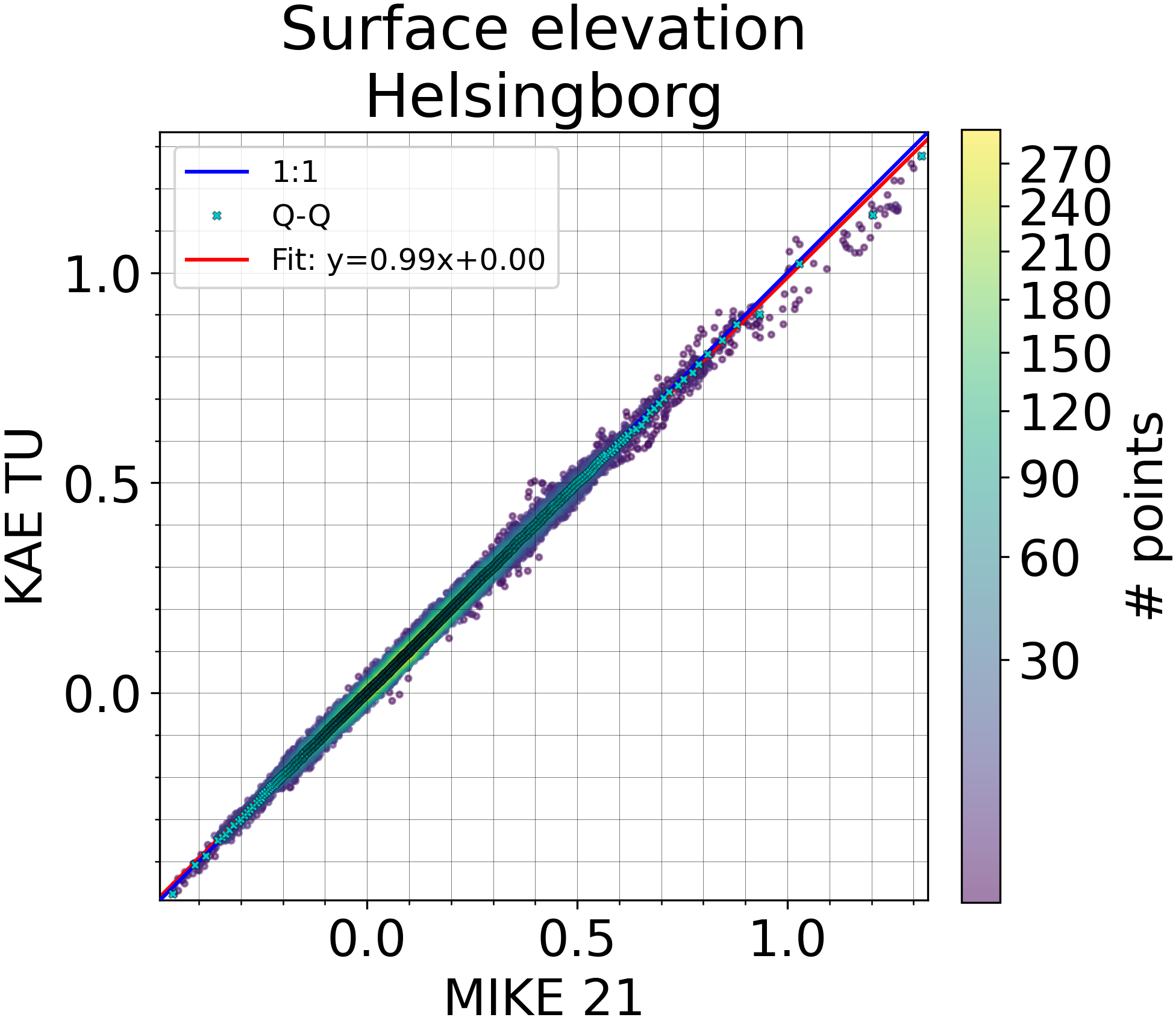}
        \caption{Øresund}
    \end{subfigure}
    \begin{subfigure}[t]{0.32\textwidth}
        \centering
        \includegraphics[width=1.0\textwidth]{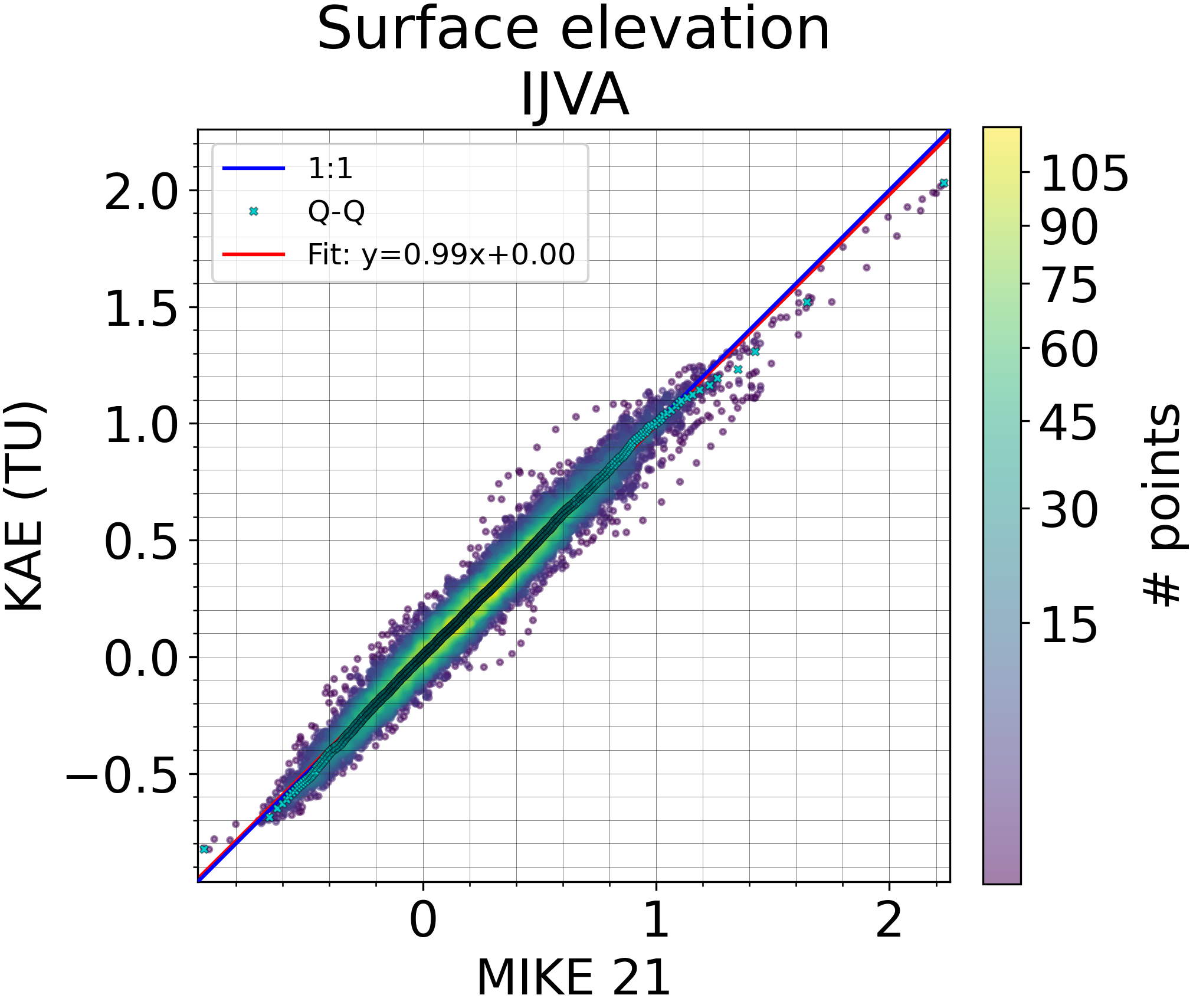}
        \caption{Southern North Sea}
    \end{subfigure}
    \begin{subfigure}[t]{0.32\textwidth}
        \centering \includegraphics[width=1.0\textwidth]{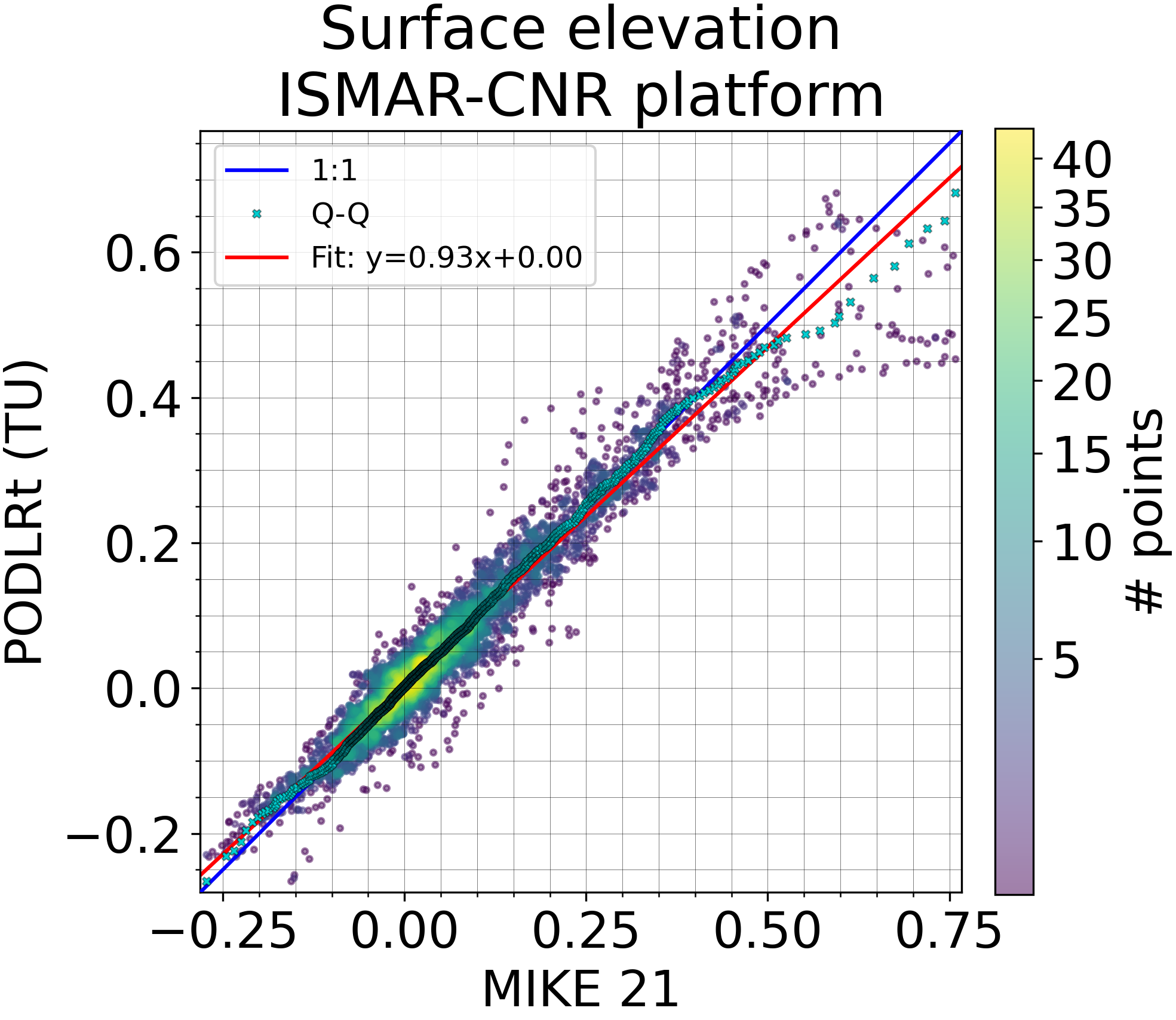}
        \caption{Adriatic Sea}
    \end{subfigure}
    \caption[Scatter $S$]{Scatter plots of the simulation data (x-axis) vs. the surrogate predictions (y-axis) in the test period for a single spatial point. Surface elevations.}\label{fig:scatter_SE}
\end{figure*}

Figure \ref{fig:scatter_SE} shows scatter plots of the predicted $S$ values on the $y$-axis compared to the MIKE 21 values on the $x$-axis. The values are shown only for a location corresponding to a measurement station for each test case. Similar figures for the other two state variables are found in \ref{app:scatter}. The leftmost figure shows that the surrogate for Øresund predicts the MIKE 21 data to a very satisfying extent with the points close to the diagonal line, also for extreme values. For Southern North Sea, there is some tendency for the surrogate to underestimate the large surface elevations and overestimate the low elevations as discussed for the Wadden Sea.

The right-most scatter plot once again showcases that the surrogate is not capturing the simulation data for Adriatic Sea to the same extent as for the other two cases. One explanation is the complex resonant oscillations in the basin, especially close to Venice where the ISMAR-CNR platform is located. Compared to the two other test-cases, there is fewer data available for training the surrogate, and the results indicate that more training and more data could be beneficial.

\subsection{Comparing surrogate performance with MIKE 21}
Since the surrogate model performance is, in principle, bounded by the physics-based model's prediction skill, the aim of the surrogate is to preserve skill. To assess the preservation level, Table \ref{tab:err_obs} shows the RMSEs of the physics-based MIKE 21 model and of the best performing surrogate when validating the predictions against observations. The bottom row indicates the percentage increase in RMSE when using the surrogate compared to MIKE 21, and a low percentage means successful retention of skill. It is confirmed that the Koopman based surrogate would be an excellent choice for emulating Øresund. For Southern North Sea, the performance is also satisfactory. Yet, at the IJVA station, there is an error increase of 12\%, but when the MIKE 21 model yields a prediction RMSE of 14.7 cm, and the KAE (TU) surrogate produces an RMSE of 16.4 cm, this marginal increase could be acceptable for practical applications.

For Adriatic Sea, two results are shown; for PODLRt (TU) and KAE (TU). While the former is slightly better at emulating MIKE 21 in terms of RMSE and the 99th percentile, it does lead to a 12 \% error increase compared to observations. In this regard, the KAE seems a better choice with an error increase of only 5.6 \%, which could be explained by better regularizing effects similar to the KAE for the Drogden station in Øresund. However, validating this in more locations remains a task for future work. 

\begin{table}[h]
\centering
\caption{RMSEs and the relative increase in RMSE of the models in the test period compared to observations in measurement stations. RMSEs given in centimeters. Surrogate models are KAE (TU) for Øresund and S. North Sea, and for Adriatic Sea, the results for both PODLRt (TU) and KAE (TU) are shown with PODLRt (TU) first. Hels.: Helsingborg, Kbh.: København, Dro.: Drogden.}\label{tab:err_obs}
\begin{tabular}{c|c|c|c|c|c|c|c}
& \multicolumn{3}{c|}{\textbf{Øresund}}&\multicolumn{3}{c|}{\textbf{S. North Sea}} & \textbf{Adr. Sea} \\
\textbf{RMSE}  & Hels. & Kbh. & Dro. & IJVA & F3 & J61 & ISMAR-CNR \\\hline
MIKE 21 [cm]    &  7.24   & 5.77 & 6.27 & 14.7 & 12.3 & 16.4 & 5.68\\
Surrogate [cm]  &  7.28   & 5.80 & 6.23 & 16.4 & 13.2 & 17.5 & 6.40, 6.00\\
Increase [\%]  &   0.51 & 0.44 &  -0.64   & 12 & 7.3 & 6.7 & 12, 5.6
\end{tabular}
\end{table}

Combining the low errors with the time spent on inference in a one-year simulation seen in Table \ref{tab:inferencetime}, there is a clear advantage of using the surrogates, namely that a factor 300-1400 speed-up is achieved. This speed-up is calculated under the conditions that the physics-based simulation is performed on a GPU, which is normal practice in industry, whereas the surrogate is simply run on a laptop CPU. However, the potential application of the surrogate exceeds the one-year simulation: with such a fast surrogate, one could perform hundreds of simulations, e.g. for ensembles for uncertainty quantification, or one could perform very long simulations, which is usually not feasible with the physics-based model.

\begin{table}[h]
\centering
\caption{Simulation time for a 1 year simulation for the three test cases. Excluding training times for surrogates and calibration times for physics-based model. Surrogate model is KAE (TU). Machine used: MIKE 21 Øresund, S. North Sea: NVIDIA A40 GPU, MIKE 21 Adriatic Sea: Nvidia Quadro M2000, surrogates: Laptop CPU (13th Gen Intel(R) Core(TM) i7-1365U with 32 GB RAM.)}\label{tab:inferencetime}
\begin{tabular}{c|c|c|c}
\textbf{Inference time}& \textbf{Øresund}& \textbf{S. North Sea} &\textbf{Adriatic} \\\hline
MIKE 21 [s]    & 2600 & 1200 & 16800 \\
Surrogate [s]  & 4  & 4 & 12\\
Speed-up factor  & 650 & 300 & 1400
\end{tabular}
\end{table}

\section{Discussion}\label{sec:discussion}
This paper has investigated reduced-order surrogate models for physics-based numerical models solving the shallow-water equation. The main assumption behind the work is that the dynamics of the system variables observed on the flexible mesh can be reduced to a limited number of dominant modes. Without this property, e.g. revealed by slow decay of the singular values when using POD, the reduced-order model cannot sufficiently resolve the data. The analysis of reconstruction errors in this work showed that the data can be resolved to a sufficient level given relatively few modes: 10-50 modes for mesh data with thousands of elements. However, there is an important trade-off when compressing data: while high simulation speed is achieved, some accuracy is lost. In two test cases, this was evident with RMSE increases up to 12\%, when using the surrogate instead of the original numerical model and validating against observations. Yet, for most practical applications, the increase in absolute numbers (on the order of a few centimeters) is generally acceptable given the factor 300-1400 speed-up. While the reported speed-ups are intended to reflect realistic operational usage rather than controlled hardware benchmarks, they work as an indicator of the potential gains from using the reduced order surrogates.

Two types of model order reductions were tested for forced coastal-ocean systems, namely POD and KAEs, as well as two types of strategies for improving long-term prediction stability and accuracy. Across all test-cases, the surrogates with temporal unrolling perform similarly with relative RMSEs of 0.0068--0.14. Generally, the KAE has slightly lower RMSEs, and they yield stable long-term predictions with learned Koopman operator eigenvalues within the unit circle. 

The key distinction between the model order reduction techniques lies in how each method handles the latent space: POD is achieved through a closed form solution, which provides an optimal linear basis for reconstruction but makes no assumptions about temporal dynamics. KAE approximates a Koopman invariant subspaces, where dynamics are linear, through an iterative non-convex optimization of parameters. The trade-off between these two approaches is that the POD is achieved fast, while the KAE relaxes some constraints such as orthogonality, allowing for the discovery of more expressive, nonlinear latent representations that can better capture complex dynamical structures.

By construction, the surrogate models are trained to perform well on average given the MSE-based loss function. While the current work includes some extreme value analysis, future work will explore whether performance can be improved for extreme value events for applications e.g. in early warning systems. 
In addition, generalization to unseen or out-of-distribution forcing conditions, could be investigated more thoroughly in future work. Another potential direction is exploring emulator superiority over the physics-based simulation used for generating training data. For instance, using the surrogate for data assimilation, thereby ensuring closer resemblance to observed values, or by exploring to what extent the surrogate can outperform the physics-based model e.g. by implicit regularization that may reduce numerical artifacts \citep{Koehler2025}.

\section{Conclusion}\label{sec:conclusion}
The proposed surrogate architectures with external forcing inputs and temporal unrolling enable accurate and computationally efficient long-term predictions across three hydrodynamic test cases. While the neural Koopman Autoencoder achieved slightly higher predictive accuracy in two out of three cases compared to the proper orthogonal decomposition based surrogates, both approaches showed strong performance. Temporal unrolling played a critical role in reducing error accumulation and improving long-horizon accuracy by an order of magnitude for some surrogates.

The best-performing surrogates achieved RMSEs of  0.0068--0.14 relative to the variability of the predicted variables, maintaining accurate autoregressive predictions over horizons of four months to one year with 30 minute time steps. Corresponding $R^2$ values ranged from 0.955 to 0.996 for water surface elevations and 0.61 to 0.995 for current velocities, indicating strong predictive skill, particularly for surface elevation dynamics. The largest errors occurred in the most complex test case, characterized by limited training data, a large computational mesh and seiching-driven dynamics, highlighting sensitivity to data availability and dynamical complexity.

Compared against in-situ observations, using the surrogate for prediction increased the error by at most 12\% relative to the underlying physics-based hydrodynamic model, corresponding to differences of only a few centimeters. In one case, the surrogate even improved predictive accuracy. Combined with inference-time speed-ups of 300 to 1400$\times$ for one-year simulations, these results indicate that the proposed reduced-order surrogates can achieve near-physics-model accuracy at a fraction of the computational cost, making them well suited for practical applications in ocean modelling.

\section*{Acknowledgements}\noindent
This work was funded in part by Innovation Fund Denmark, project no. 4297-00048B and in part by DHI’s research contract with the Danish Ministry of Higher Education and Science. The operational model setup for the Adriatic Sea is funded by Ministero delle Infrastrutture e dei Trasporti, Provveditorato Interregionale alle OO. PP. del Veneto, Trentino Alto Adige, Friuli Venezia Giulia.

\section*{Data and code availability}\noindent
Two of the three datasets used in this study are publicly available via a persistent repository (Zenodo). The third dataset is subject to third-party and/or operational constraints and cannot be shared publicly.

The research code used to develop and train the surrogate models is not publicly available at this stage due to intellectual property considerations and ongoing work towards potential commercialization. The paper provides a complete description of the methods, model formulations, and experimental setup to allow independent reproduction of the approach.

\section*{Declaration of generative AI use}\noindent
During the preparation of this work the author(s) made limited use of ChatGPT and the use was restricted to improving language and flow of the text, and not for any conceptual work. After using this tool/service, the author(s) reviewed and edited the content as needed and take(s) full responsibility for the content of the published article.

\newpage\clearpage
\bibliographystyle{elsarticle-harv} 
\bibliography{bibliography_new.bib}

\newpage
\appendix
\section{Test cases}\label{app:obs_overview}
This section presents information relevant to the three test cases.
\subsection{Forcing variables and dimensions}
Table \ref{tab:forcings_overview} gives an overview of the forcing variables and their dimensions in each case. In all cases, the boundary conditions are given as all three state variables, $S$, $U$, and $V$, therefore there is a factor 3 on the dimension. The wind and air pressure are given on 2D grids, and consist of a total of three variables as well (u- and v-component of wind velocities, air pressure). In the Adriatic Sea case, the wind and air pressure grid has been reduced by removing all grid cells that correspond to over-land values, such that only over-sea values are included. Therefore only 1448 out of the 4200 grid cells are active. The total number of forcing dimensions, $N_u$, is given in the first column.

The forcing data for the Øresund and Southern North Sea data is publicly available together with the MIKE 21 model setups and the simulation data used for this paper \citep{dhi_2024_14160710,dhi_2025_14929387}.

\begin{table}[H]
\centering
\caption{Overview of forcing variables in the three test cases.}\label{tab:forcings_overview}
\begin{tabular}{c|cc}
\textbf{Case} & \textbf{Forcing} & \textbf{Dimension}  \\ \hline
\multirow{3}{*}{\makecell{Øresund \\ $N_u=168$}}    & North BC & $3\cdot 13$\\ 
                            & South BC & $3\cdot 27$\\
                            & Wind and Air pressure & $3\cdot (4\cdot4)$\\ \hline
 \multirow{3}{*}{\makecell{Southern\\North Sea\\ $N_u=7275$}}  & North BC & $3\cdot 16$\\
                                & South BC & $3\cdot 9$\\
                                & Wind and Air pressure & $3\cdot (48\cdot50)$ \\\hline
\multirow{3}{*}{\makecell{Adriatic\\Sea\\$N_u=4404$}}   &BC & $3\cdot 20$\\
                                & Wind and Air pressure & \makecell{$3\cdot (75\cdot56)$\\(only $3\cdot 1448$ active)}\\\hline
\end{tabular}
\end{table}

\subsection{Observation stations}
Observations from in-situ measurement stations are used for comparison to the MIKE 21 and surrogate predictions. The stations together with their coordinates and the sources of the data are listed in Table \ref{tab:observations_sources}.
\begin{table}[H]
\centering
\caption{Overview of measurement stations used for validating the surrogate model. Coordinates (longitude/latitude) and reference to the source.}\label{tab:observations_sources}
\begin{tabular}{cc|ccp{2.7cm}}
\textbf{Case} & \textbf{Station} & \textbf{Longitude} & \textbf{Latitude} & \textbf{Source} \\ \hline
\multirow{3}{*}{Øresund}    & Helsingborg &12.6845&56.0412&\multirow{3}{*}{\makecell{\citet{dhi_2024_14160710} /\\ \citet{CMEMS_baltic}}}\\ 
                            & København & 12.65&55.7&\\
                            & Drogden & 12.7117&55.5358&\\ \hline
 \multirow{3}{*}{\makecell{Southern\\North Sea}}  & IJVA &3.71044&52.88381&  \multirow{3}{*}{\makecell{\citet{dhi_2025_14929387} / \\ \citet{IJVA} / \\\citet{CMEMS_NWS}}}\\
                                & F3 platform &4.72000 &54.84999& \\
                                & J61 &2.94380&53.82325& \\\hline
\multirow{1}{*}{\makecell{Adriatic\\Sea}}   & \makecell{ISMAR-\\CNR} &12.51472&45.32306 & \citet{ISMARCNRplatform}\\
\end{tabular}
\end{table}

\section{Parameters}
This section provides an insight into the parameters of the surrogate models. Specifically the number of latent dimensions used in each test case, as well as all hyper-parameters used for the surrogate models. For the analyses and hyper-parameter tuning a subset of the training data is used for training and validation. The periods are shown in Table \ref{tab:hyperparam_train_val}.

\begin{table}[h]
\centering
\caption{Training and validation periods for hyper-parameter tuning}\label{tab:hyperparam_train_val}
\begin{tabular}{c|c|c|c}
\textbf{} & \textbf{Øresund} & \textbf{S, North Sea} & \textbf{Adriatic Sea} \\ \hline
Train period & Jan-Apr 2021 & Jan-Feb 2022 & Jan-Feb 10 2021\\
Val. period & May-Jun 2021 & Mar 2022 & Feb 11-Feb 28 2021
\end{tabular}
\end{table}

\subsection{Latent dimensions and reconstruction errors}\label{app:reconstruction_errs}
Across surrogate models the same number of latent dimensions is used regardless of whether it is Koopman-based, POD-based, concatenated or separated autoencoders. Table \ref{tab:PCsandreconerrs_state} shows an overview of the dimensions and reconstruction errors for each type of model. If only one latent dimension is given, it means that the autoencoder is concatenated. E.g. for Southern North Sea the $U$ and $V$ variables are concatenated in one autoencoder with latent dimension 20. The latent dimensions are chosen based on an analysis as described in Section \ref{sec:latentdim}.

The RMSE Koopman column corresponds to the reconstruction of the states with the best performing Koopman-based surrogate. The last columns shows the value range (mean across spatial elements) for each variable. Table \ref{tab:PCsandreconerrs_forcing} shows corresponding results for the forcing variables. However, since the forcings are only encoded and not decoded in the Koopman surrogates, it is only the POD reconstruction, which is shown. 

\begin{table}[H]
\centering
\caption{Number of modes and the resulting reconstruction errors for each state variable in the three test cases Evaluated on the training period. $S$: Surface elevation, $U$: U-component of velocity, $V$: V-component of velocity. }\label{tab:PCsandreconerrs_state}
\begin{tabular}{c|c|c|c|c|c|c}
\textbf{Case} & \textbf{State} & \makecell{\textbf{Latent }\\\textbf{dim.}} & \makecell{\textbf{RMSE}\\\textbf{POD}} & \makecell{\textbf{RMSE}\\\textbf{Koopman}} & \textbf{Unit} & \makecell{\textbf{Value}\\\textbf{range}}\\ \hline
\multirow{3}{*}{Øresund} & $S$ & \multirow{3}{*}{15} & $5.8\cdot10^{-3}$  & $3.2\cdot10^{-3}$ & m & 1.4 \\
 & $U$ &  & $4.6\cdot10^{-3}$  & $3.0\cdot10^{-3}$ & m/s & 0.32\\
 & $V$ &  & $6.5\cdot10^{-3}$ & $4.5\cdot10^{-3}$  & m/s & 0.63\\ \hline
 \multirow{3}{*}{S. North Sea}  & $S$ & 20                  & $1.7\cdot10^{-2}$ & $2.7\cdot10^{-2}$ &  m & 3.9 \\
                                & $U$ &  \multirow{2}{*}{20} & $1.8\cdot10^{-2}$ & $1.1\cdot10^{-2}$ &  m/s &1.0\\
                                & $V$ &                      & $1.6\cdot10^{-2}$ & $1.1\cdot10^{-2}$ & m/s & 0.98\\\hline
\multirow{3}{*}{Adriatic Sea}   & $S$ &  10                  & $6.2\cdot10^{-3}$ & $1.5\cdot10^{-2}$ & m & 0.84\\
                                & $U$ &  \multirow{2}{*}{50}  & $9.1\cdot10^{-3}$ & $1.1\cdot10^{-2}$ & m/s & 0.21\\
                                & $V$ &                       & $1.0\cdot10^{-2}$ & $1.2\cdot10^{-2}$ & m/s & 0.22\\
\end{tabular}
\end{table}

\begin{table}[H]
\centering
\caption{Number of components and the resulting reconstruction errors for each forcing in the three test cases.}\label{tab:PCsandreconerrs_forcing}
\begin{tabular}{c|c|c|c|c|c}
\textbf{Case} & \textbf{Forcing} & \makecell{\textbf{Latent }\\\textbf{dim.}} &\makecell{\textbf{RMSE}\\\textbf{POD}} & \textbf{Unit} & \makecell{\textbf{Value}\\\textbf{range}}\\ \hline
\multirow{9}{*}{Øresund} & North BC $S$  & \multirow{6}{*}{50} & $1.3\cdot10^{-3}$ & m& 1.4 \\
 & North BC $U$         &  & $1.9\cdot10^{-3}$ & m/s & 1.2 \\
 & North BC $V$         &  & $1.8\cdot10^{-3}$ & m/s & 0.90 \\
 & South BC $S$        &  & $1.5\cdot10^{-3}$ & m   & 1.6\\
 & South BC $U$         &  & $4.0\cdot10^{-4}$ & m/s &0.22  \\
 & South BC $V$         &  & $1.6\cdot10^{-3}$ & m/s & 0.96  \\
 & Air pressure       &  &  $1.1\cdot10^{-1}$&  hPa & 69\\ 
 & Wind $U$             &  &  $5.2\cdot10^{-2}$&  m/s & 22 \\
 & Wind $V$             &  &  $4.5\cdot10^{-2}$&  m/s & 22 \\\hline
 \multirow{9}{*}{S. North Sea}  & North BC $S$ & \multirow{3}{*}{10} & $3.9\cdot10^{-2}$ & m &2.9\\
 & North BC $U$                   &                                   & $1.2\cdot10^{-2}$ & m/s&0.62 \\
 & North BC $V$                   &                                   & $1.8\cdot10^{-2}$ & m/s& 0.84\\
 & South BC $S$                  &    \multirow{3}{*}{10}            & $8.0\cdot10^{-3}$ & m& 7.4\\
 & South BC $U$                   &                                   & $4.7\cdot10^{-3}$ & m/s&2.2\\
 & South BC $V$                &                                   & $1.9\cdot10^{-3}$ & m/s &1.0\\
 & Wind $U$                       &   \multirow{3}{*}{10}             & $1.4$   & m/s & 32\\
 & Wind $V$                       &                                   & $1.3$   & m/s & 28\\
 & Air pressure                 &                                   & $0.64$  & hPa & 69\\ \hline
\multirow{6}{*}{Adriatic Sea}   & BC $S$ & \multirow{3}{*}{10} & $1.6\cdot10^{-3}$ & m &0.56\\
 & BC $U$                                                      &      & $1.0\cdot10^{-3}$ & m/s & 0.065\\
 & BC $V$                                                       &     & $3.3\cdot10^{-3}$ & m/s & 0.21\\
 & Air pressure         & \multirow{3}{*}{60}                 & $0.26$ & hPa & 43.7\\ 
 & Wind $U$                                                     &      & $0.80$ &  m/s & 25\\
 & Wind $V$                                                     &      & $0.84$ &  m/s & 28\\
\end{tabular}
\end{table}

\subsection{Hyper-parameters}\label{app:hyperparams}

A number of hyper-parameters are used when fitting the surrogate models. An overview of the hyper-parameters used in each test case is given in tables \ref{tab:hyperparams} (Øresund), \ref{tab:hyperparams_Adriatic} (Adriatic Sea), and \ref{tab:hyperparams_SNS} (Southern North Sea). For the POD-based models, the hyper-parameters only concerns the temporal propagator in the latent space, since the POD itself is computed using a closed form solution. Since the PODMLP is not linear in the latent space, an eigenvalue regularized version does not exist and the middle row is therefore irrelevant. For the PODGRU, there is only one version, and no version with eigenvalue regularization nor temporal unrolling.

All models except the GRU are trained with a learning rate scheduler which is the PyTorch \texttt{ReduceLROnPlateau} \citep{ReduceLROnPlateauTorch}. Unless otherwise specified in the tables, the methods use the default variables, e.g. momentum variables for the Adam optimizers. 

The TU steps and TU window stride are two important parameters in the implementation of temporal unrolling. The temporal unrolling is implemented in a batch-parallelized manner such that within one batch, different unrollings of length 'TU steps' are initiated at the same time with 'TU window stride' space. E.g. if 'TU window stride' is 1, the unrolled sequences would be
\begin{align*}
    \text{Thread 1: } &x_1, x_2, ..., x_{\text{TU steps}}\\
    \text{Thread 2: } &x_2, x_3, ..., x_{\text{TU steps}+1}\\
    &\quad\vdots
\end{align*}
If 'TU window stride' is $w$, the sequences would be:
\begin{align*}
    \text{Thread 1: } &x_1, x_2, ..., x_{\text{TU steps}}\\
    \text{Thread 2: } &x_{1+w}, x_{2+w}, ..., x_{\text{TU steps}+w}\\
    &\quad\vdots
\end{align*}
Hence, the parameter 'TU steps' needs to be smaller than the batch size. 

In the following sections, the hyper parameters will be presented. A few common abbreviations are used to compress the table: Optim.=Optimizer, LR=Learning rate, W.D.=weight decay, LR sche.=Learning rate scheduler, pat.=patience, fac.=factor, tol.=tolerance, TU win.=temporal unrolling window. Note that all autoencoders are symmetric, meaning that there are the same number of parameters (layers and hidden dimensions) in the encoder and decoder.

\subsubsection{Øresund}
For PODMLP, the MLP propagator has two hidden layers of dimension 115, corresponding to the output dimension from the autoencoder. For PODGRU, some extra hyper-parameters specific to the architecture and to the Darts forecasting models \citep{darts} are: \texttt{input\_chunk\_length=24}, \texttt{training\_length=35}, \texttt{hidden\_dim=180}, \texttt{n\_rnn\_layers=2}, and \texttt{dropout=0.05}. Interestingly, the hyper-parameter optimization found that an input length of 24 time-steps, corresponding exactly to two tidal-periods of 12 hours the Øresund area, was optimal for the GRU. For the KAE, the state autoencoder is linear, while the forcing encoder is nonlinear with layers of dimension 120, 120, 50. 

\begin{table}[H]
\centering
\caption{Hyper-parameters for each surrogate for the Øresund case.}\label{tab:hyperparams}
\begin{tabular}{c|l|c|c|c|c}
\textbf{} & & \textbf{PODLRt} & \textbf{PODMLP } &\textbf{PODGRU } & \textbf{KAE} \\ \hline
\multirow{8}{*}{Simple} & Batch size & 40 & 40  & 128 & 128\\
                        & Optimizer & Adam & Adam & Adam & AdamW\\
                        & Optim. LR & $10^{-3}$ &  $10^{-3}$ &$10^{-3}$ & $1.4\cdot 10^{-4}$\\
                        & Optim. W.D. & 0  & 0  & 0 & $10^{-4}$ \\
                        & LR sche. pat. & 10 & 10 & - &   10\\
                        & LR sche. fac. & 0.1 & 0.1 & - &  0.5\\
                        & Early stop. tol.   & $10^{-2}$ & $10^{-3}$  & $10^{-4}$ (abs.) &   $10^{-2}$\\
                        & Early stop pat. & 20 & 40  &  10 & 40\\\hline
 \multirow{8}{*}{Eig}   & Batch size & 40 & - &- & 128\\
                        & Optimizer & Adam & - &- &  AdamW\\
                        & Optim. LR & $10^{-3}$  & - &- &  $1.4 \cdot 10^{-4}$\\
                        & Optim. W.D. & 0  & - &- &    $10^{-4}$\\
                        & LR sche. pat. & 10 & - & - & 10\\
                        & LR sche. fac. & 0.1 & - &  - & 0.5\\
                        & Early stop. tol. & $10^{-2}$ &-  & - & $10^{-2}$ \\
                        & Early stop pat. & 20 & - &  - & 40\\\hline
\multirow{10}{*}{TU}     & Batch size & 128 & 64 &  - & 128\\
                        & Optimizer & Adam & Adam & - &AdamW\\
                        & Optim. LR & $10^{-2}$ & $3\cdot 10^{-3}$ & - & $10^{-3}$ \\
                        & Optim. W.D. & 0  & 0 &   - &$10^{-4}$ \\
                        & LR sche. pat. & 10 & 10 &  - &10 \\
                        & LR sche. fac. & 0.1 & 0.1 &    - &0.5\\
                        & Early stop. tol. & $10^{-4}$ & $5 \cdot 10^{-4}$ & - &$10^{-2}$\\
                        & Early stop pat. & 20 & 40 &  - &40 \\
                        & TU steps  & 10 & 9 &  - &10\\
                        & TU win. stride & 4 & 3 & - & 4\\
\end{tabular}
\end{table}

\subsubsection{Southern North Sea}
For PODMLP, the MLP propagator has two hidden layers of dimension 70, corresponding to the output dimension from all autoencoders (20+20+10+ 10+10). For PODGRU, the extra hyper-parameters are: \texttt{input\_chunk\_length=16}, \texttt{training\_length=26}, \texttt{hidden\_dim=180}, \texttt{n\_rnn\_layers=2}, and \texttt{dropout=0.05}.  In the KAE, the autoencoders are concatenated with latent dimension of 40 for the states and of 30 for the forcings. Both the state autoencoder and forcing encoder are linear. 
\begin{table}[H]
\centering
\caption{Hyper-parameters for each surrogate for the Southern North Sea case.}\label{tab:hyperparams_SNS}
\begin{tabular}{c|l|c|c|c|c}
\textbf{} & & \textbf{PODLRt} & \textbf{PODMLP} &\textbf{PODGRU} & \textbf{KAE}  \\ \hline
\multirow{8}{*}{Simple} & Batch size & 40 & 40 & 128 & 256  \\
                        & Optimizer & Adam & Adam &Adam & Adam  \\
                        & Optim. LR & $10^{-3}$ & $10^{-3}$ & $10^{-3}$  &  $10^{-4}$ \\
                        & Optim. W.D. & 0  & 0 & 0& $10^{-4}$  \\
                        & LR sche. pat. & 10 & 10 &  -&  10 \\
                        & LR sche. fac. & 0.1 & 0.1 & -&   0.1\\
                        & Early stop. tol.   & $10^{-2}$ & $10^{-2}$  & $10^{-4}$ (abs.) & $10^{-2}$ \\
                        & Early stop pat. & 20 & 40  &  10 & 30 \\\hline
 \multirow{8}{*}{Eig}   & Batch size & 40 & - & - & 256 \\
                        & Optimizer & Adam & - & - &Adam\\
                        & Optim. LR & $10^{-3}$  & - &  - & $10^{-4}$\\
                        & Optim. W.D. & 0  & - &   - &$10^{-4}$\\
                        & LR sche. pat. & 10 & - & - & 10 \\
                        & LR sche. fac. & 0.1 & - &  - & 0.1 \\
                        & Early stop. tol. & $10^{-2}$ & - &-   \\
                        & Early stop pat. & 40 & - & - & 30 \\\hline
\multirow{10}{*}{TU}     & Batch size & 64 & 64 & - &20 \\
                        & Optimizer & Adam & Adam &- & Adam \\
                        & Optim. LR & $10^{-2}$ & $10^{-3}$ & - & $10^{-2}$   \\
                        & Optim. W.D. & 0  & 0 & - &  $10^{-5}$  \\
                        & LR sche. pat. & 10 & 10 &  - &10  \\
                        & LR sche. fac. & 0.1 & 0.1 & - &  0.1 \\
                        & Early stop. tol. & $10^{-3}$ & $10^{-3}$ & - & $10^{-3}$\\
                        & Early stop pat. & 40 & 40 & - & 30 \\
                        & TU steps  & 12 & 12 & - & 12 \\
                        & TU win. stride & 4 & 4 &  - & 4 \\
\end{tabular}
\end{table}

\subsubsection{Adriatic Sea}
For PODMLP, the MLP propagator has two hidden layers of dimension 130, corresponding to the output dimension from all autoencoders (10+50+10+60). For PODGRU, the extra hyper-parameters are: \texttt{input\_chunk\_length=48}, \texttt{training\_length=83}, \texttt{hidden\_dim=190}, \texttt{n\_rnn\_layers=3}, and \texttt{dropout=0.05}.  For the KAE, the state autoencoders and forcing encoders are linear.
\begin{table}[H]
\centering
\caption{Hyper-parameters for each surrogate for the Adriatic Sea case.}\label{tab:hyperparams_Adriatic}
\begin{tabular}{c|l|c|c|c|c}
\textbf{} & & \textbf{PODLRt} & \textbf{PODMLP} &\textbf{PODGRU} & \textbf{KAE} \\ \hline
\multirow{8}{*}{Simple} & Batch size & 40 & 128  & 128 & 128\\
                        & Optimizer & Adam & Adam & Adam & AdamW  \\
                        & Optim. LR & $7 \cdot 10^{-3}$ & $7 \cdot10^{-3}$  &$10^{-3}$  & $5 \cdot10^{-3}$ \\
                        & Optim. W.D. & 0  & 0 & 0& $10^{-6}$\\
                        & LR sche. pat. & 10 & 10 & -&  10 \\
                        & LR sche. fac. & 0.1 & 0.1 & -&  0.5 \\
                        & Early stop. tol.   & $10^{-5}$ & $10^{-5}$  & $10^{-4}$ (abs.) &  $10^{-2}$ \\
                        & Early stop pat. & 20 & 40  &  10 & 50 \\\hline
 \multirow{8}{*}{Eig}   & Batch size & 40 & - &  - &  128 \\
                        & Optimizer & Adam & - &  - & AdamW \\
                        & Optim. LR & $7 \cdot10^{-3}$  & - &   - &  $5\cdot10^{-3}$ \\
                        & Optim. W.D. & 0  & - &   - &  $10^{-6}$ \\
                        & LR sche. pat. & 10 & - &  - &  10 \\
                        & LR sche. fac. & 0.1 & - &  - &   0.5 \\
                        & Early stop. tol. & $10^{-5}$ &-  &   - &  $10^{-2}$ \\
                        & Early stop pat. & 20 & - &  - &  50 \\\hline
\multirow{10}{*}{TU}     & Batch size & 40 & 64 &   - & 64\\
                        & Drop out rate & - & 0.3 & - &  -\\
                        & Optimizer & Adam & Adam &  - & AdamW\\
                        & Optim. LR & $6 \cdot10^{-3}$ & $10^{-4}$ &  - &  $10^{-4}$  \\
                        & Optim. W.D. & 0  & 0 &   - &  $10^{-3}$ \\
                        & LR sche. pat. & 10 & 10 &  - &  10 \\
                        & LR sche. fac. & 0.1 & 0.5 &   - &  0.5\\
                        & Early stop. tol. & $8.5\cdot10^{-5}$ & $10^{-5}$ &  - &  $10^{-2}$ \\
                        & Early stop pat. & 30 & 40 &  - &  50  \\
                        & TU steps  & 10 & 10 &   - & 10 \\
                        & TU win. stride & 5 & 3 &  - &  4 \\
\end{tabular}
\end{table}

\section{Results}
The main text shows results mainly for the surface elevations. This appendix shows the complementary figures for the other two state variables, u- and v-components of the current velocities. 

\subsection{Surrogate model comparisons}\label{app:surr_comparisons}
\begin{figure}[h]
\centering
\includegraphics[width=1\textwidth]{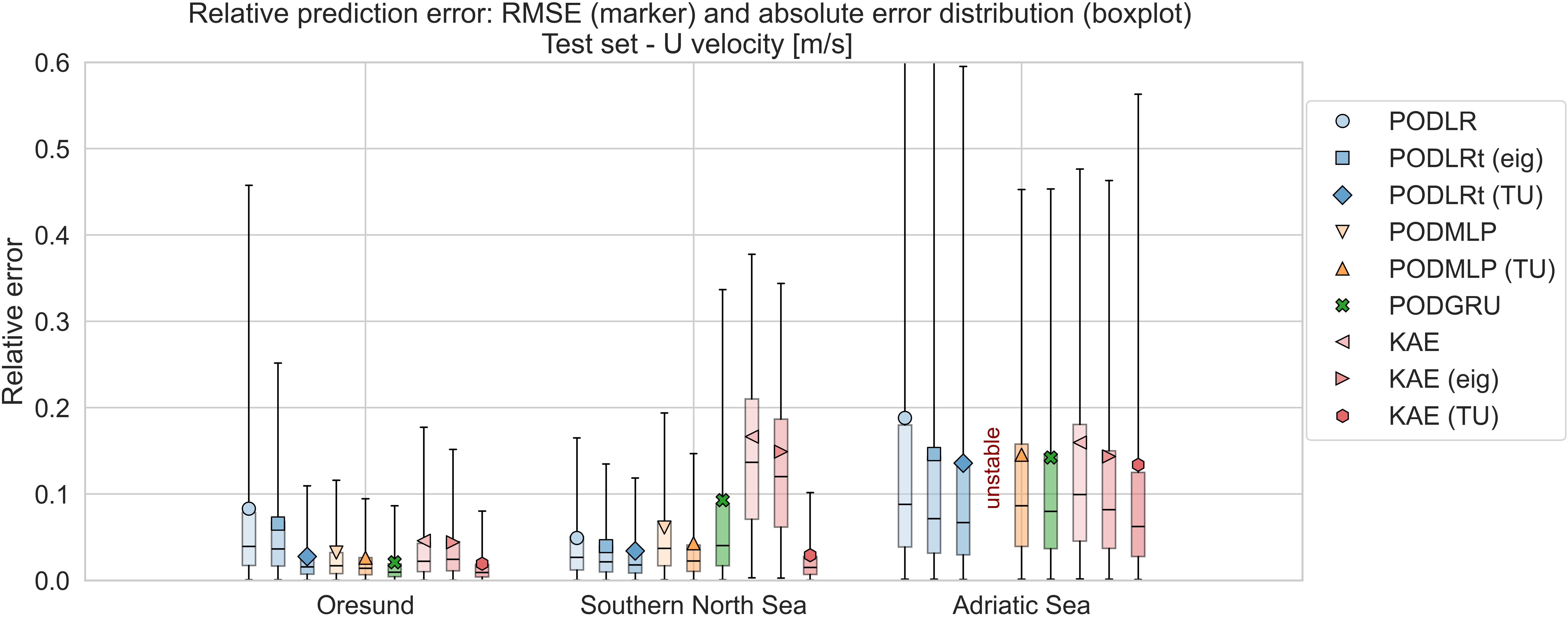}
\caption[Surrogate RMSEs]{Test errors of U component of current velocity relative to the range of values. The boxplots show the quantiles and median of absolute errors with 1st and 99th percentile whiskers. Markers are RMSE.}\label{fig:com_test_U}
\end{figure}

\begin{figure}[h]
\centering
\includegraphics[width=1\textwidth]{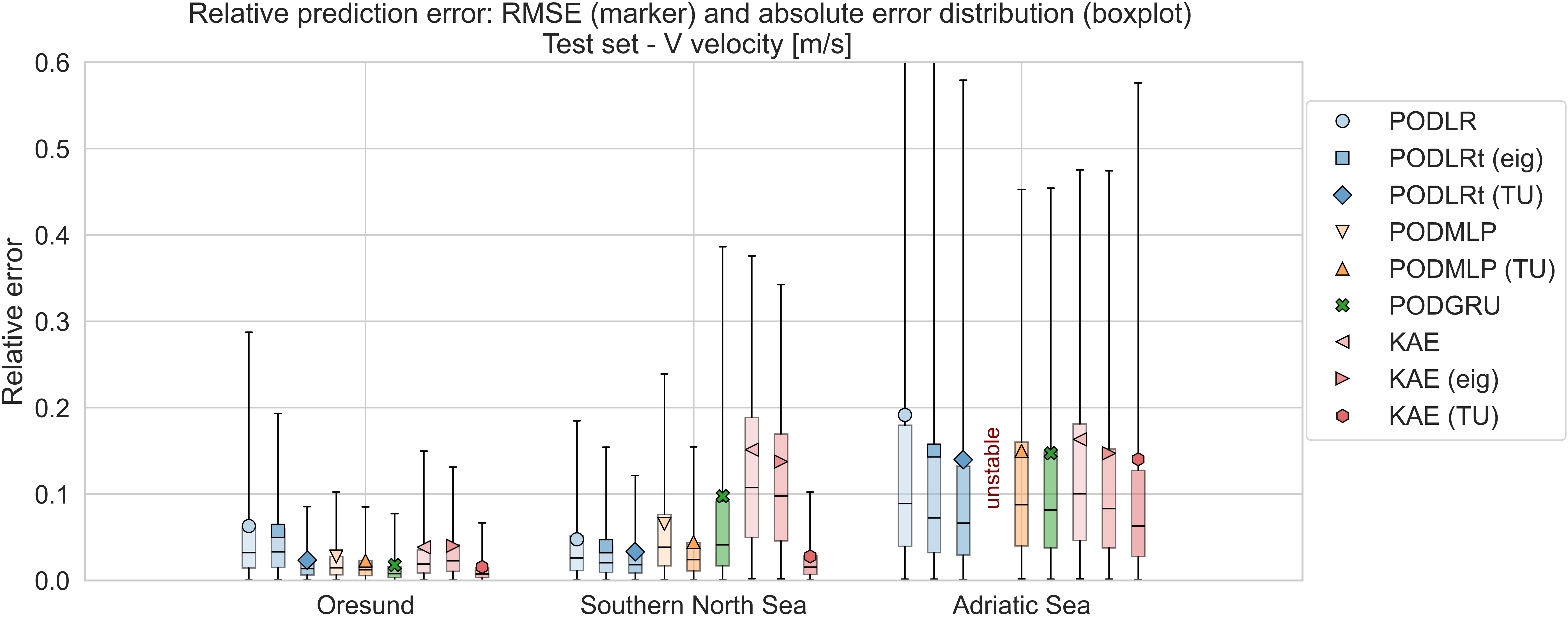}
\caption[Surrogate RMSEs]{Test errors of V component of current velocity relative to the range of values. The boxplots show the quantiles and median of absolute errors with 1st and 99th percentile whiskers. Markers are RMSE.}\label{fig:com_test_V}
\end{figure}

Figure \ref{fig:com_test_U} shows the relative RMSEs for all surrogate models for $U$, and Figure \ref{fig:com_test_V} shows the errors for $V$. KAE (TU) still performs best for Øresund and Southern Noth Sea, whereas the PODLRt (TU) performs slighly worse than the KAE(TU) for Adriatic Sea in terms of RMSE and percentiles up to the 75th. 

\subsection{Scatter plots}\label{app:scatter}
\begin{figure}[h]
    \centering
    \begin{subfigure}[t]{0.32\textwidth}
        \centering
        \includegraphics[width=1.0\textwidth]{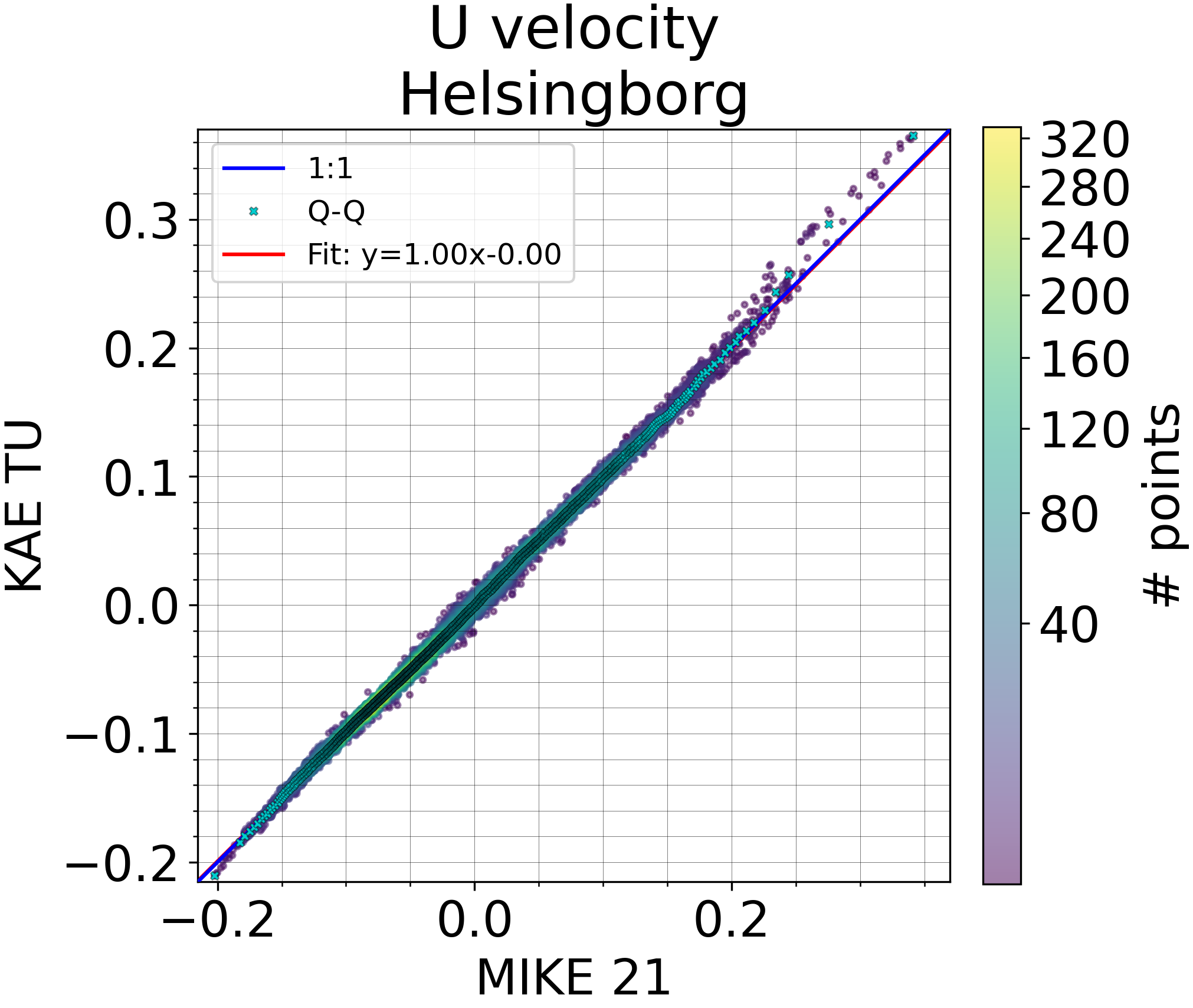}
    \end{subfigure}
    \begin{subfigure}[t]{0.32\textwidth}
        \centering \includegraphics[width=1.0\textwidth]{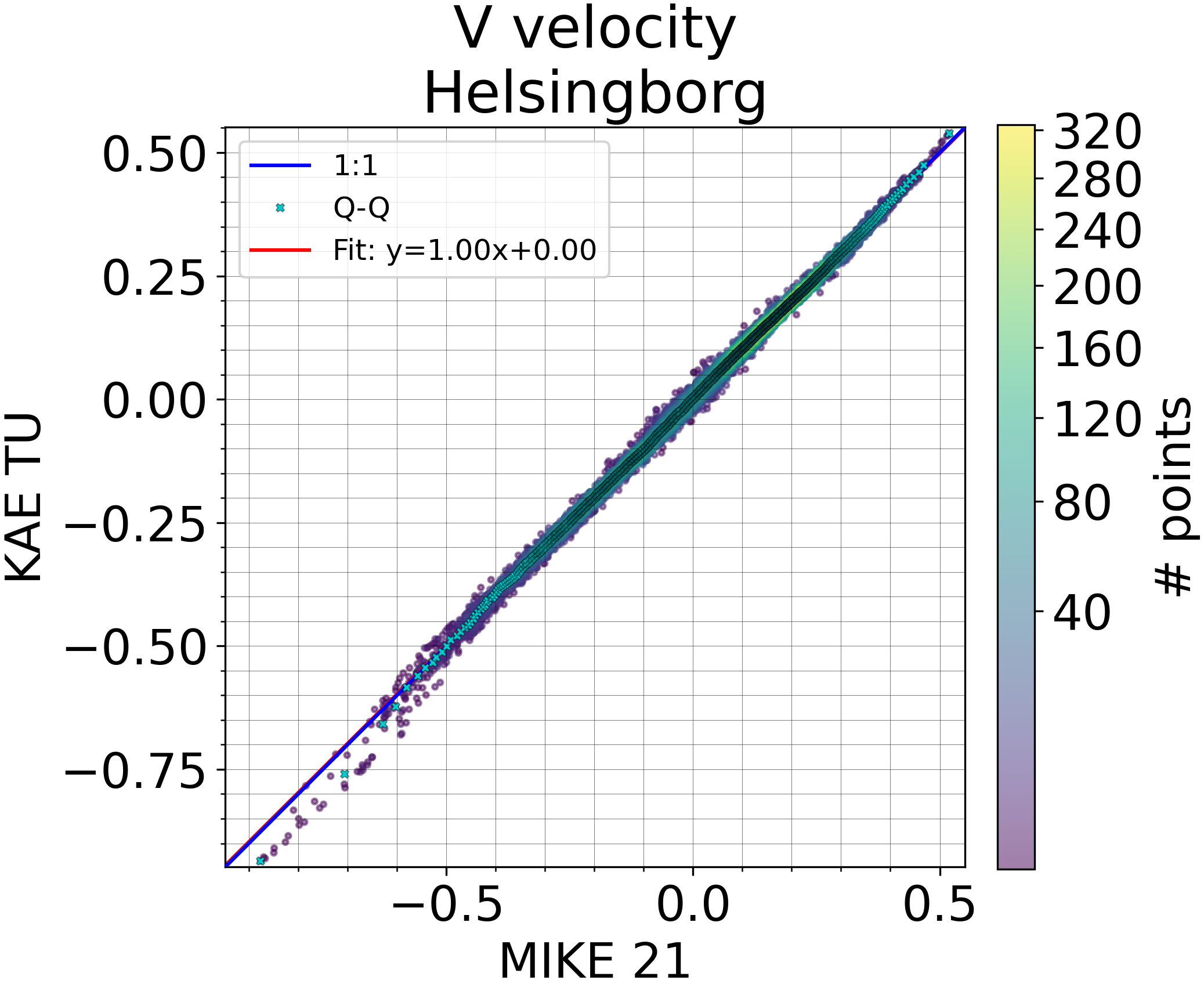}
    \end{subfigure}
    \caption[Scatter Øresund]{Scatter plots of the simulation data (x-axis) vs. the surrogate predictions (y-axis) in the test period. U and V for the Øresund case.}\label{fig:scatter_Oresund}
\end{figure}

\begin{figure}[h]
    \centering
    \begin{subfigure}[t]{0.32\textwidth}
        \centering
        \includegraphics[width=1.0\textwidth]{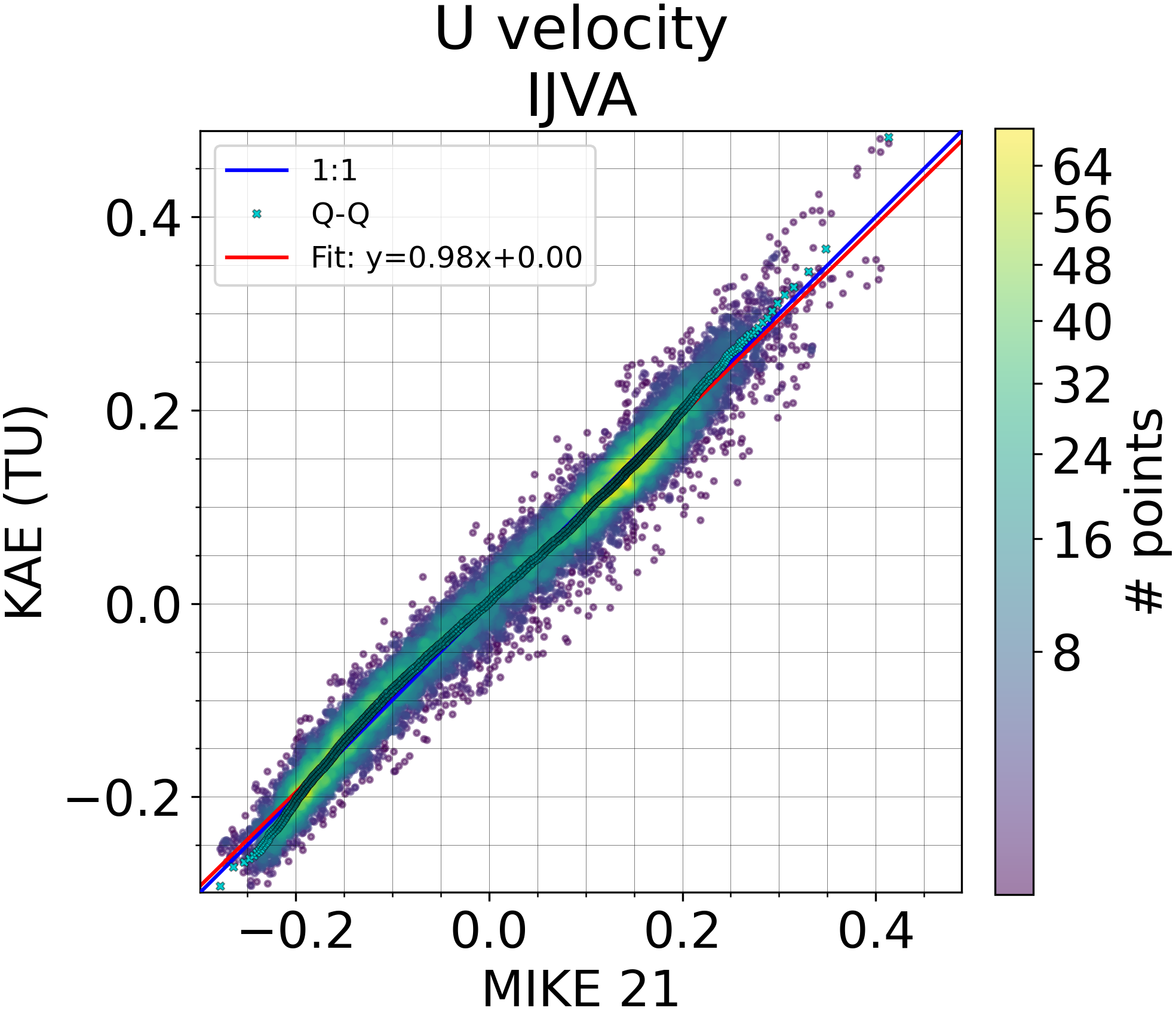}
    \end{subfigure}
    \begin{subfigure}[t]{0.32\textwidth}
        \centering \includegraphics[width=1.0\textwidth]{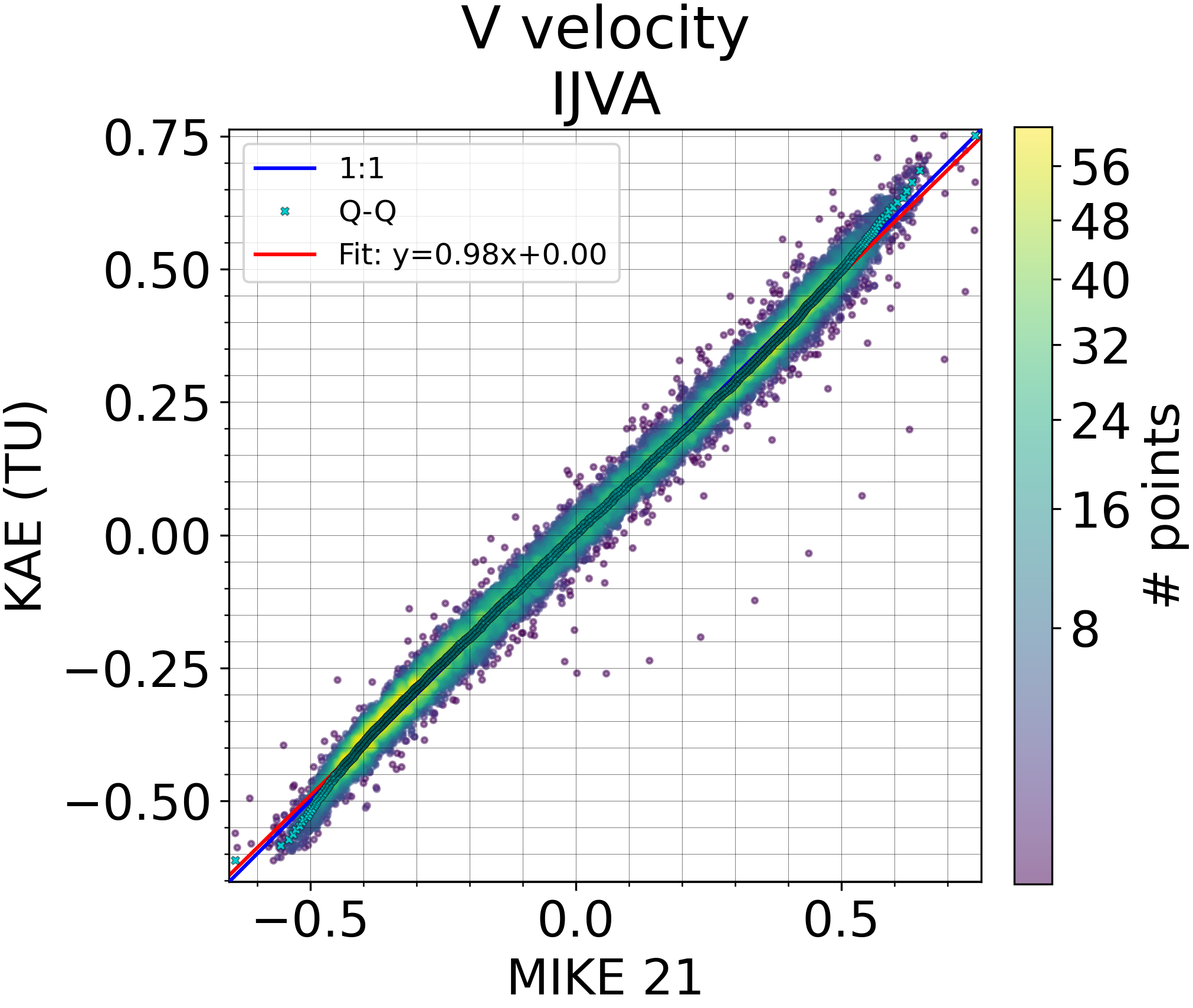}
    \end{subfigure}
    \caption[Scatter Southern North Sea]{Scatter plots of the simulation data (x-axis) vs. the surrogate predictions (y-axis) in the test period. U and V for the Southern North Sea case.}\label{fig:scatter_SNS}
\end{figure}

\begin{figure}[h]
    \centering
    \begin{subfigure}[t]{0.32\textwidth}
        \centering
        \includegraphics[width=1.0\textwidth]{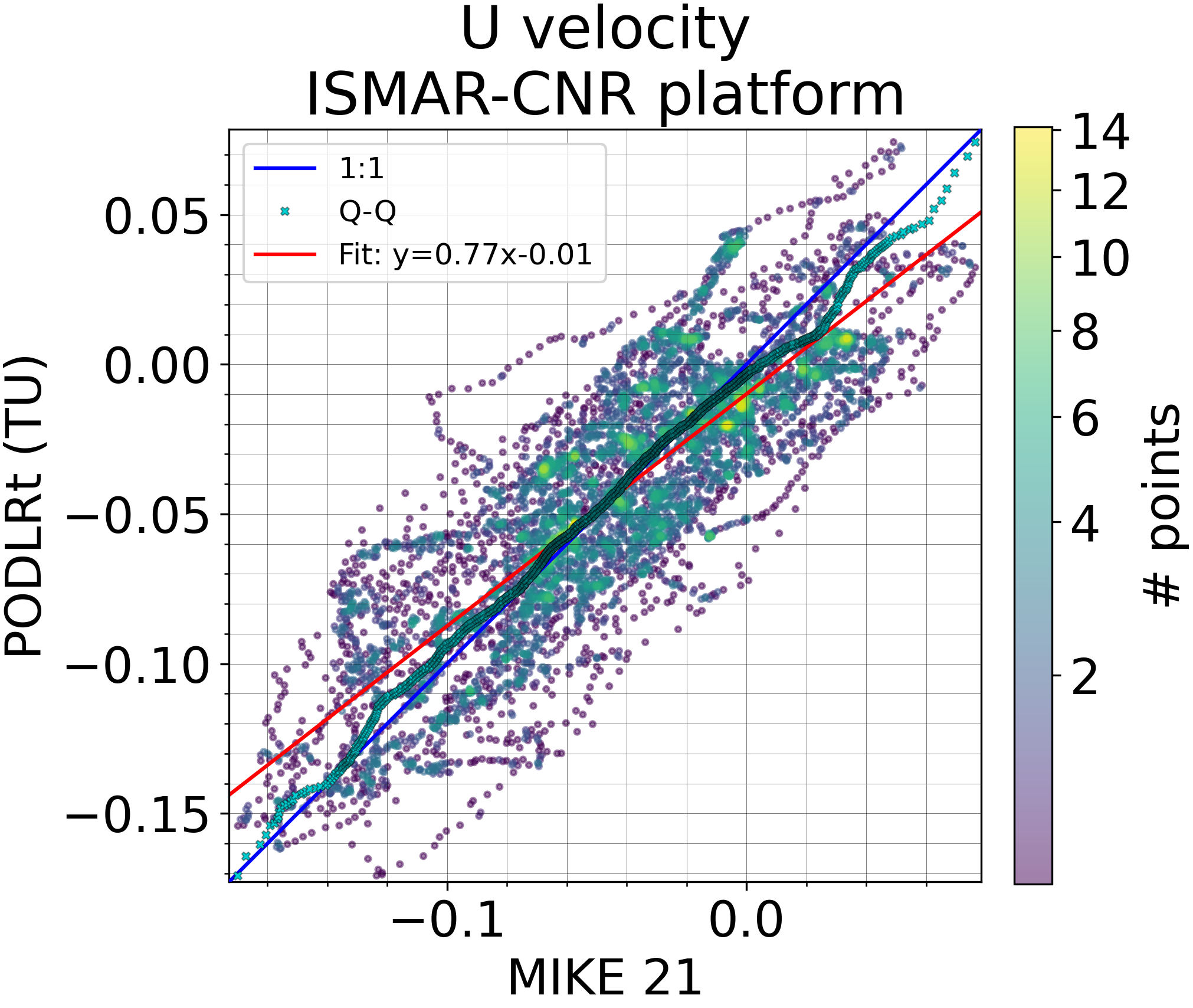}
    \end{subfigure}
    \begin{subfigure}[t]{0.32\textwidth}
        \centering \includegraphics[width=1.0\textwidth]{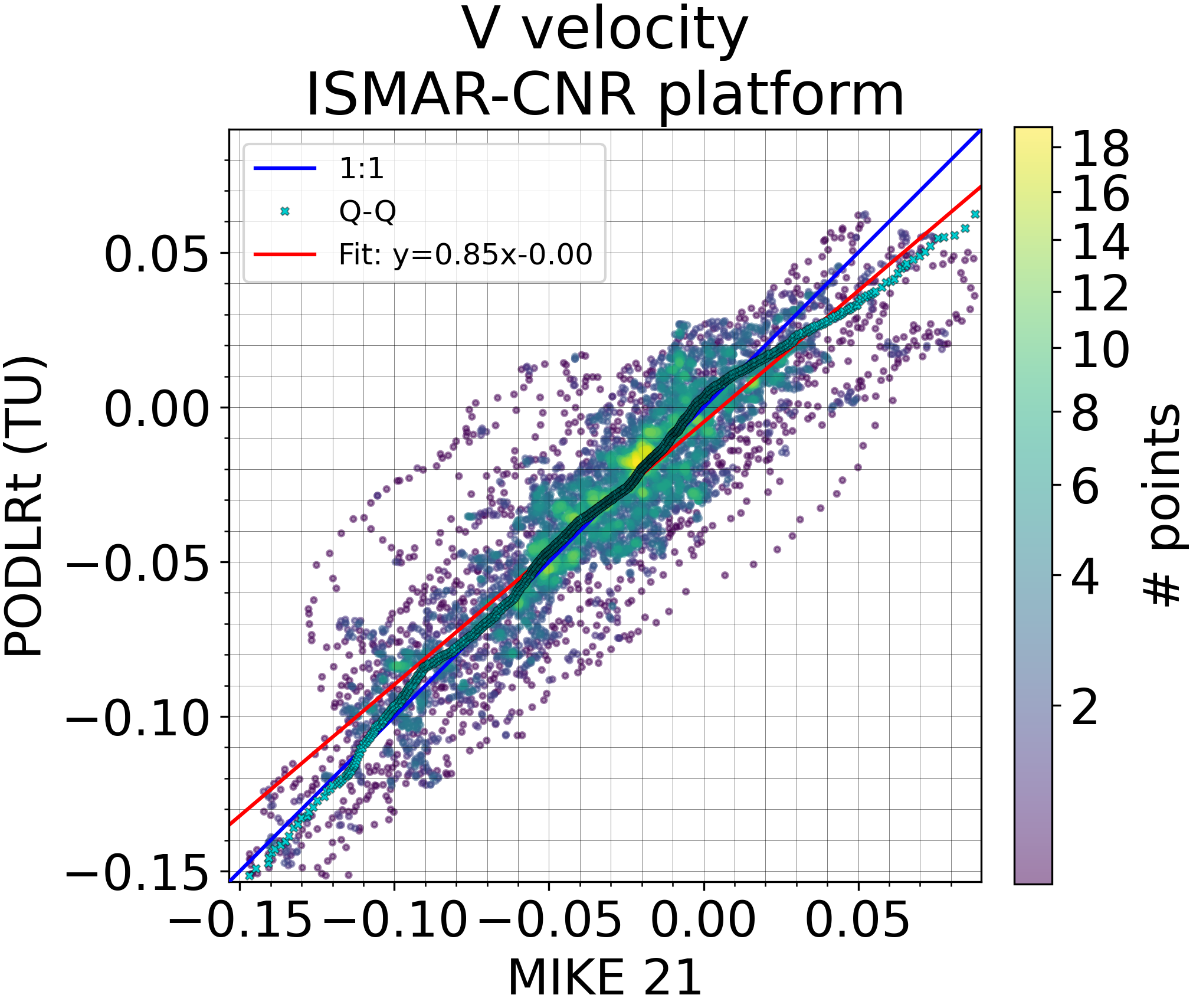}
    \end{subfigure}
    \caption[Scatter Adriatic Sea]{Scatter plots of the simulation data (x-axis) vs. the surrogate predictions (y-axis) in the test period. U and V for the Adriatic Sea case.}\label{fig:scatter_Adri}
\end{figure}

Figures \ref{fig:scatter_Oresund}, \ref{fig:scatter_SNS} and \ref{fig:scatter_Adri} show the scatter plots of the predictions in the test period for the U and V variables. The conclusions are mainly the same as for the surface elevations. Some differences include that the surrogate for Øresund slightly under-estimates large $U$-values and slightly over-estimates small $V$-values. For Adriatic Sea, the predictions are further from the 1:1 line, indicating a worse fit of the current velocities than of the surface elevations.

\subsection{Spatial RMSEs}\label{app:rmsemaps}
\begin{figure}[h]
    \centering
    \begin{subfigure}[t]{0.32\textwidth}
        \centering
        \includegraphics[height=3.2cm]{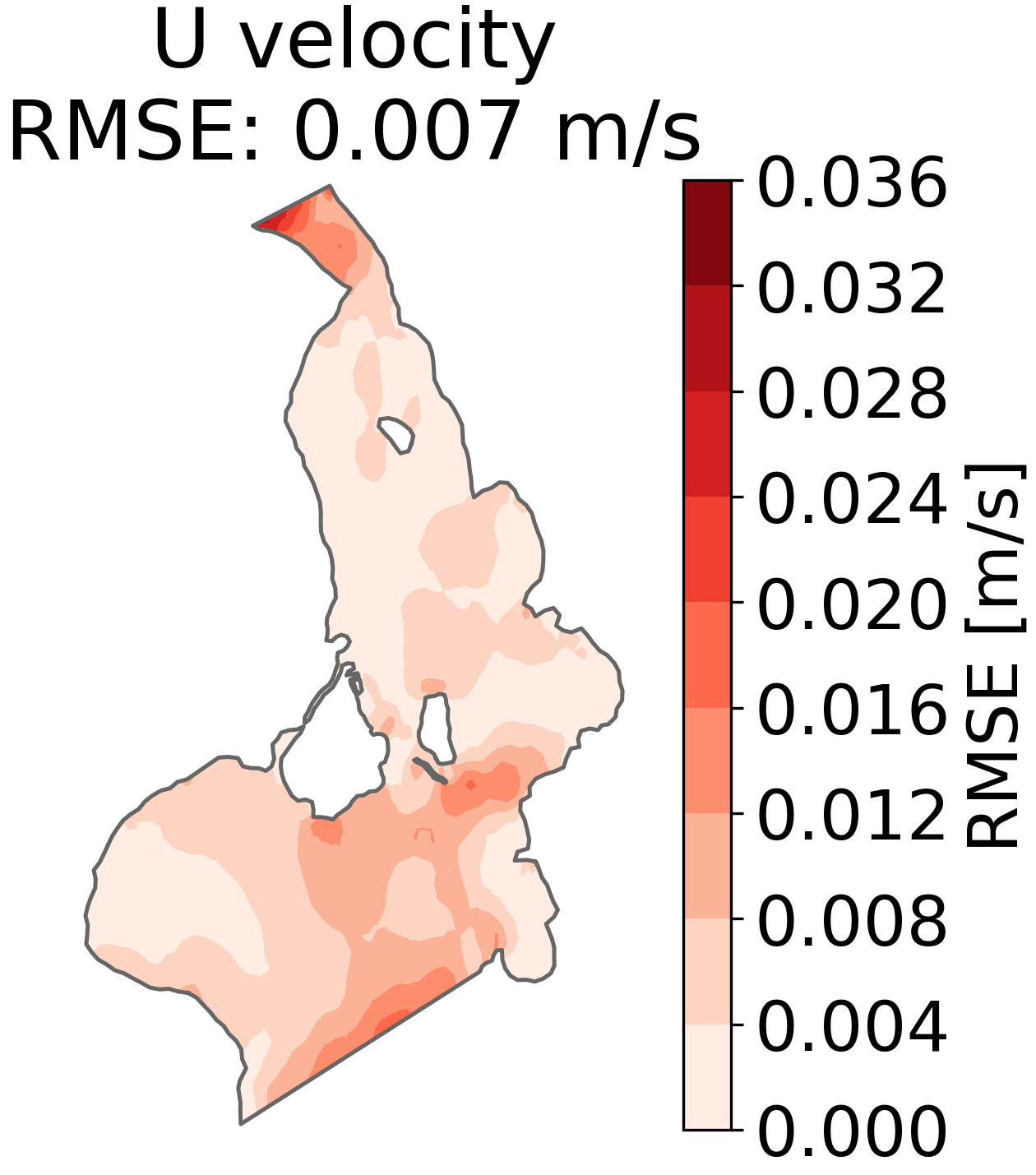}
        \caption{Øresund}
    \end{subfigure}
    \begin{subfigure}[t]{0.32\textwidth}
        \centering
        \includegraphics[height=3.2cm]{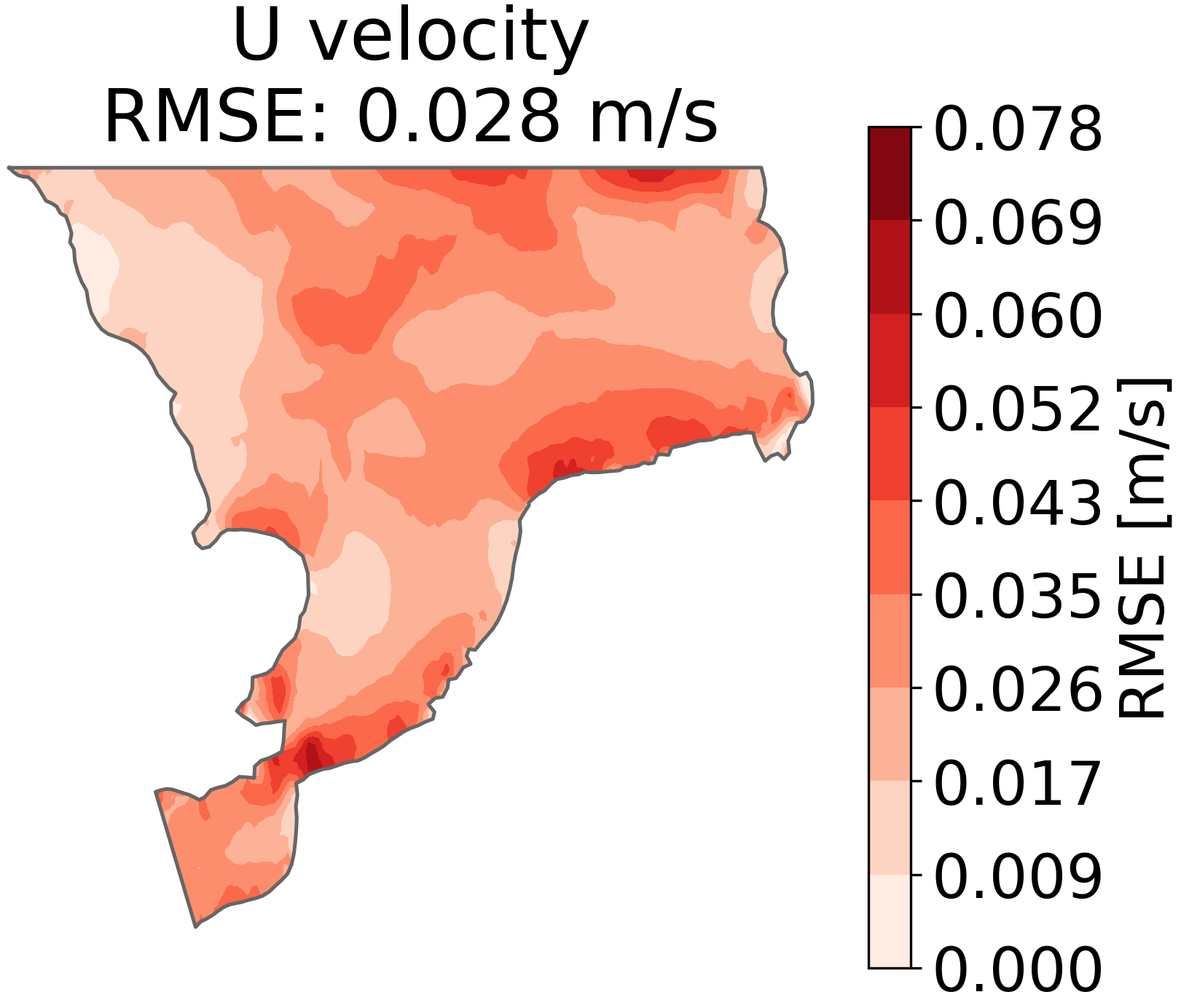}
        \caption{Southern North Sea}
    \end{subfigure}
    \begin{subfigure}[t]{0.32\textwidth}
        \centering
        \includegraphics[height=3.2cm]{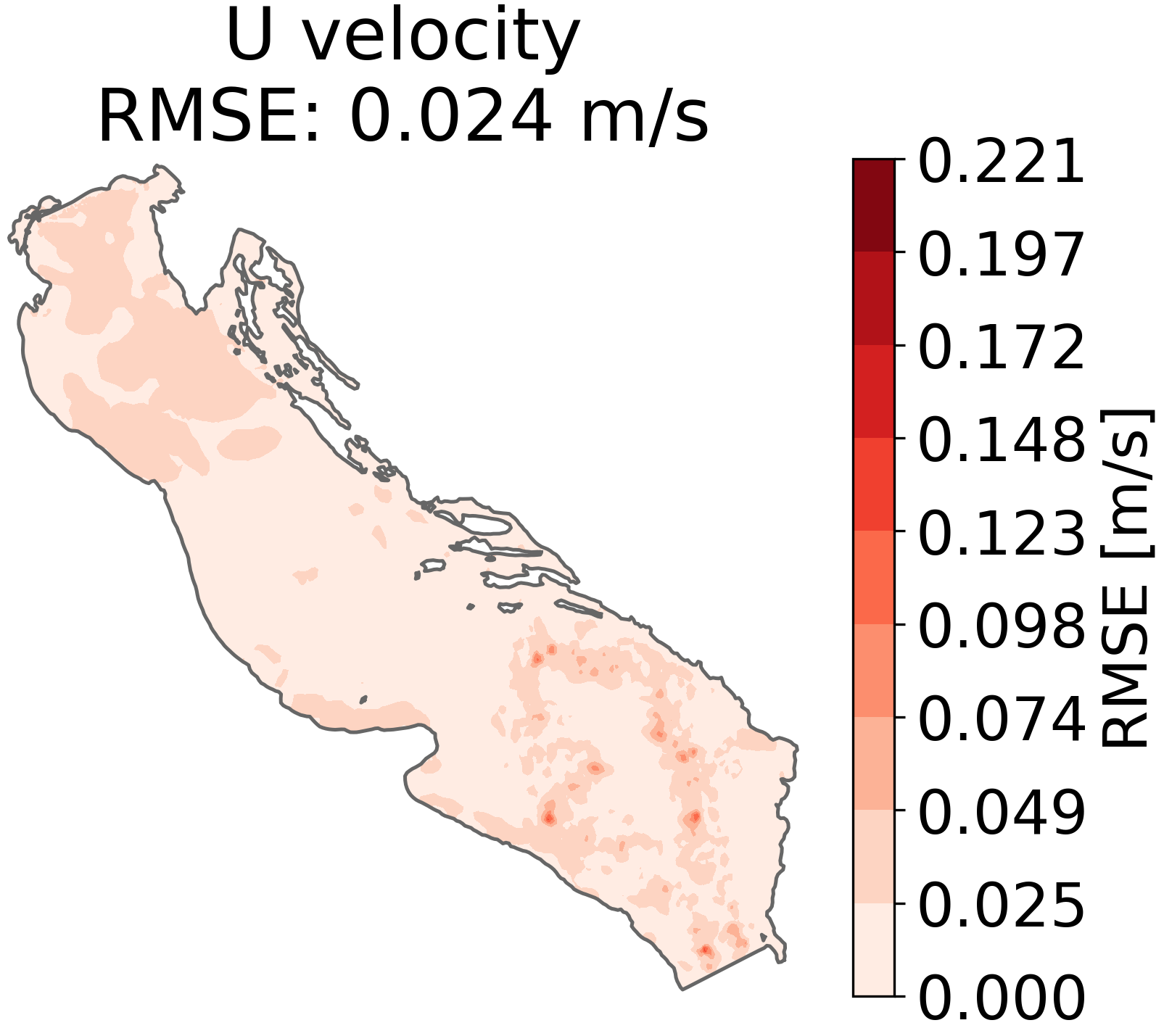}
        \caption{Adriatic Sea}
    \end{subfigure}
    \caption[U velocity RMSE map]{Spatial RMSEs across the test period of the u-component of the current velocities.}\label{fig:rmse_map_U}
\end{figure}

\begin{figure}[h]
    \centering
    \begin{subfigure}[t]{0.32\textwidth}
        \centering
        \includegraphics[height=3.2cm]{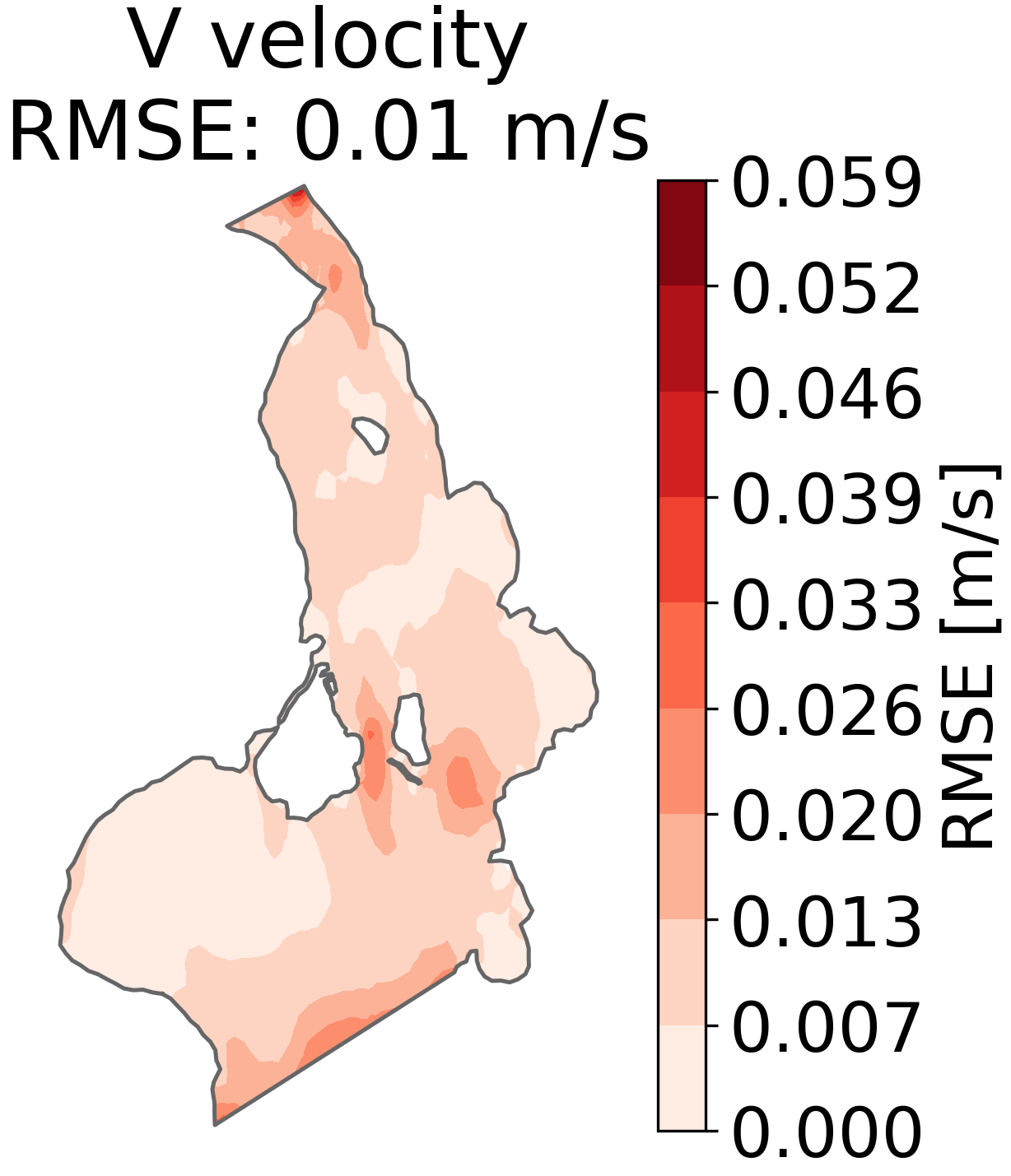}
        \caption{Øresund}
    \end{subfigure}
    \begin{subfigure}[t]{0.32\textwidth}
        \centering
        \includegraphics[height=3.2cm]{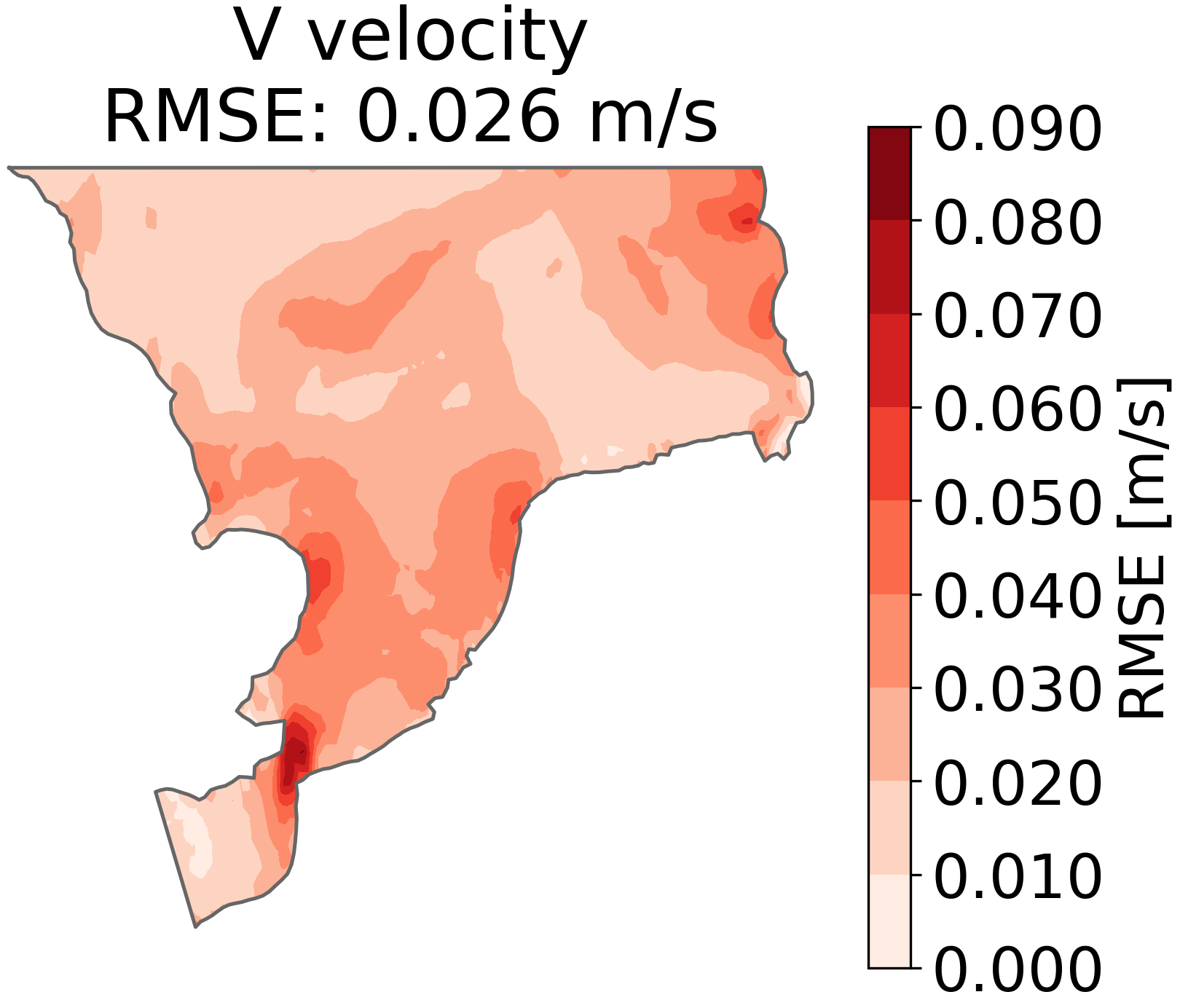}
        \caption{Southern North Sea}
    \end{subfigure}
    \begin{subfigure}[t]{0.32\textwidth}
        \centering
        \includegraphics[height=3.2cm]{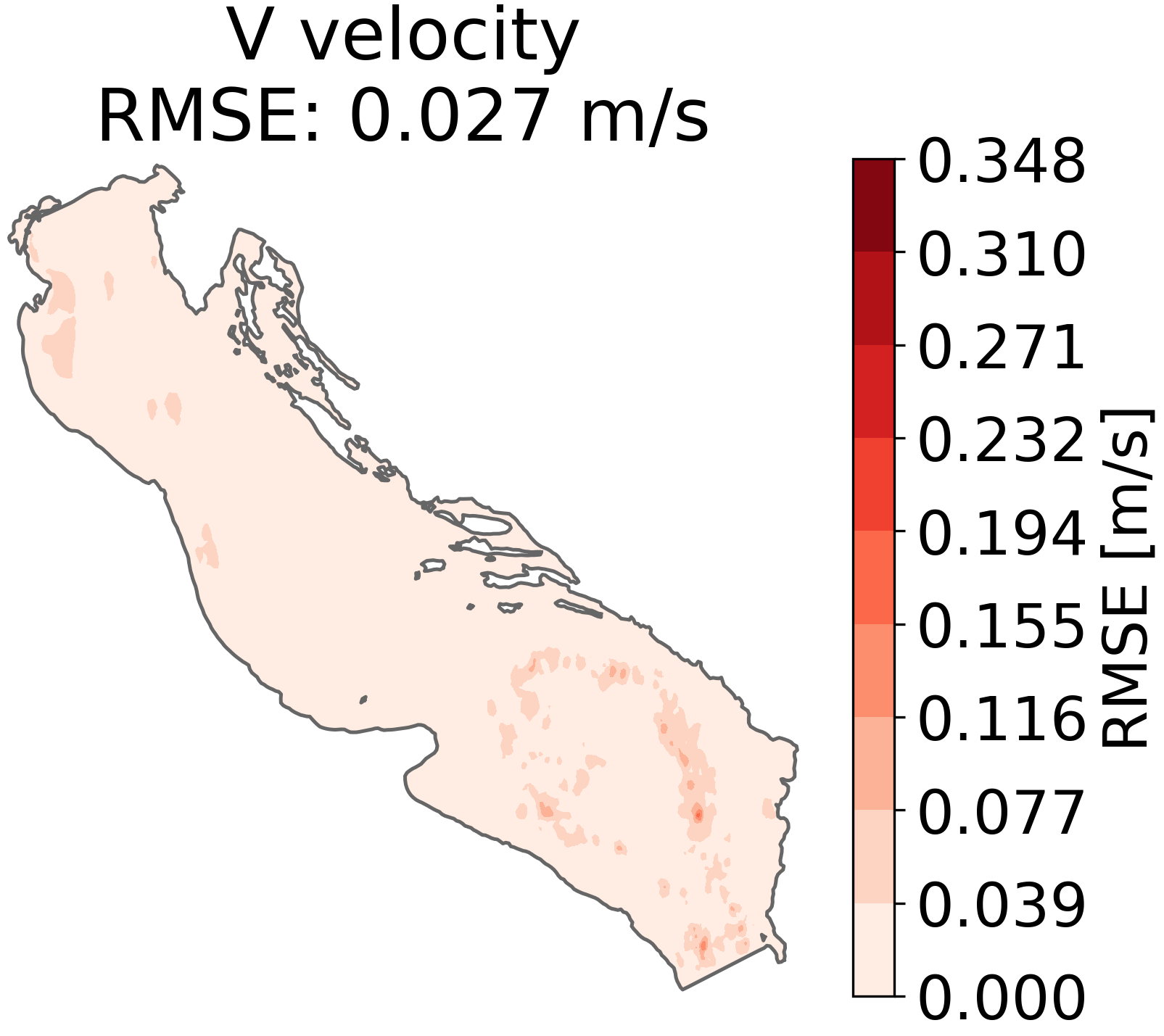}
        \caption{Adriatic Sea}
    \end{subfigure}
    \caption[V velocity RMSE map]{Spatial RMSEs across the test period of the v-component of the current velocities.}\label{fig:rmse_map_V}
\end{figure}

Figure \ref{fig:rmse_map_U} and \ref{fig:rmse_map_V} show the spatial RMSEs for U and V, respectively. The conclusions from these figures are similar to those of the surface elevations. 

\end{document}